\documentclass{egpubl}
\usepackage{eg2024}
\usepackage{comment}
\usepackage{booktabs} 
\usepackage{amsmath}
\usepackage{amsfonts}
\usepackage{bm}
\usepackage{tabularx}
\usepackage{subcaption}
\usepackage{multirow}
\usepackage{siunitx}
\usepackage{graphicx}
\usepackage{dfadobe} 
\usepackage{tikz}

\newcommand{\ud}{\,\mathrm{d}}

\newcolumntype{b}{>{\centering \arraybackslash}X}
\newcolumntype{s}{>{\centering \arraybackslash \hsize=.4\hsize}X}

\biberVersion
\BibtexOrBiblatex
\usepackage[backend=biber,bibstyle=EG,citestyle=alphabetic,backref=true,maxnames=20]{biblatex}
\electronicVersion
\PrintedOrElectronic

\usepackage{xcolor}
\usepackage{xparse,xcoffins}

\ExplSyntaxOn
\NewCoffin\imagecoffin
\NewCoffin\labelcoffin

\keys_define:nn { tg/label }
 {
  label   .tl_set:N = \l_tg_label_tl,
  labelbox .bool_set:N = \l_tg_label_box_bool,
  labelbox .default:n = true,
  fontsize .tl_set:N = \l_tg_label_size_tl,
  fontsize .initial:n = \footnotesize,
  pos .choice:,
  pos/nw .code:n = \tl_set:Nn \l_tg_label_pos_tl { left,up },
  pos/ne .code:n = \tl_set:Nn \l_tg_label_pos_tl { right,up },
  pos/sw .code:n = \tl_set:Nn \l_tg_label_pos_tl { left,down },
  pos/se .code:n = \tl_set:Nn \l_tg_label_pos_tl { right,down },
  pos/n .code:n = \tl_set:Nn \l_tg_label_pos_tl { hc,up },
  pos/w .code:n = \tl_set:Nn \l_tg_label_pos_tl { left,vc },
  pos/s .code:n = \tl_set:Nn \l_tg_label_pos_tl { hc,down },
  pos/e .code:n = \tl_set:Nn \l_tg_label_pos_tl { right,vc },
  pos .initial:n = nw,
  unknown .code:n   = \clist_put_right:Nx \l_tg_label_clist
                       { \l_keys_key_tl = \exp_not:n { #1 } }
 }
\clist_new:N \l_tg_label_clist
\box_new:N \l_tg_label_box
\box_new:N \l_tg_label_image_box

\NewDocumentCommand{\xincludegraphics}{O{}m}
 {
  \group_begin:
  \tl_clear:N \l_tg_label_tl
  \clist_clear:N \l_tg_label_clist
  \keys_set:nn { tg/label } { #1 }
  \tl_if_empty:NTF \l_tg_label_tl
   {
    \tg_includegraphics:Vn \l_tg_label_clist { #2 }
   }
   {
    \SetHorizontalCoffin\imagecoffin
     {
      \tg_includegraphics:Vn \l_tg_label_clist { #2 }
     }
    \SetHorizontalCoffin\labelcoffin
     {
      \raisebox{\depth}
       {
        \bool_if:NTF \l_tg_label_box_bool
         { \fcolorbox{white}{white}{\l_tg_label_size_tl\l_tg_label_tl} }
         { \l_tg_label_size_tl\l_tg_label_tl }
       }
     }
    \SetVerticalPole\imagecoffin{left}{3pt+\CoffinWidth\labelcoffin/2}
    \SetVerticalPole\imagecoffin{right}{\Width-3pt-\CoffinWidth\labelcoffin/2}
    \SetHorizontalPole\imagecoffin{up}{\Height-3pt-\CoffinHeight\labelcoffin/2}
    \SetHorizontalPole\imagecoffin{down}{3pt+\CoffinHeight\labelcoffin/2}
    \use:x{\JoinCoffins\imagecoffin[\l_tg_label_pos_tl]\labelcoffin[vc,hc]} 
    \TypesetCoffin\imagecoffin
   }
   \group_end:
 }
\NewDocumentCommand{\setlabel}{m}
 {
  \keys_set:nn { tg/label } { #1 }
 }

\cs_new_protected:Nn \tg_includegraphics:nn
 {
  \includegraphics[#1]{#2}
 }
\cs_generate_variant:Nn \tg_includegraphics:nn { V }

\ExplSyntaxOff
\title{Neural BSSRDF: Object Appearance Representation Including Heterogeneous Subsurface Scattering}
\author[Thomson TG et al.]
{\parbox{\textwidth}{\centering Thomson TG$^1$,
        Jeppe Revall Frisvad$^1$,
        Ravi Ramamoorthi$^2$,
        and Henrik Wann Jensen$^3$
}
\\
{\parbox{\textwidth}{\centering $^1$Technical University of Denmark\\
         $^2$University of California, San Diego\\
         $^3$Luxion
       }
}}
\begin{document}

\include{teaser}

\maketitle

\begin{abstract}
%
Monte Carlo rendering of translucent objects with heterogeneous scattering properties is often expensive both in terms of memory and computation. If the scattering properties are described by a 3D texture, memory consumption is high. If we do path tracing and use a high dynamic range lighting environment, the rendering easily becomes computationally heavy. We propose a compact and efficient neural method for representing and rendering the appearance of heterogeneous translucent objects. The neural representation function resembles a bidirectional scattering-surface reflectance distribution function (BSSRDF). However, conventional BSSRDF models assume a planar half-space medium and only surface variation of the material, which is often not a good representation of the appearance of real-world objects. Our method represents the BSSRDF of a full object taking its geometry and heterogeneities into account. This is similar to a neural radiance field, but our representation works for an arbitrary distant lighting environment. In a sense, we present a version of neural precomputed radiance transfer that captures all-frequency relighting of heterogeneous translucent objects. We use a multi-layer perceptron (MLP) with skip connections to represent the appearance of an object as a function of spatial position, direction of observation, and direction of incidence. The latter is considered a directional light incident across the entire non-self-shadowed part of the object. We demonstrate the ability of our method to store highly complex materials while having high accuracy when comparing to reference images of the represented object in unseen lighting environments. As compared with path tracing of a heterogeneous light scattering volume behind a refractive interface, our method more easily enables importance sampling of the directions of incidence and can be integrated into existing rendering frameworks while achieving interactive frame rates.
\end{abstract}


\section{Introduction}

Many rendering techniques do not capture the bleeding of light through a translucent object. Surface scattering models like those based on the bidirectional reflectance distribution function (BRDF) are local and cannot capture the effect; the models by Fan et al.~\cite{is_neuralbrdf} and Zeltner et al.~\cite{zeltner2023real} are recent examples. Volume path tracing is needed to capture how light entering the object at one surface location scatters and emerges at another. A bidirectional texture function (BTF)~\cite{rainer2020unified,kuznetsov2021neumip} is able to capture some subsurface scattering because it includes two spatial dimensions for surface texturing. A BTF is however meant to model a surface patch and not the global bleeding of light through an object. The bidirectional scattering-surface reflectance distribution function (BSSRDF) includes the full scattering of light incident at one location to light emerging in another object surface location. This function captures the same effects as full volume path tracing except that we cannot place the camera inside the object. Volume path tracing is a local formulation where we consider each surface and subsurface scattering event; the BSSRDF is a global formulation considering only incident and emergent light~\cite{preisendorfer2014radiative}.

When using a BSSRDF, we most commonly employ an approximate analytic model~\cite{jensen_01}. Models exist that generalize to include dependency on the directions of incidence and observation~\cite{frederickx2017forward} or on the local object geometry~\cite{vicini2019}. These models are however for rendering of homogeneous translucent materials.

Rendering of a heterogeneous translucent object is computationally expensive because the scattering events happen behind a refractive interface. This means that rays cannot be traced directly toward the light making importance sampling more difficult when volume path tracing is used for rendering~\cite{deng2020practical}. An often used solution is to approximate the subsurface scattering using a factored BSSRDF representation~\cite{peers_06,kurt2021gensss} or by solving the diffusion equation for the object volume~\cite{wang2008modeling,wang2010real,arbree_11}. These approaches however rely on approximate representations of the surface and subsurface scattering.

An alternative way to capture the appearance of a translucent object is by precomputed radiance transfer (PRT)~\cite{sloan2005local,laurent_22}, but PRT methods assume low-frequency variation of the radiance as a function of the directions of incidence and observation. Our method is conceptually similar to PRT, but we provide a neural representation of the subsurface scattering that adapts to an arbitrary distant lighting environment which is not necessarily of low frequency. We also do not assume that the subsurface scattering leads to diffuse emergent light, which is a common assumption in PRT models based on analytic BSSRDF models~\cite{wang2005all}.

Another approach to capturing the full appearance of a translucent object is by using a neural radiance field (NeRF)~\cite{mildenhall2021nerf}. This is however a representation of the object appearance in one particular lighting environment. Derived methods~\cite{boss_nerd_21,zhangnerf_21} enable arbitrary incident illumination by jointly optimizing a model for shape, BRDF, and illumination. By assuming a BRDF model, these methods however sacrifice their ability to represent the appearance of a translucent object. A method for relighting of a captured NeRF has been achieved by assuming a specific type of lighting environment~\cite{li2022neulighting}, in this case an outdoor sun and sky lighting model was assumed.

We represent the appearance of a heterogeneous translucent object by assuming known surface geometry and distant illumination. We care only about light emerging at the surface of the object and therefore use the global BSSRDF representation of the appearance. Our BSSRDF depends on the full object geometry, not only one point of incidence. In this way, we obtain a method capable of capturing the appearance of a heterogeneous translucent object from an arbitrary view and illuminated by an arbitrary distant lighting environment. Our use of the global formulation makes it easier for us to importance sample the incident illumination as we do not need to evaluate the light reaching scattering events behind a refractive interface. We demonstrate good rendering performance both in terms of accuracy and rendering time as compared with volume path tracing.
We compare the accuracy of our model with that of classic appearance representation in spherical harmonics as well as factored NeRF~\cite{zhangnerf_21} and a neural BTF model~\cite{kuznetsov2021neumip}.

\section{Theory}


For a translucent object, the radiance leaving a point of observation $\bm{x}_o$ in the direction $\vec{\omega}_o$ is given by the reflected radiance equation~\cite{preisendorfer2014radiative,jensen_01}
\begin{equation}
\label{Render_equation}
L_r(\bm{x}_o, \vec{\omega}_o) = 
\int_{A_i} \int_{2\pi} S(\bm{x}_i, \vec{\omega}_i; \bm{x}_o, \vec{\omega}_o) L_i(\bm{x}_i, \Vec{\omega}_i) \cos\theta_i \ud\omega_i \ud{A_i} \,,
\end{equation}
where $L_i$ is the incident radiance from the direction $\vec{\omega}_i$ at position $\bm{x}_i \in A_i$, so that the angle of incidence is $\theta_i$, and $S$ is the BSSRDF. We let $A_i$ denote the surface area of the object, and we use arrow overline for unit length direction vectors (e.g.~$\vec{\omega}_o$). We can evaluate this expression (\ref{Render_equation}) if we have a known BSSRDF or by a random walk (volume path tracing) if we have an object with known surface and volume scattering properties~\cite{preisendorfer2014radiative,pharr2000monte,frisvad2020survey}.

Assuming distant illumination, incident radiance is the same for positions in the object surface that are not in shadow. Letting $V$ denote visibility, we have
\begin{equation}
L_i(\bm{x}_i, \Vec{\omega}_i) = V(\bm{x}_i, \Vec{\omega}_i)L_i(\vec{\omega}_i) \,,
\end{equation}
and we can write up a BSSRDF for distant illumination:
\begin{equation} \label{eq:distantS}
\hat{S}(\vec{\omega}_i; \bm{x}_o, \vec{\omega}_o) = \int_{A_i} S(\bm{x}_i, \vec{\omega}_i; \bm{x}_o, \vec{\omega}_o) V(\bm{x}_i, \Vec{\omega}_i) (\vec{\omega}_i\cdot\vec{n}_i) \ud{A_i} \,,
\end{equation}
where $\vec{n}_i$ is the surface normal at $\bm{x}_i$. Assuming $A_i$ to be the surface of a closed object, light can be incident from all directions. We thus need to change our integration in Eq.~\ref{Render_equation} to be across the unit sphere ($4\pi$ solid angles), and we note that when the cosine term is negative we always have zero visibility ($\vec{\omega}_i\cdot\vec{n}_i < 0 \Rightarrow V = 0$). The reflected radiance equation~(\ref{Render_equation}) then becomes
\begin{equation} \label{eq:distrefrad}
\boxed{
 L_r (\bm{x}_o, \vec{\omega}_o) = \int_{4\pi} L_i(\vec{\omega}_i) \,\hat{S}(\vec{\omega}_i; \bm{x}_o, \vec{\omega}_o) \ud{\omega_i} \,.}
\end{equation}
With an MLP representation of $\hat{S}$, we can use this equation for computing reflected radiance directly as a shader in a rendering system.

\subsection{Neural BSSRDF}
Let $\psi$ denote the parameters of a multilayer perceptron (MLP) $f_\psi$. The purpose of the MLP is to represent the appearance of a translucent object. To this end, we optimize the parameters to account for a vast set of observations captured in rendered images illuminated by directional lights of unit radiance. A directional light of direction $\vec{\omega}_e$ illuminates the object in each image.
Letting $\bm{x}_{o,k},\vec{\omega}_{o,k}$ denote the position and direction of observation for a ray through pixel $k$, we find parameters $\psi*$ for the MLP to represent the observation by
\begin{equation}
    \psi* = \arg\min_\psi \mathbb{E}_k[(f_\psi(-\vec{\omega}_e; \bm{x}_{o,k}, \vec{\omega}_{o,k}) - L_{r,k})^2] \,,
\end{equation}
where $L_{r,k} \approx \hat{S}(-\vec{\omega}_e; \bm{x}_{o,k}, \vec{\omega}_{o,k})$ is the reflected radiance observed along the ray through pixel $k$. To make the network useful for rendering with arbitrary placement of the object in a scene, the position and direction of observation is given to the network in object space coordinates. The MLP $f_{\psi*}$ will serve as a representation of $\hat{S}$ in Eq.~\ref{eq:distrefrad}.

Our concept is a variation of neural PRT~\cite{rainer2022neural}, but we do not separate our model into a single environment encoded in spherical harmonics and a transport MLP which are combined using a learned operator. Instead, our model is designed for rendering of a translucent object in arbitrary environments, and we include the direction of a distant source directly as an input for the network. This eases the integration of our model into rendering of the represented translucent object in a scene where we can assume that sources of incident illumination are distant. 

The idea of using a factored transport in PRT is similar to our method. Our representation, like theirs, is decomposed into two components, specular and subsurface. We use traditional algorithms for the specular component and our neural BSSRDF for the subsurface scattering approximation.

\subsection{Importance Sampling}
\label{importance_sampling}
When using our model in a renderer, we evaluate Eq.~\ref{eq:distrefrad} with our neural BSSRDF $f_{\psi*}$ in place of $\hat{S}$. For Monte Carlo integration of Eq.~\ref{eq:distrefrad}, we sample directions of incidence $\vec{\omega}_i$ according to a probability density function $\mathrm{pdf}(\vec{\omega}_i)$. The $N$th sample estimator of the integral is then
\begin{equation}
L_{r}(\bm{x}_o, \vec{\omega}_o) \approx \frac{1}{N}\sum_{j=1}^N \frac{L_i(\vec{\omega}_{i,j}) \,f_{\psi*}(\vec{\omega}_{i,j}; \bm{x}_o, \vec{\omega}_o)}{\mathrm{pdf}(\vec{\omega}_{i,j})} \,,
\end{equation}
and importance sampling is performed by choosing the pdf. One option is importance sampling of the $L_i$ term by importance sampling an environment map~\cite{pharr2023physically}, for example. This works well with our technique, but it has the drawback of sampling directions that are not necessarily important for the object appearance in a given point of observation $\bm{x}_o$. As an alternative, we would like an importance sampling technique with a pdf roughly proportional to the value of the neural BSSRDF $\hat{S}$ for different directions of incidence across the unit sphere.

As in existing work on importance sampling for neural representations of spatially varying BRDFs~\cite{is_neuralbrdf}, we sample the direction of incidence $\vec{\omega}_i$ using spherical coordinates ($\theta_i,\phi_i$) and a normalized 2D Gaussian kernel with standard deviation $\sigma$ for the pdf. To simplify the importance sampling module, we assume that the normal direction in object space is the most important and use a Gaussian kernel with zero mean. The main difference from importance sampling for a spatially varying BRDF is that light incident from any direction on the unit sphere can be important ($\theta \in [0,\pi]$). We use the Box-Muller transform~\cite{howes2007efficient} to sample direction vectors according to the $\sigma$ retrieved from the network.

\begin{figure}
    \includegraphics[width=\linewidth, trim=2cm 17.8cm .5cm 2cm, clip]{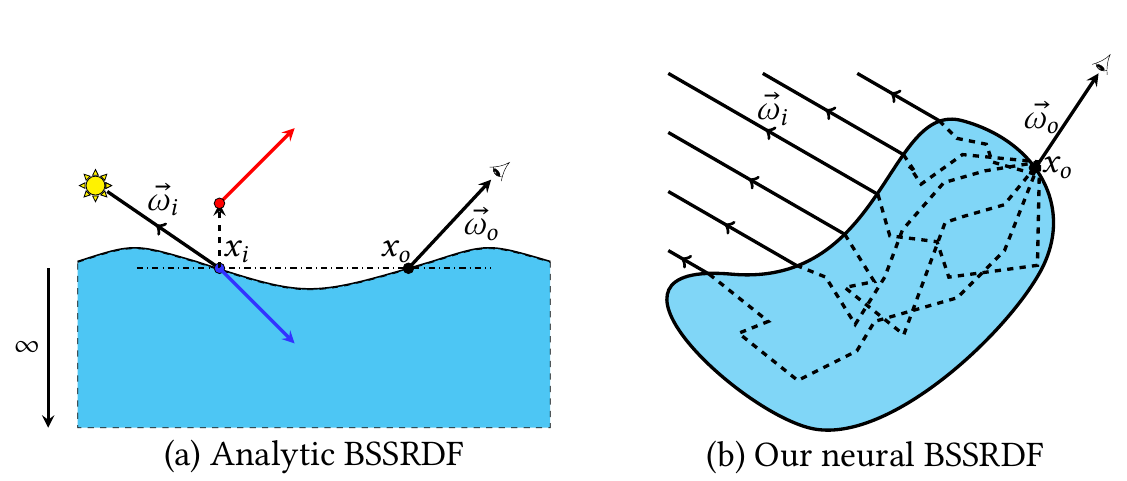} \\[-3.5ex]
    \caption{
        The technical difference between conventional analytic BSSRDF models and our neural BSSRDF model. Our model estimates the full appearance of an object illuminated by a directional light, in contrast to the conventional assumption of a ray incident on a half-space medium. This enables us to represent objects with intricate heterogeneous surface and subsurface scattering with high accuracy. Conventional models would have to violate many assumptions to represent such objects.
    }
    \label{directional}
\end{figure}

%

 
\section{Implementation}
\label{Neural Materials}



        
        Our neural BSSRDF is an MLP representation of $\hat{S}$ (Eq.~\ref{eq:distantS}) that takes as input $\vec{\omega}_i, \vec{\omega}_o, \bm{x}_o$ (7-vector input: 2 dimensions for each direction in spherical coordinates and 3 for the position in Cartesian coordinates) and returns relative outgoing radiance (3-vector if RGB)
        for unit radiance incident at every object surface position (see Figure~\ref{directional}). The result is then scaled to the contribution from a distant light to get the contribution in a Monte Carlo renderer. Our neural BSSRDF only captures the subsurface scattering. Surface reflection effects (Fresnel effects) are computed separately.

\begin{figure}
            \includegraphics[width =.75\linewidth,  trim = 3.8cm 2cm 8.8cm 2cm]{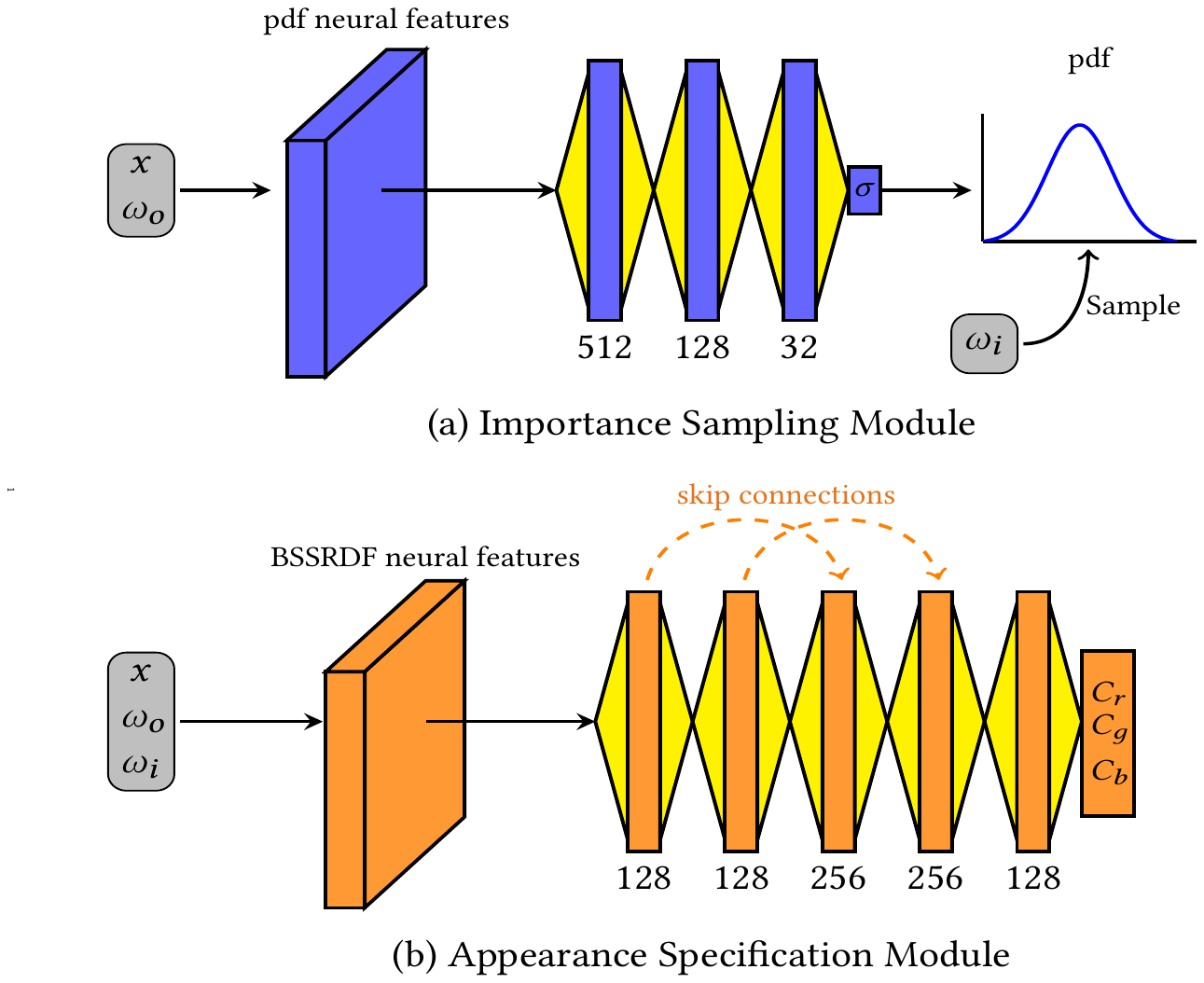}
            \caption{Our Network Architecture consists of 2 parts: Importance Sampling Module and Appearance Specification Module. }
            \label{network_architecture}
        \end{figure}

        For a given point of observation $\bm{x}_o$, light incident along the normal $\vec{n}_o$ would often have high importance when we only consider subsurface scattering. This is due to Fresnel reflectance increasing with increasing angle of incidence so that less light refracts into the medium for more oblique incidence. We thus define our Gaussian lobe in the tangent space of $\bm{x}_o$ and make it symmetric around the normal direction. To improve the expressivity of the function, we use the symmetric generalized Gaussian distribution function~\cite{varanasi1989parametric}.
        In practice, we uniformly sample $32\times64$ directions (32 inclination $\theta_i$ by 64 azimuthal $\phi_i$ angles) across the unit sphere for every position and direction of observation ($\bm{x}_o, \vec{\omega}_o$) in the neural BSSRDF, see Figure~\ref{network_architecture}. We convert this into a distribution and minimise the distance between the generated distribution and our pdf representation.

    \subsection{Network Architecture}
        \label{architecture}
        
        Our architecture consists of two major components: an importance sampling module and an appearance specification module, see Figure~\ref{network_architecture}.
        Due to the known inability of MLPs to learn high frequency functions in low dimensional domains~\cite{tancik2020fourier} and due to the bias towards low-frequency functions, we cast our problem into a higher dimensional space before passing it to the MLP. Both the modules therefore have two components: (a) feature extraction and (b) neural network. The feature extraction module is similar in both the modules where we use Fourier feature maps. Our Fourier feature map is defined by
        \begin{equation}
            f(x) = [ \cos(2\pi G x), \sin(2\pi G x)]^T\,,
        \end{equation}
        where $G$ is a matrix of size $256 \times n$ representing a 2D Gaussian distribution with a mean of $\mu=18$ and standard deviation of $\sigma=1$. Note that being a randomly sampled isotropic distribution, the Gaussian $G$ performs frequency modulation on the feature map. Furthermore, the standard deviation ($\sigma$) is a significant factor in extracting the features, however $\sigma = 1$ worked well in all our tests, presumably because it sustains the bell curve of the distribution.

        The neural network in the importance sampling module is a 3-layers MLP with 512, 128 and 32 neurons respectively with ReLU (Rectified Linear Unit) activations between each layer. This module is responsible for predicting the standard deviation ($\sigma$) for the Gaussian distribution on the appearance specification module. Since the aim is to reduce the difference of two distributions, we use Kullback-Leibler divergence (KLD) loss for fitting our distribution prediction. Furthermore, we normalize the spherical coordinates for out of bound values while drawing samples.

        The appearance specification module, on the other hand, has a 5-layers MLP with 128 neurons each in every layer with two skip connections: from layer 1 to 3 and layer 2 to 4, respectively. The skip connections reduce the memory footprint of the network to approximately 649KB on disk, where the weights are well optimized at every layer in the neural architecture. We used ReLU as activation after every layer, except for the last layer after which we used no activation, and we used L2 as the loss function between our prediction and ground truth.
        
    \subsection{Training and Dataset}
        \label{training}
        Our model is trained on data produced by path tracing a scene consisting of the object of interest and a directional light source with unit radiance. The direction of the light source is randomly changed and the camera is rotated and translated at a uniform distance from the center of the geometry. To guarantee convergence, each rendered image had a high sample count. We also filtered the data to only include ray hits on the geometry.

        We use an online learning method to train our model for the appearance specification module. In this method, the renderer and the neural network run in parallel. The task of the renderer is to keep rendering data at all times. When new data becomes available, the renderer pushes it into the buffer used by the network for learning. We use the running L2 loss and the L2 loss on a small test set (produced prior to training) to check the network convergence. Another approach to do the same would be an offline method which essentially means that we collect and store the necessary data into the disk and train the model independently. This can be done with a general purpose renderer like Mitsuba \cite{nimier2019mitsuba} or PBRT \cite{pharr2023physically}. For quantitative analysis, we have tested an offline learning method where we collect data from a renderer and train the model as a secondary step. We keep 5\% of the data as test set and another 5\% was kept for validation. We found that the collected data $\hat{S}(\vec{\omega_i}; \bm{x}_o, \vec{\omega_o})$ from approximately $\sim$835 images was adequate to have the network represent a heterogeneous medium. Note that the images do not directly correspond to the amount of data acquired. Every view might have a different number of hits on the geometry.

        The importance sampling module is trained on top of the appearance specification module. Thus, it is considered to be in the post-processing stage, and since it is trained on inference data from the appearance specification module, no explicit data are generated to train this module. We randomly take the position $\bm{x}_o$ on the geometry, $\vec{\omega}_o$ on the hemisphere along with a batch of $32 * 64$ uniformly distributed $\vec{\omega}_i$ on the sphere. The resultant data are collected into bins which form our ground truth distribution. This distribution is expected to be approximately Gaussian (discussed in Sec.~\ref{importance_sampling}), why we fit a network that predicts the standard deviation ($\sigma$) of the resultant distribution.

    \subsection{Precomputed Radiance Transfer (PRT)}
        \label{PRT}
        By employing a straightforward MLP, we demonstrate that our neural representation can be executed interactively on a modern graphics processing unit (GPU). Furthermore, our approach involves simplifying the traditional BSSRDF by reducing its dimensionality to be on par with a BRDF function. This makes our model extensible to a wide range of applications. To demonstrate the effectiveness of our methodology, we use a spherical harmonics-based PRT technique and obtain real-time rendering. We consider this a variant of our material representation, which achieves real-time performance at the cost of accuracy. Traditional PRT techniques rely on efficient sampling of spherical functions such as environment illumination and diffuse irradiance to construct signals in the form of basis vectors and coefficients. Due to the lower dimensionality requirements, PRT has not so far been very effective in addressing higher dimensional problems, such as a BSSRDF. 
        
        The lighting coefficients are computed at run-time in the standard way, 
        \begin{equation}
        b_l^m = \int_{4\pi} L_i(\vec{\omega}_i) Y_l^m(\vec{\omega}_i)\,\ud\omega_i \,,
        \end{equation}
        where $L_i$ is the incident environment map and $Y_l^m$ are the (real) spherical harmonic basis functions.

        We also precompute the spherical harmonic coefficients of the light transport in terms of the neural BSSRDF, 
        \begin{equation}
        a_l^m(\bm{x}_o, \vec{\omega_o}) = \int_{4\pi} \hat{S}(\vec{\omega_i};\bm{x}_o;\vec{\omega}_o) Y_l^m(\vec{\omega}_i)\,\ud\omega_i \,, 
        \end{equation}
        where the subsurface scattering function $\hat{S}$ is integrated with the (real) spherical harmonic $Y_l^m$ and we use the neural BSSRDF 
        MLP model $f_{\psi^{\ast}}$ for $\hat{S}$. We use degree 2 spherical harmonics $l\leq 2$ with 9 coefficients to represent our coefficients $a_l^m$ and $b_l^m$. In addition to its lower dimensionality, our neural BSSRDF enables inexpensive low variance estimates of the integral above for coefficients $a_l^m$.  In contrast, path tracing does not provide reliable low variance estimates, often necessitating an exceptionally high number of samples to achieve an acceptable level of variance.
        
        It is important to highlight that the computation of $a_l^m$ occurs at each vertex within the object geometry for every linearly distributed angle $\vec{\omega_o}$ on the hemisphere. In practice, we distribute $\vec{\omega}_o$ into equal solid angles on the hemisphere for every vertex and Monte Carlo sample $\vec{\omega}_i$ over the sphere to construct the coefficients $a_l^m$ for each solid angle. At render time, we interpolate the coefficients for the observed $\vec{\omega}_o$ and compute the result as:
        \begin{equation}
            L_r(\bm{x}_o, \Vec{\omega}_o) = \sum_{l=0}^{\infty} \sum_{m=-l}^l a_l^m(\bm{x}_o,\vec{\omega}_o) b_l^m \,.
        \end{equation}
        
        It is important to note that a spherical harmonic representation has inherent limitations in accurately representing high-frequency functions. As a consequence, the performance gains achieved through this conversion are accompanied by a trade-off in accuracy.

        In order to surpass the limitations of spherical harmonics in capturing intricate details with precision, it is possible to explore alternative representations like wavelets. Wavelets are well-known for their superior ability to represent high-frequency functions, but capturing full view and light dependent effects remains a significant challenge. Utilizing wavelets, or similar representations can achieve improved fidelity with better accuracy in representing complex materials that exhibit intricate high-frequency behavior.
        
    \begin{table*}
    \centering
    \caption{Performance and accuracy analysis for our method.}
    \label{fig:table_comp_ours}
    \begin{tabular}{c c r c S c S S} 
    &&&&&&\\[-4.5ex]
    \hline
    \rule{0pt}{2.5ex}\textbf{Experiment Name} & \textbf{Model} & \textbf{SPP} & \textbf{Render Time (sec.)$\downarrow$} & \textbf{Storage (MB)$\downarrow$} & \textbf{MSE$\downarrow$} & \textbf{SSIM$\uparrow$}& \textbf{\reflectbox{F}LIP mean$\downarrow$}\\[0.5ex] 
    \hline
    \multirow{3}{*}{Grape} & \rule{0pt}{2.5ex}Path Tracing & 10000 & 1846.44 &  &  &  &\\ 
      & Ours(Net) & 256 & 2.038 & 1.31 & $1.6\cdot10^{-4}$ & 0.975 & 0.051\\
    & Ours(SH) & 1 & $3.2\cdot10^{-4}$ & 21.1 & $2.8\cdot10^{-3}$ & 0.983 & 0.082\\[0.5ex]
    \hline
     
    \multirow{3}{*}{Dragon} & \rule{0pt}{2.5ex}Path Tracing & 10000 & 2412.56 &  &  &  &\\ 
    & Ours(Net) & 256 & 1.813 & 1.31 & $1.4\cdot10^{-5}$ & 0.990 & 0.009\\
    & Ours(SH) & 1 & $1.2\cdot10^{-3}$ & 779.6 & $6.2\cdot10^{-4}$ & 0.971 & 0.045\\[0.5ex]
    \hline
    
    \multirow{3}{*}{Paperweight} & \rule{0pt}{2.5ex}Path Tracing & 10000 & 2503.83 &  &  &  &\\ 
    & Ours(Net) & 256 & 3.950 & 3.77 & $1.1\cdot10^{-3}$ & 0.987 & 0.079\\
    & Ours(SH) & 1 & $3.6\cdot10^{-4}$ & 46.1 & $3.4\cdot10^{-3}$ & 0.953 & 0.115\\[0.5ex]
    \hline
     
    \multirow{3}{*}{Drink with ice} & \rule{0pt}{2.5ex}Path Tracing & 10000 & 1496.24 &  &  &  &\\ 
    & Ours(Net) & 256 & 3.231 & 1.31 & $3.0\cdot10^{-4}$ & 0.870 & 0.043\\
    & Ours(SH) & 1 & $2.8\cdot10^{-4}$ & 19.2 & $1.4\cdot10^{-3}$ & 0.969 & 0.070\\[0.5ex]
    \hline
      
    \multirow{3}{*}{Lucy} & \rule{0pt}{2.5ex}Path Tracing & 10000 & 1198.36 &  &  &  &\\ 
    & Ours(Net) & 256 & 2.189 & 0.65 & $4.7\cdot10^{-4}$ & 0.986 & 0.035\\
    & Ours(SH) & 1 & $3.7\cdot10^{-4}$ & 64.2 & $1.1\cdot10^{-3}$ & 0.984 & 0.382\\[0.5ex]
    \hline
     

     \hline
    \end{tabular}
    \end{table*}

    \subsection{Framework and Rendering}
        \label{rendering}
        During network training, we employ our in-house GPU rendering framework that leverages NVIDIA OptiX \cite{optix} for efficient data collection. The neural network is trained on a C++ distribution of PyTorch \cite{pytorch} (popularly known as libTorch). However, for inference, weights from the PyTorch model are extracted and computed on our implementation of network inference. Our framework can be easily integrated into Monte Carlo renderers since we do not batch queries at inference time while each shading point is evaluated independently.

\begin{figure}
        
        \subcaptionbox*{Grape Model}[.45\linewidth]{
        \includegraphics[width = \linewidth]{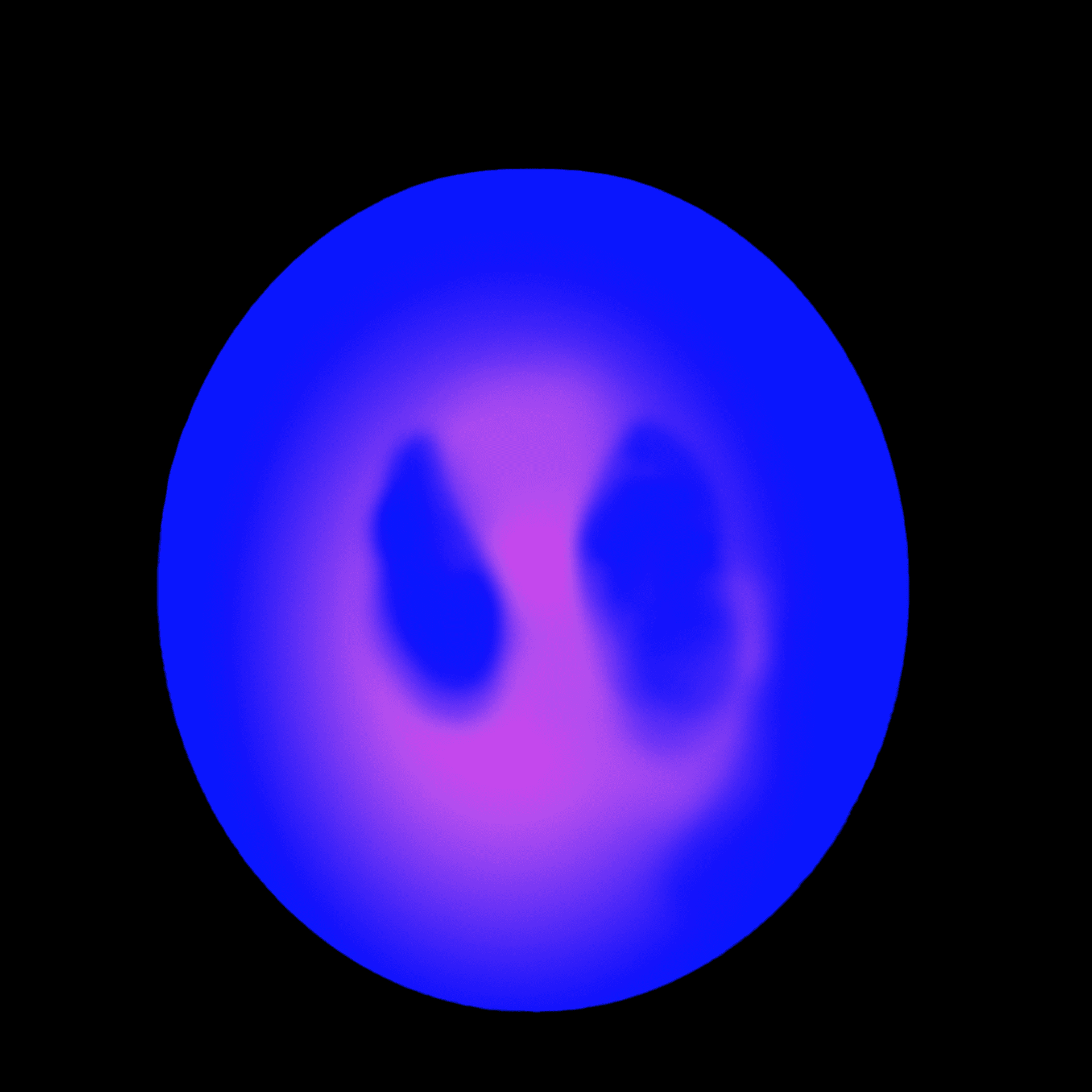}
        }
        \subcaptionbox*{Paperweight Model}[.45\linewidth]{
        \includegraphics[width = \linewidth]{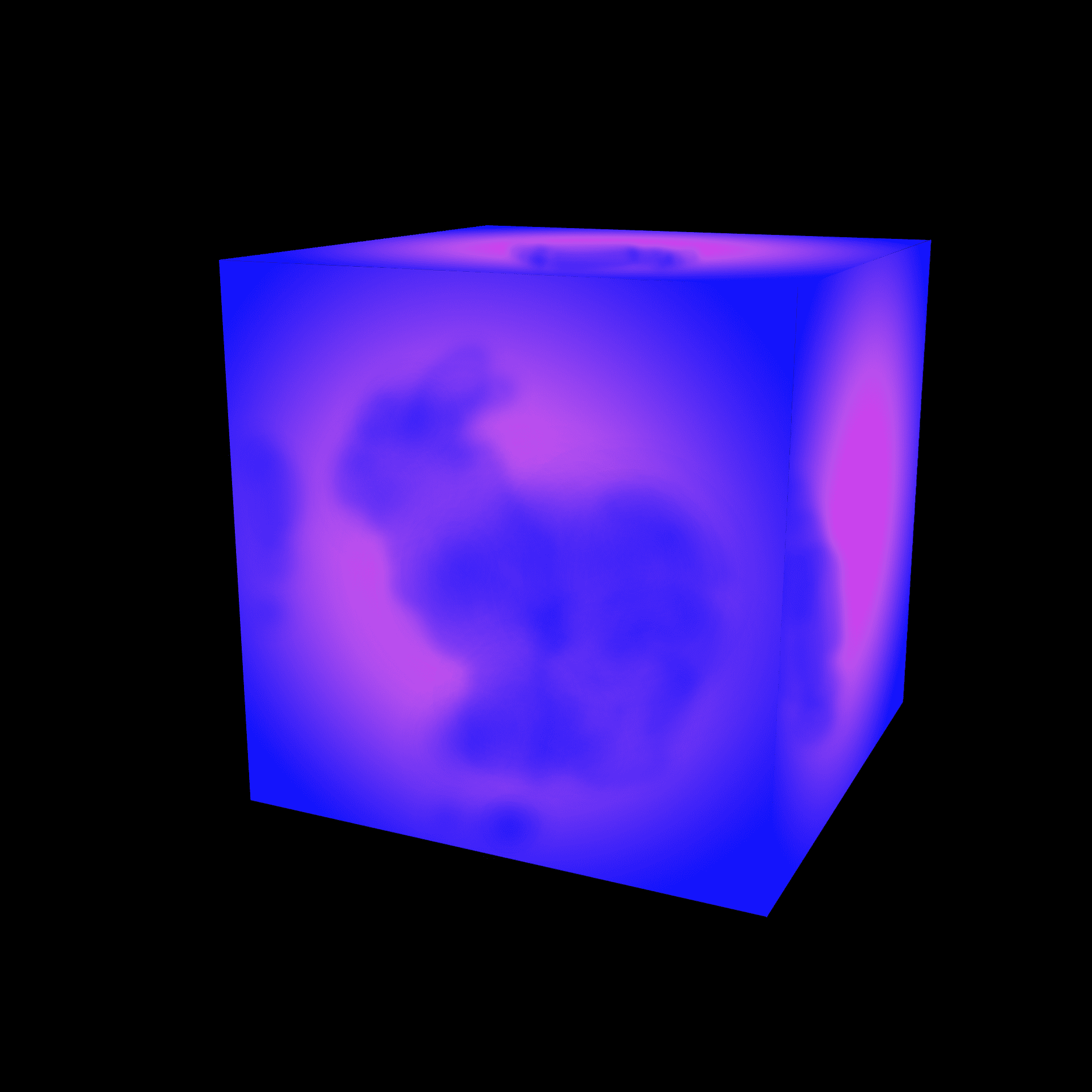}
        } 
        \includegraphics[height = .45\linewidth]{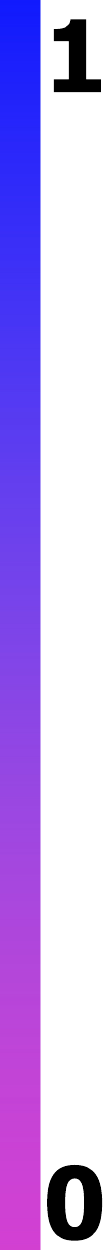}\\[-1.5ex]
        \caption{Visualizations of the standard deviation $\sigma$ predicted by our importance sampling module. The standard deviation is generated once per pixel and subsequently $\vec{\omega}_i$ is sampled for the appearance specification module using the generated distribution.}
    \label{fig:stan_dev}
\end{figure}
    
        Rendering is done inside a shader, where our importance sampling module is used for inference. This returns the standard deviation ($\sigma$) of the expected Gaussian pdf, see examples in Figure~\ref{fig:stan_dev}. A sample direction $\vec{\omega}_{i,j}$ is drawn from the predicted Gaussian pdf. Our appearance specification module is inferred for a relative contribution based on the inputs (i.e.~$\vec{\omega}_o, \bm{x}, \vec{\omega}_i$), which in turn is scaled for the actual contribution from the distant illumination. Furthermore, surface reflection (Fresnel effect) is computed separately.

        In our PRT simulation, as part of our alternative material representation, we evenly distribute $16\times64$ $\vec{\omega}_o$ directions at each vertex in object space. Subsequently, we perform a precomputation step to calculate spherical harmonic coefficients for each $\vec{\omega}_o$ by employing Monte Carlo sampling of $\vec{\omega}_i$ at every vertex on the shading geometry. These coefficients are then used for inference of our appearance specification module and construct a signal representing the subsurface scattering phenomenon. Similarly, during the precomputation phase, we compute spherical harmonic coefficients for environment lighting. When it comes to rendering, the coefficients are convolved together to get the desired output.
        

    \begin{figure}
    \centering
     \begin{subfigure}{0.33\linewidth}
         \centering
         \xincludegraphics[width=\linewidth, label = \color{white}(a), labelbox = false]{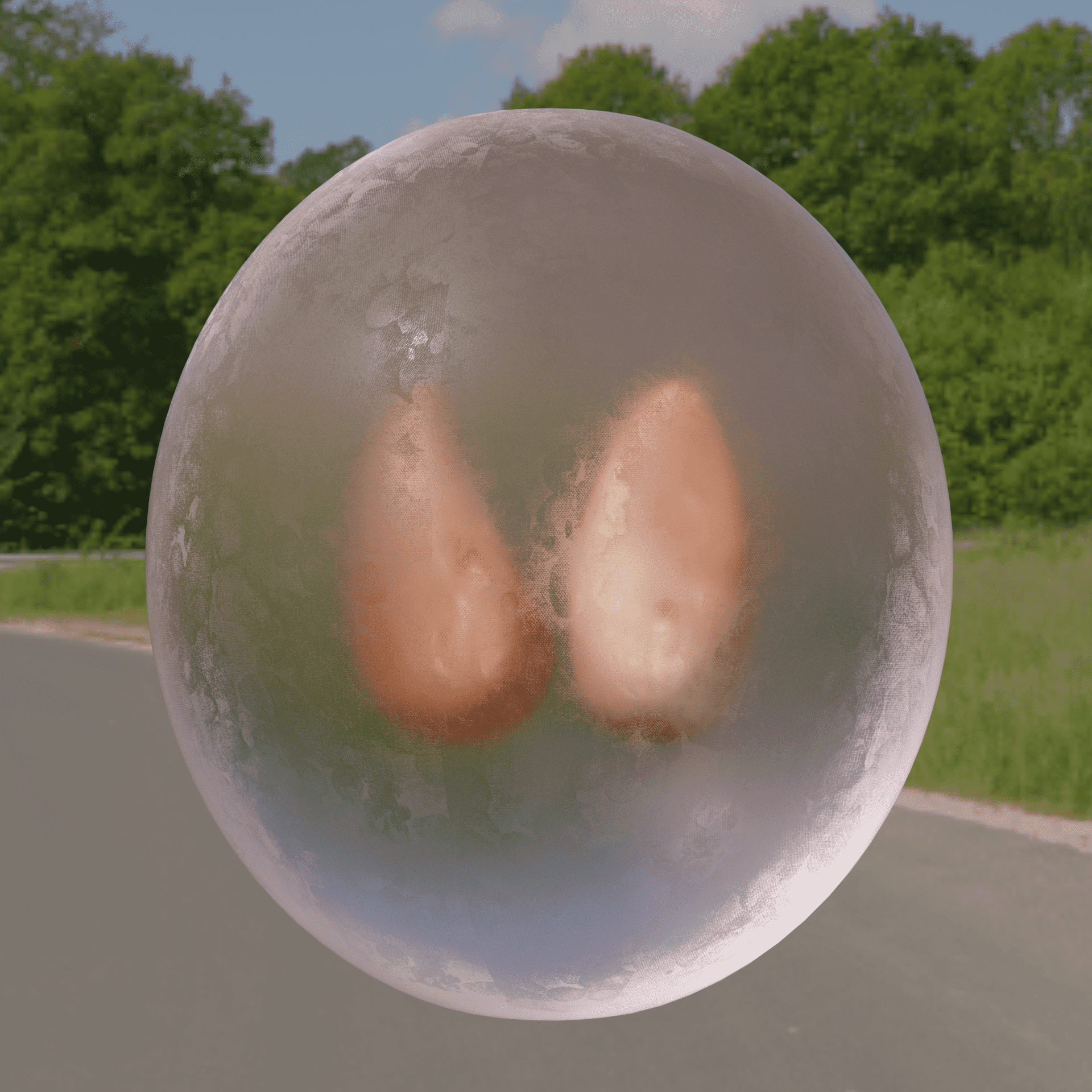}\llap{\includegraphics[height=.8cm]{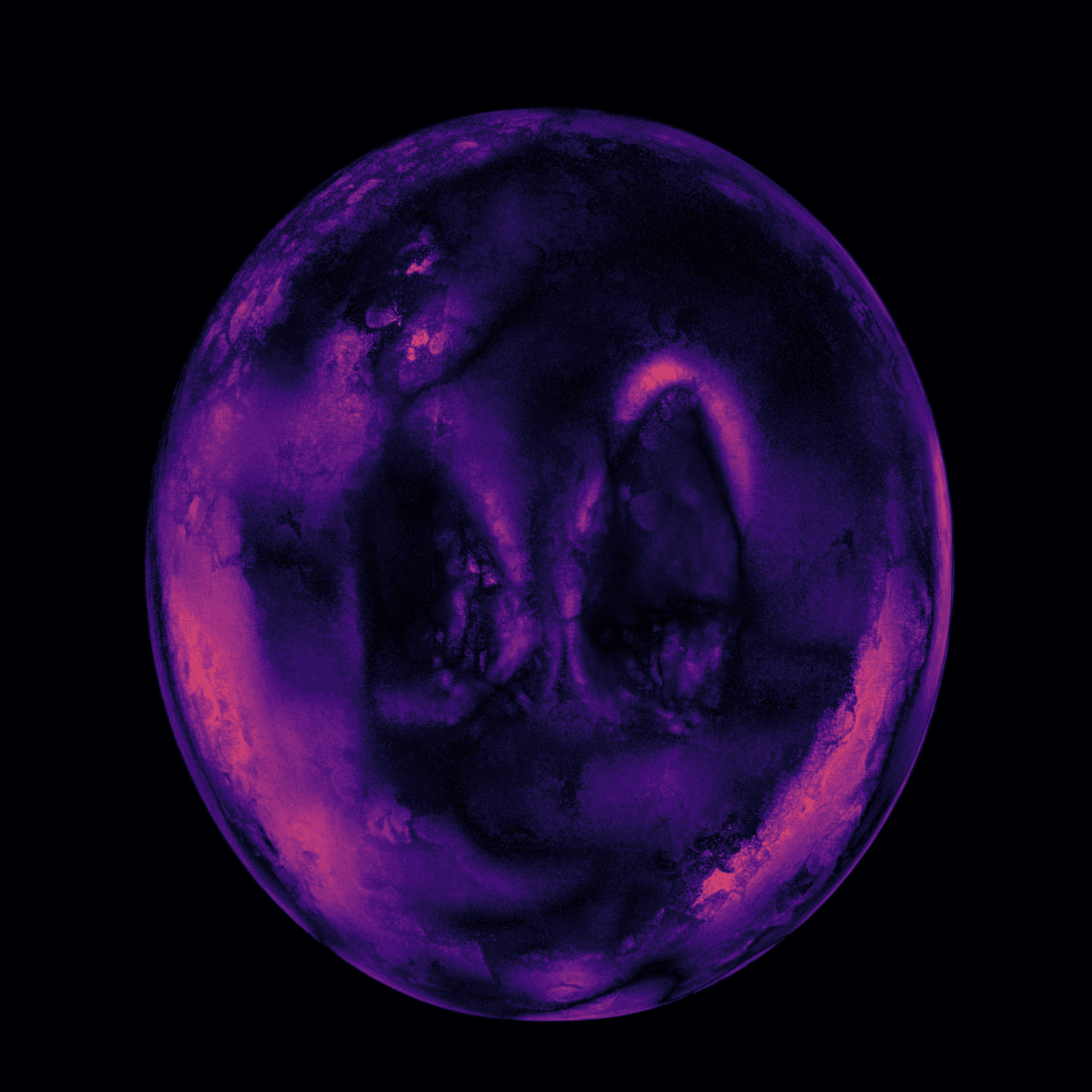}\hspace{0cm}}
         \put(-50,70){\color{white}SSIM : 0.924}
     \end{subfigure}%
     \begin{subfigure}{0.33\linewidth}
         \centering
         \xincludegraphics[width=\linewidth, label = \color{white}(b), labelbox = false]{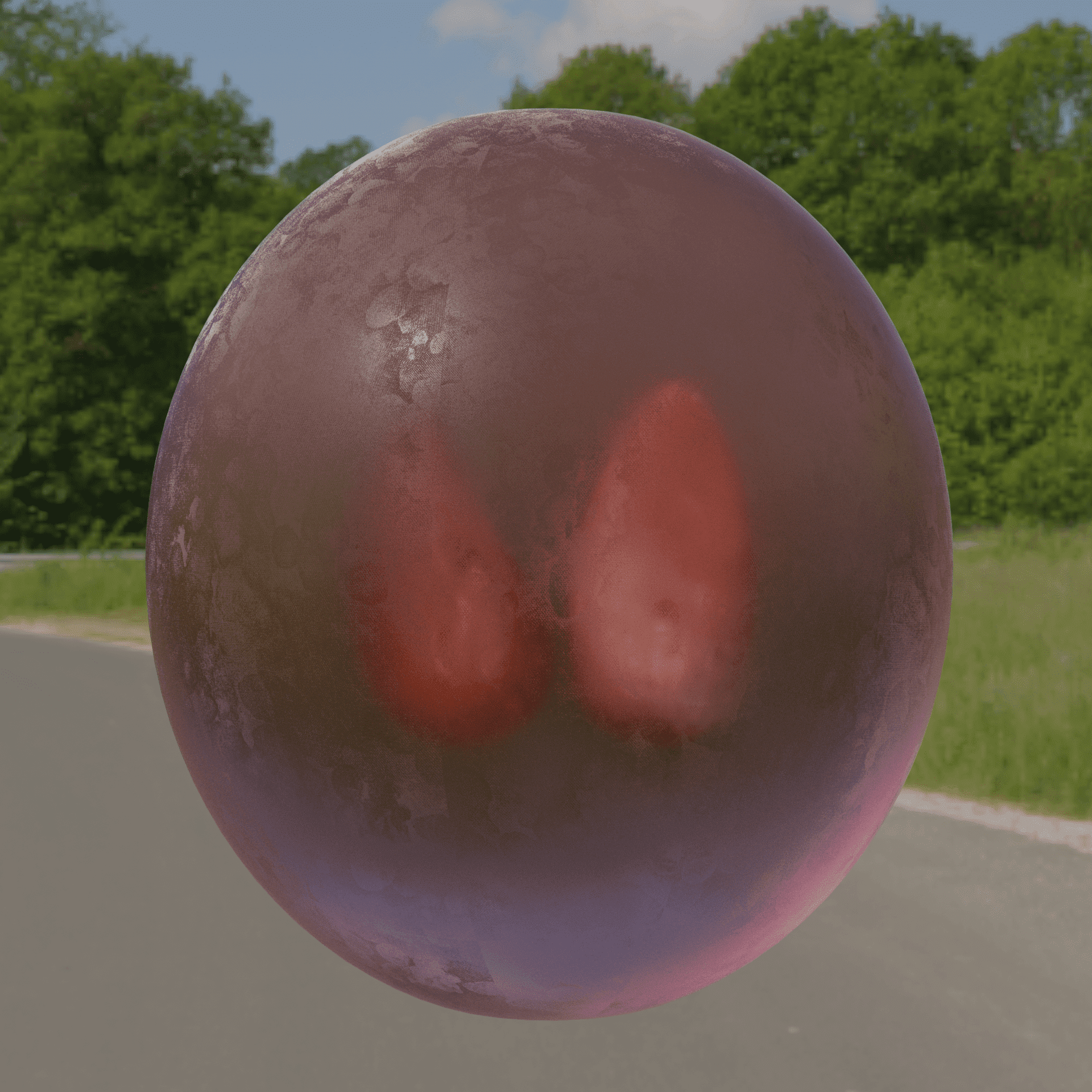}\llap{\includegraphics[height=.8cm]{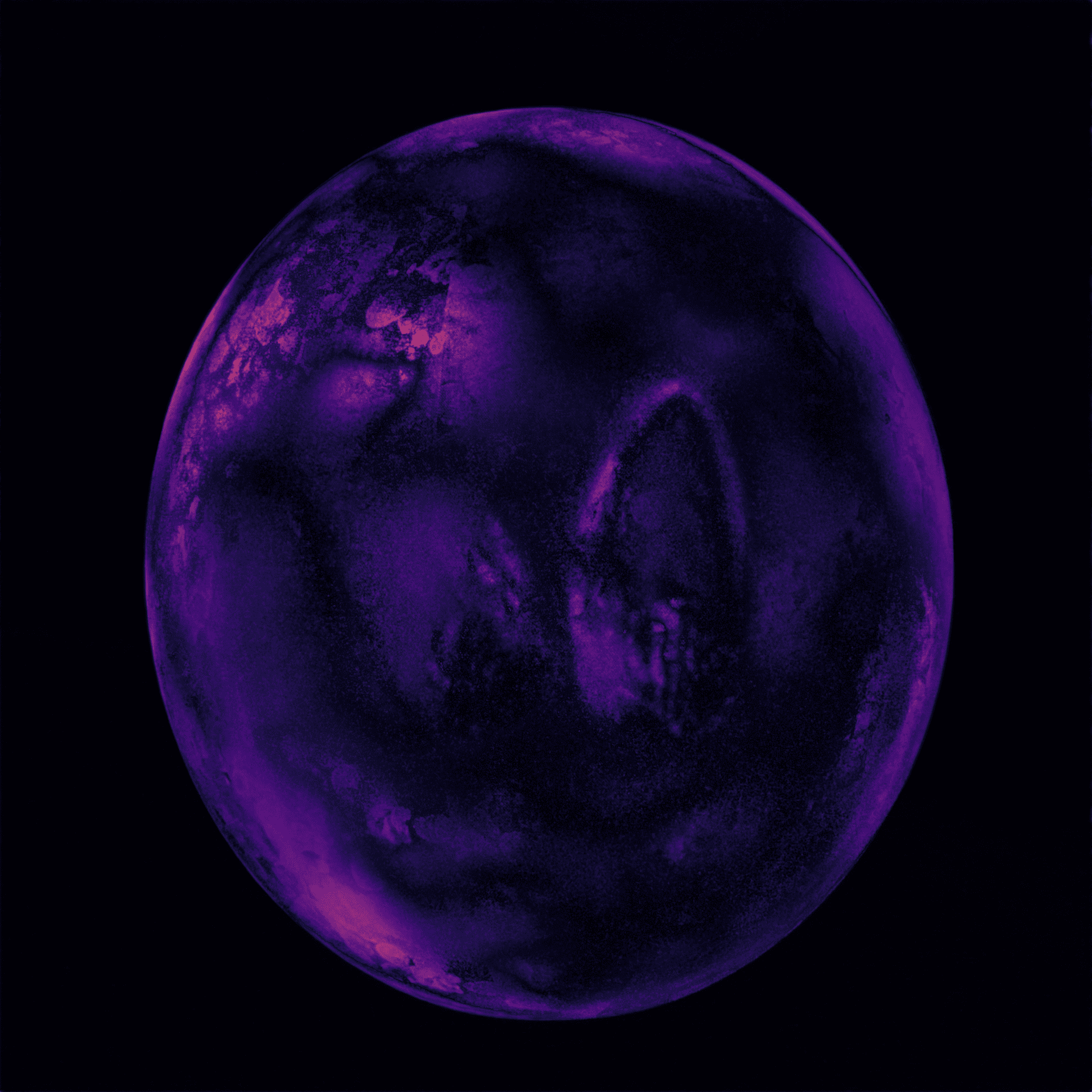}\hspace{0cm}}
         \put(-50,70){\color{white}SSIM : 0.929}
     \end{subfigure}%
     \begin{subfigure}{0.33\linewidth}
         \centering
         \xincludegraphics[width=\linewidth, label = \color{white}(c), labelbox = false]{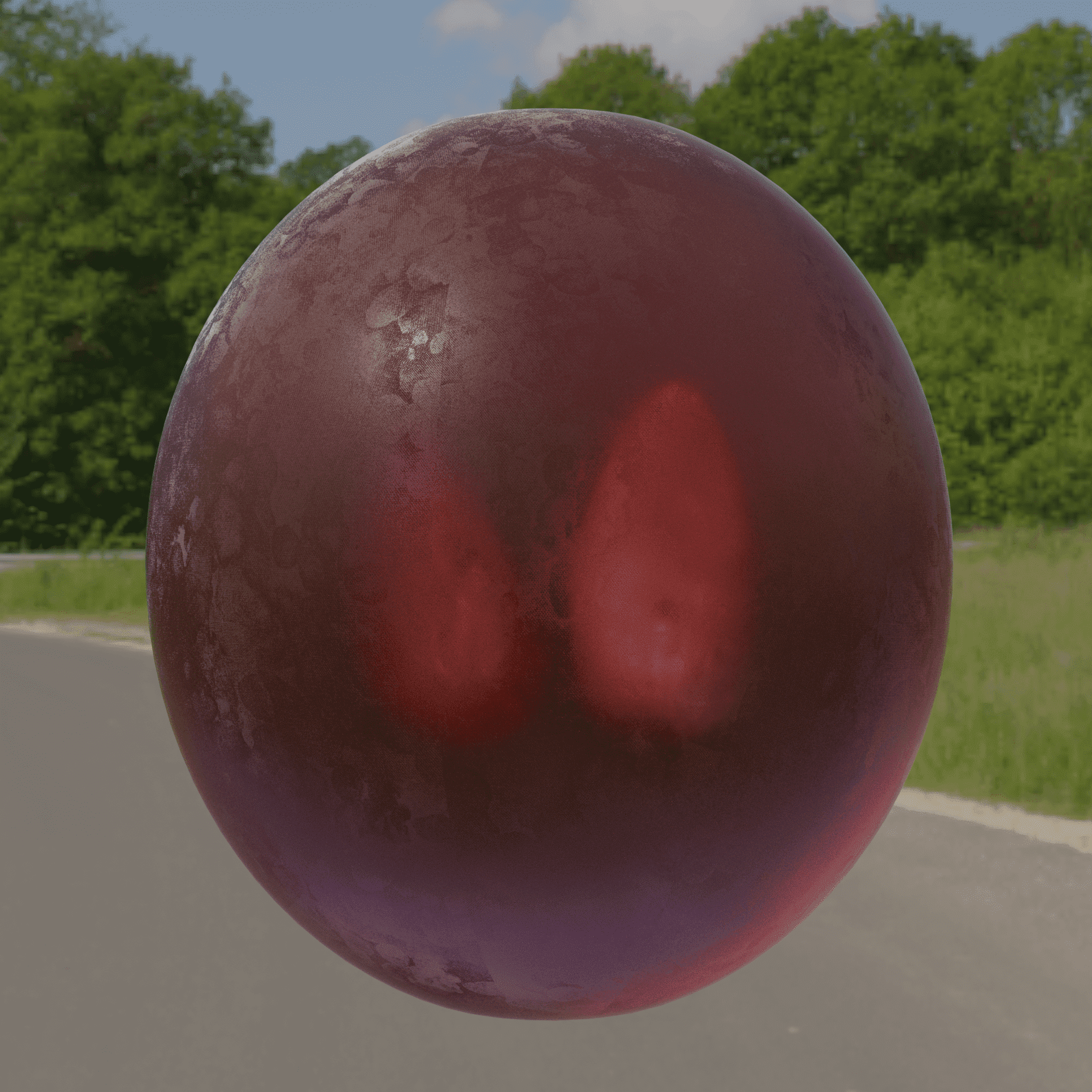}\llap{\includegraphics[height=.8cm]{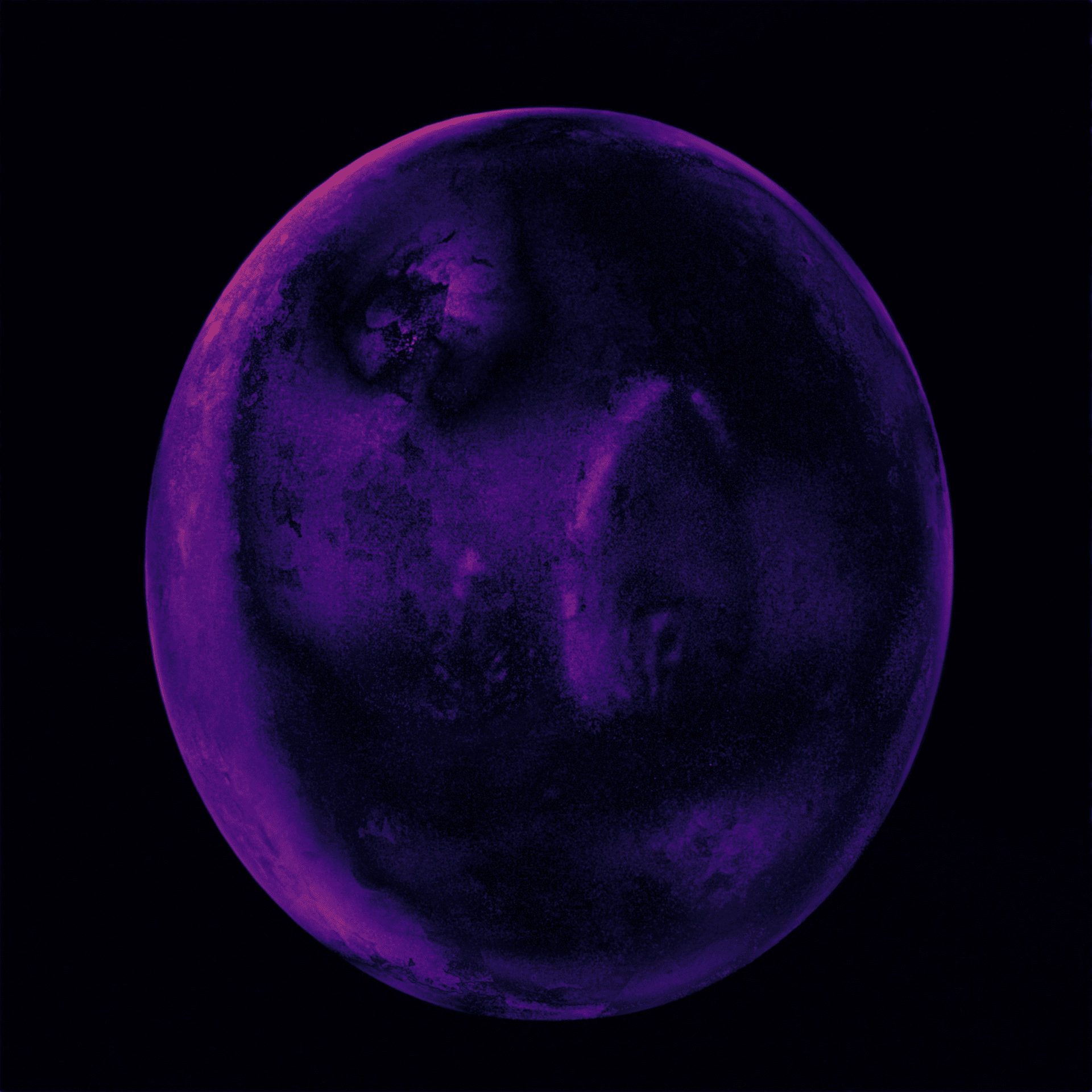}\hspace{0cm}}
         \put(-50,70){\color{white}SSIM : 0.937}
     \end{subfigure}\\[-.2ex]
     \begin{subfigure}{0.33\linewidth}
         \centering
         \xincludegraphics[width=\linewidth, label = \color{white}(d), labelbox = false]{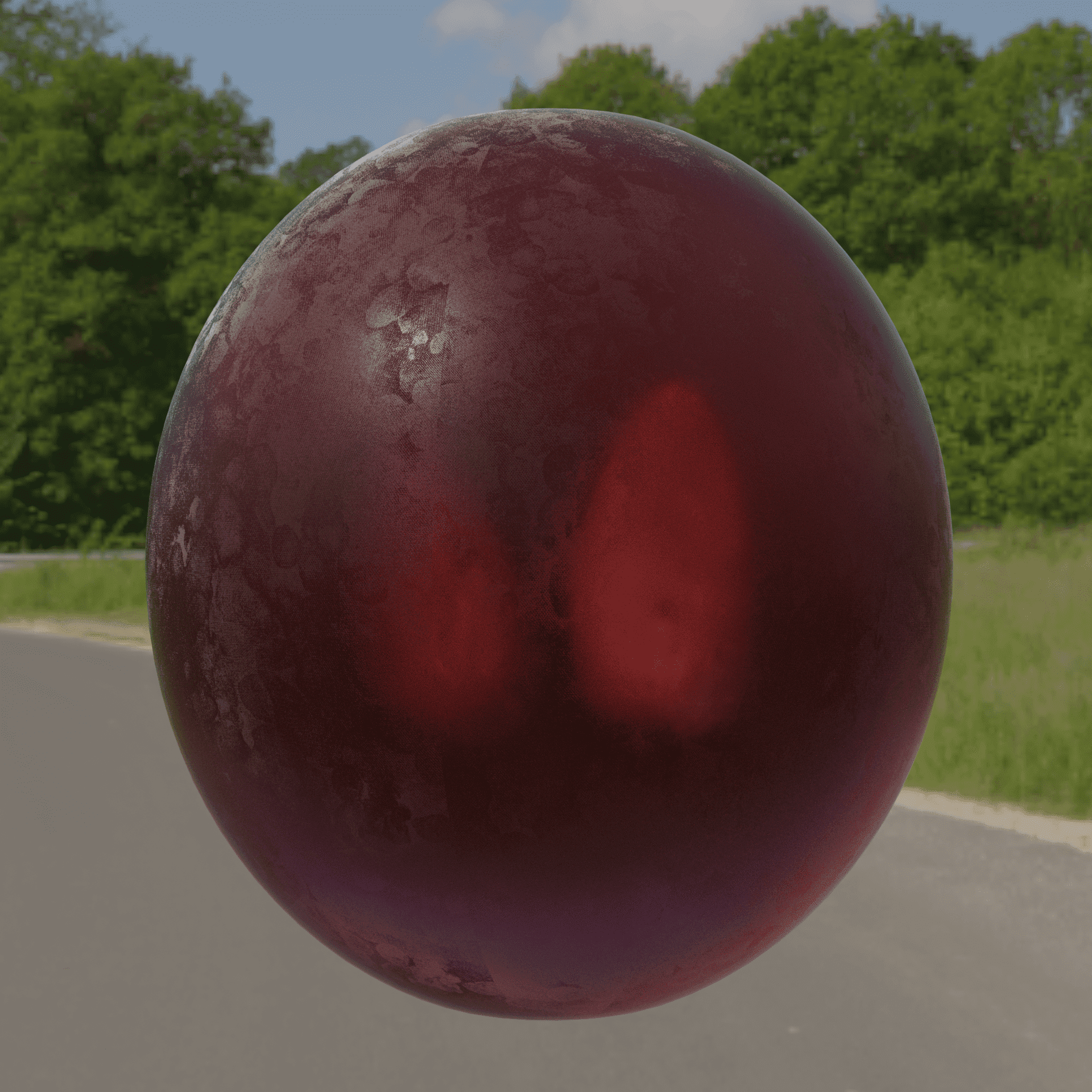}\llap{\includegraphics[height=.8cm]{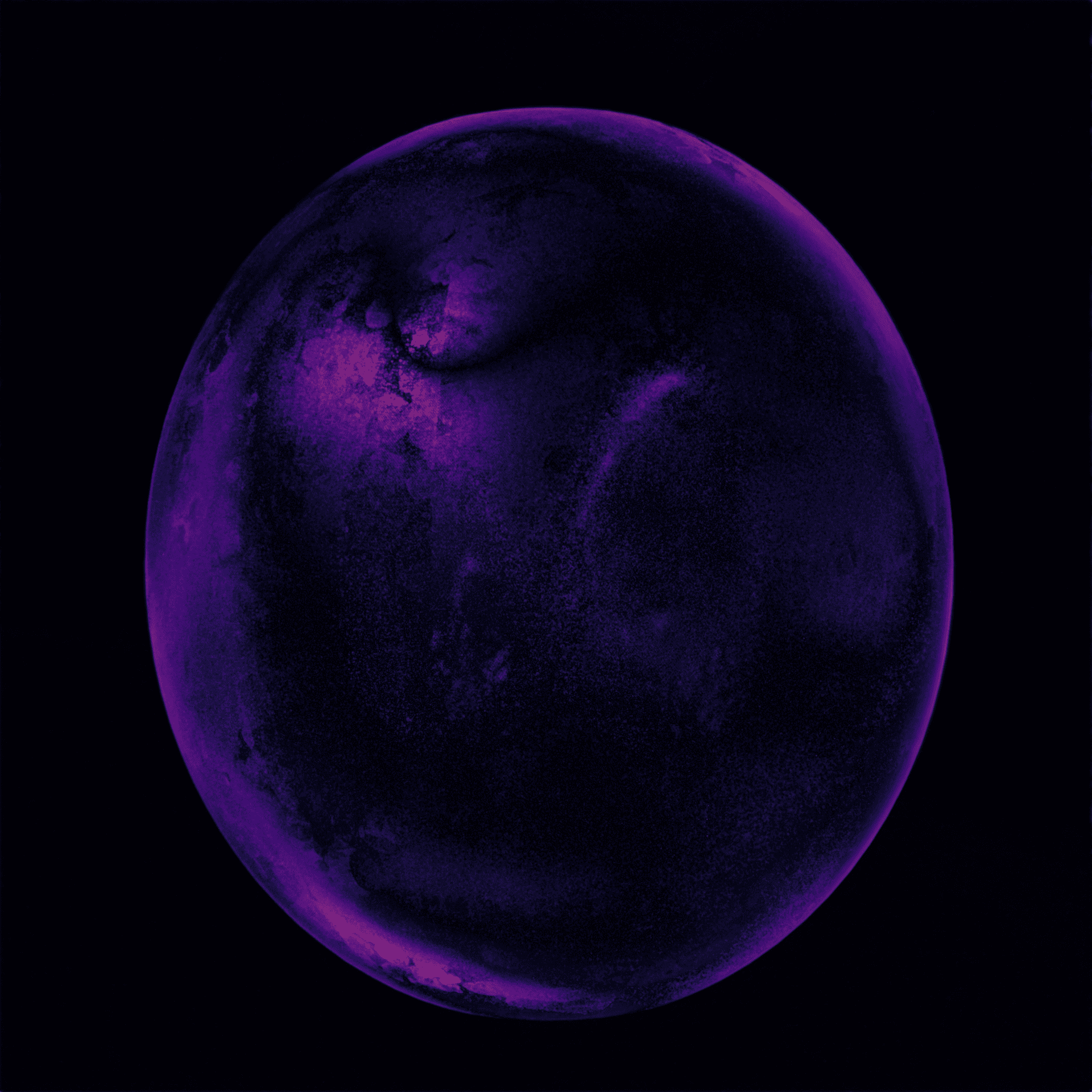}\hspace{0cm}}
         \put(-50,70){\color{white}SSIM : 0.956}
     \end{subfigure}%
     \begin{subfigure}{0.33\linewidth}
         \centering
         \xincludegraphics[width=\linewidth, label = \color{white}(e), labelbox = false]{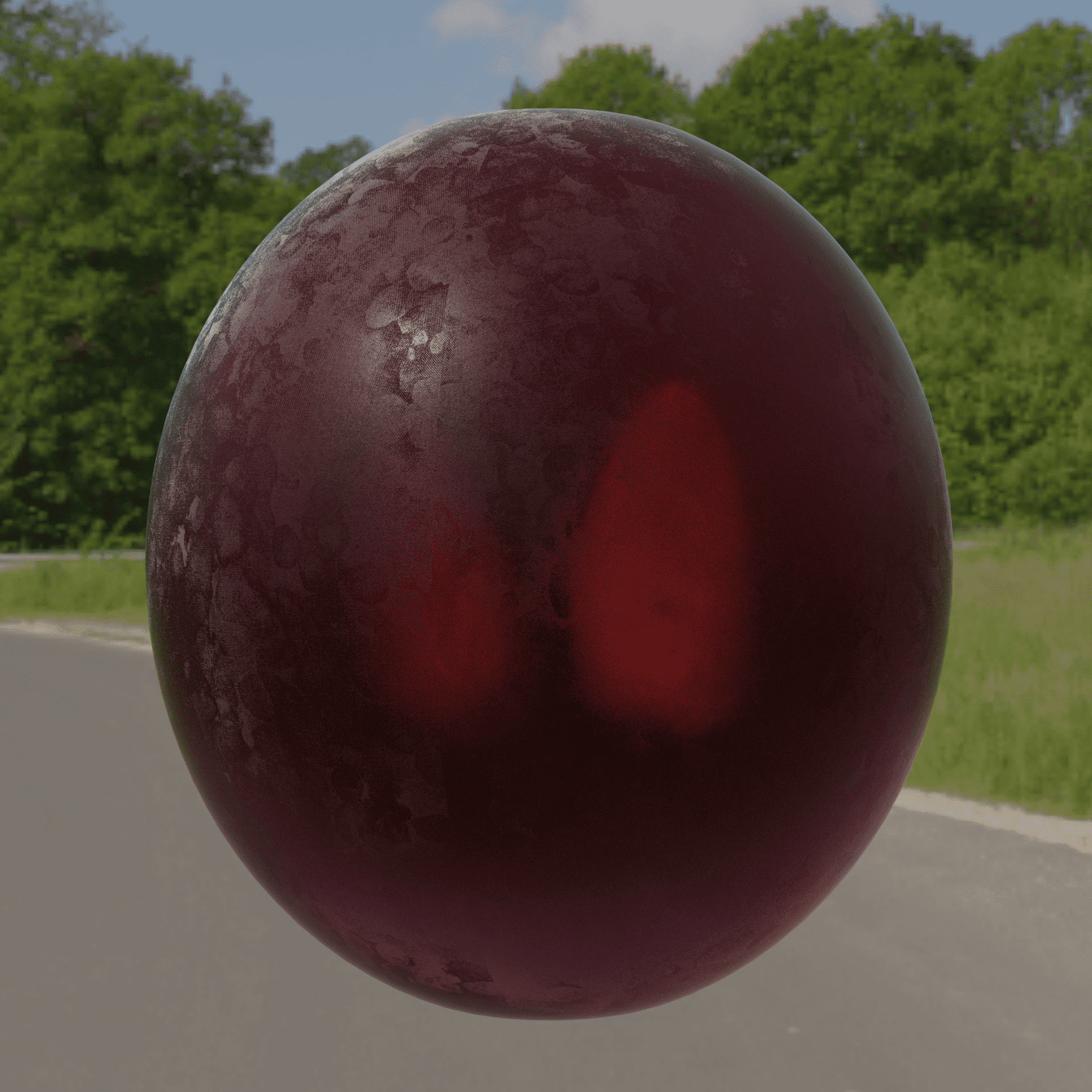}\llap{\includegraphics[height=.8cm]{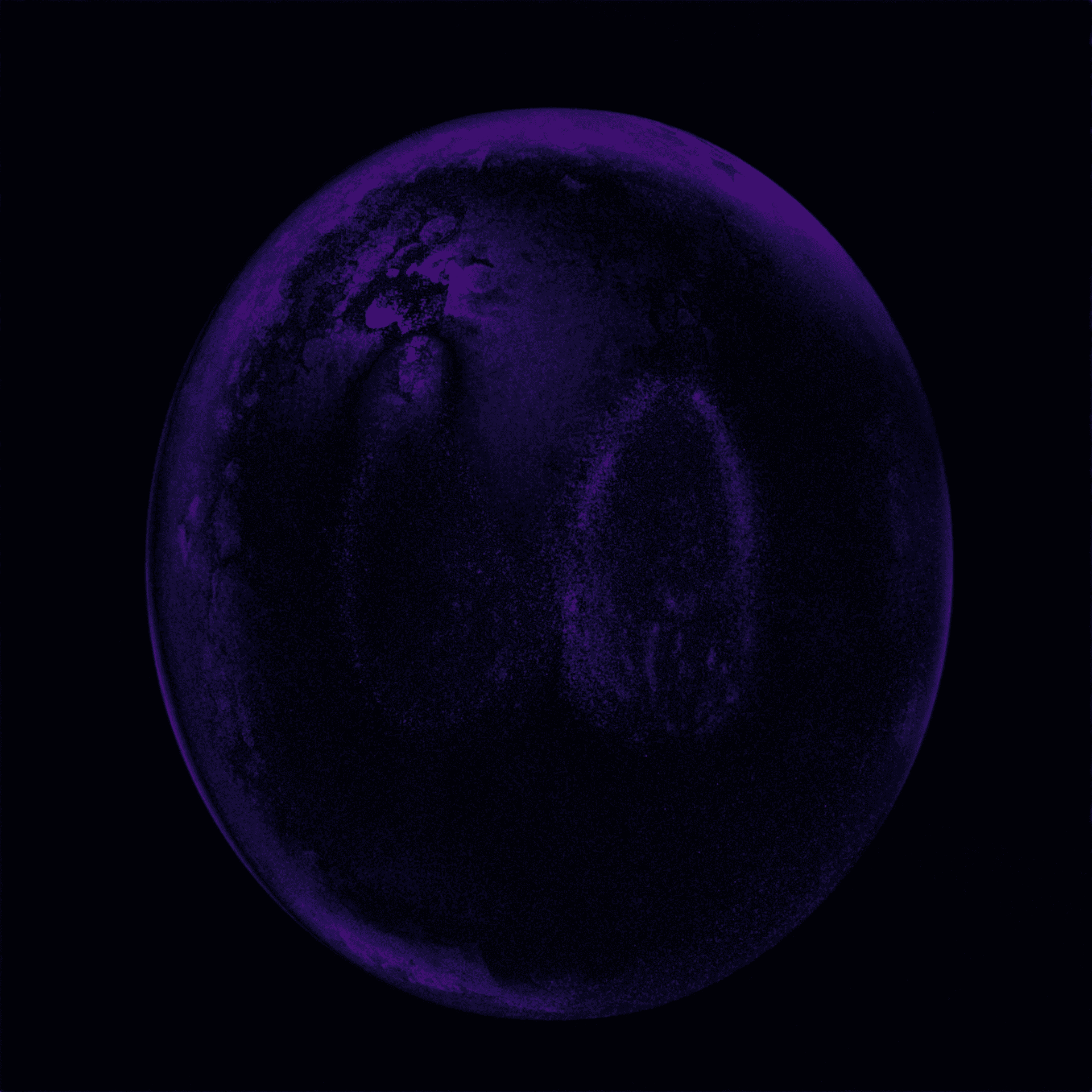}\hspace{0cm}}
         \put(-50,70){\color{white}SSIM : 0.972}
     \end{subfigure}%
     \begin{subfigure}{0.33\linewidth}
         \centering
         \xincludegraphics[width=\linewidth, label = \color{white}(f), labelbox = false]{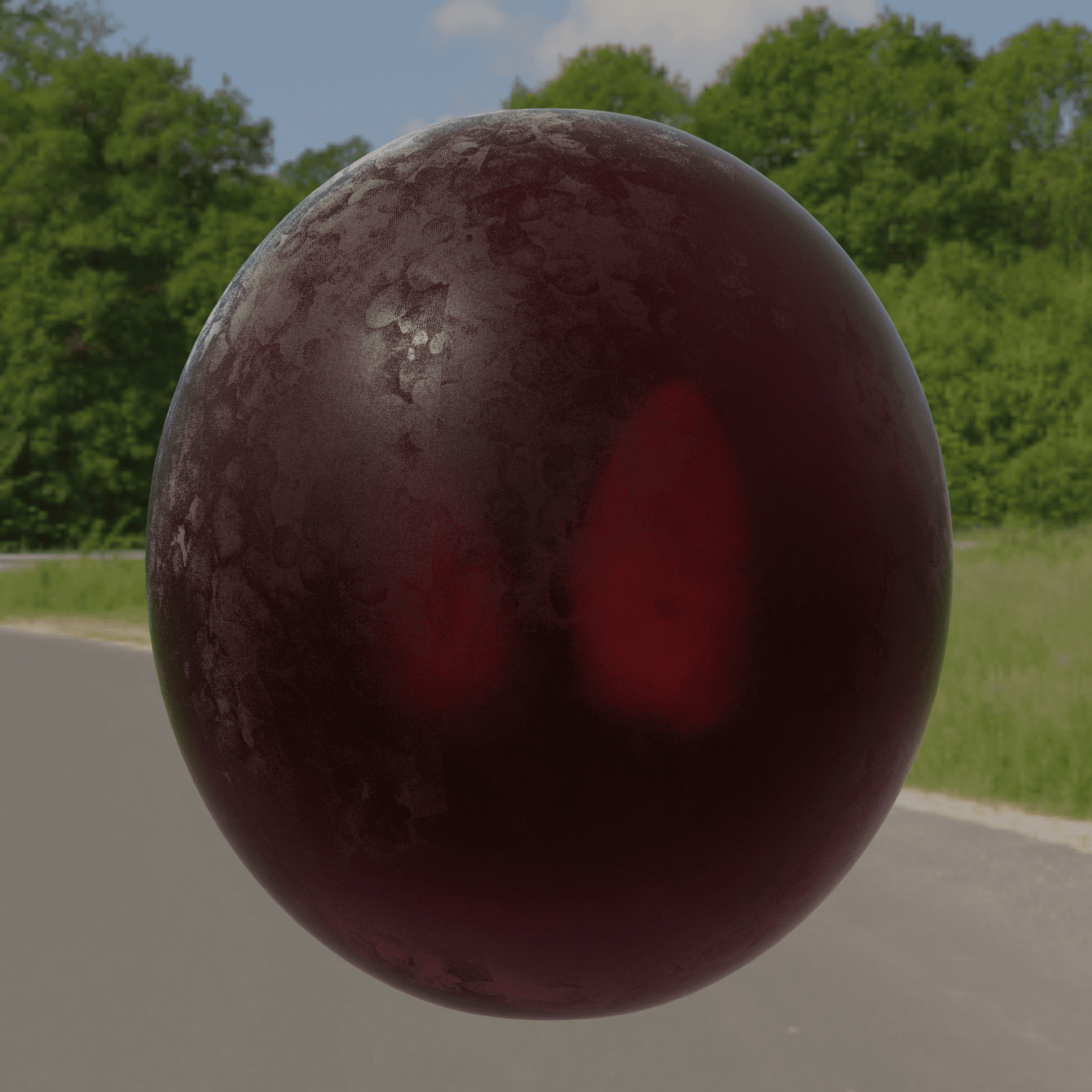}\llap{\includegraphics[height=.8cm]{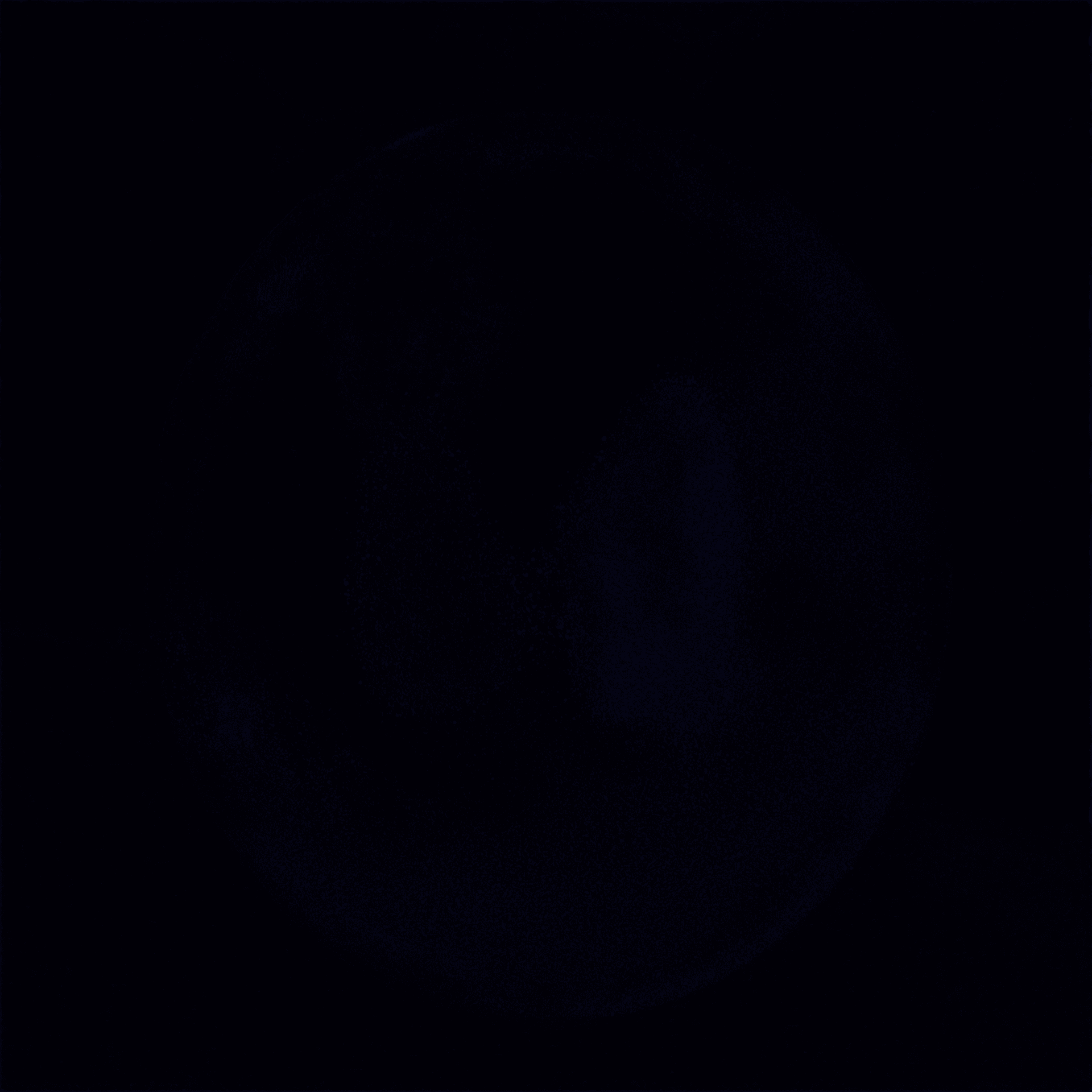}\hspace{0cm}}
         \put(-50,70){\color{white}SSIM : 0.985}
     \end{subfigure}%
     \\[-1ex]
  \caption{Our method applied to objects with different levels of translucency. The insert at the bottom right shows the \reflectbox{F}LIP mean between the images rendered using path tracing and our neural BSSRDF. The SSIM (top-right) demonstrates high accuracy compared to the ground truth.} 
  \label{fig:translucency} 
\end{figure}

\begin{figure}[h]
  \centering
  \begin{minipage}{0.748\linewidth} 
    \includegraphics[width=\linewidth, trim = {5cm 5cm 5cm 5cm}, clip]{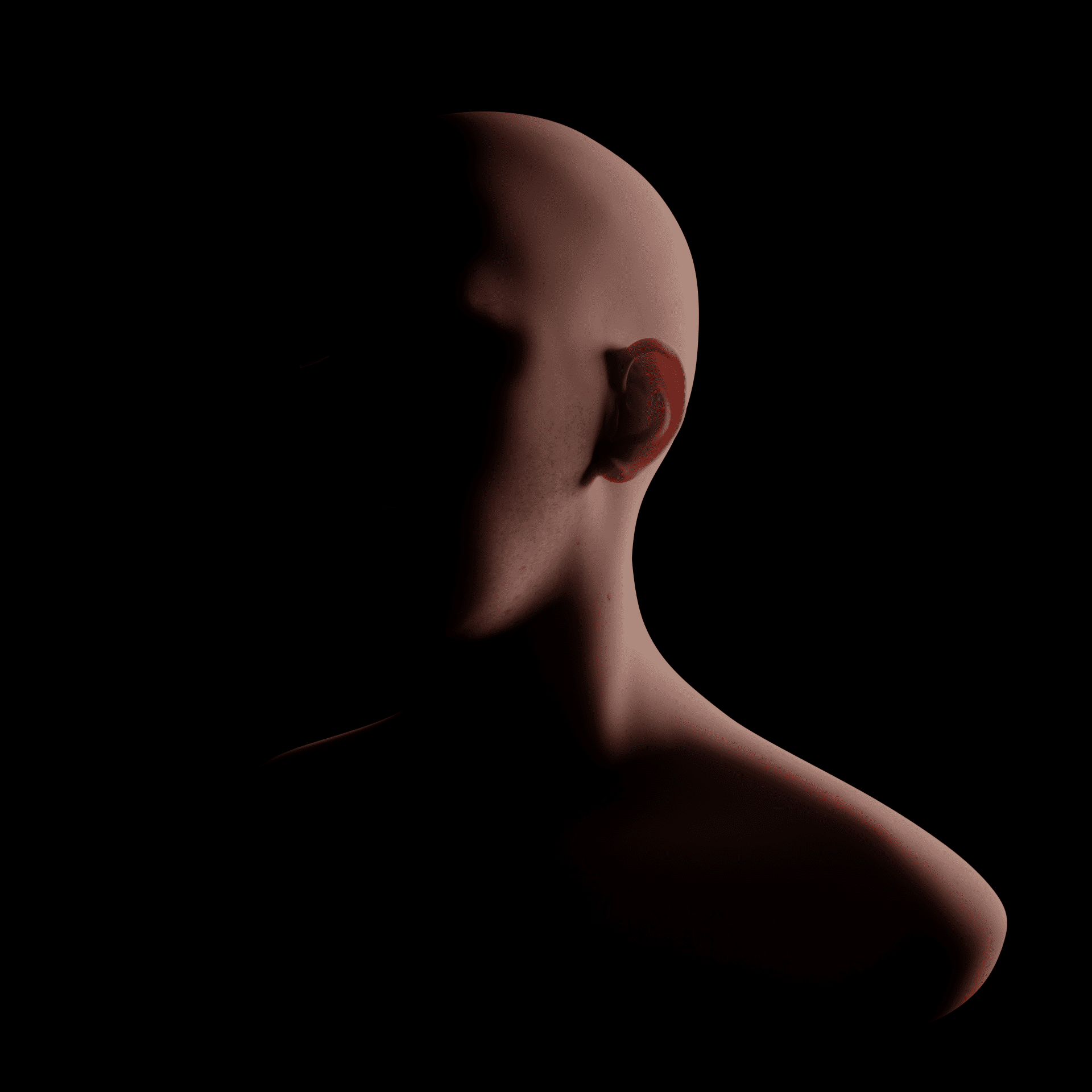}
  \end{minipage}%
  \begin{minipage}{0.25\linewidth} 
    \xincludegraphics[width=\linewidth, trim = {32cm 36cm 24cm 20cm}, clip, label = \color{white}(ref), pos = ne, labelbox = false]{figures/head/head_r.png}\par
    \vspace{-.3ex}
    \xincludegraphics[width=\linewidth, trim = {32cm 36cm 24cm 20cm}, clip, label = \color{white}(ours), pos = ne, labelbox = false]{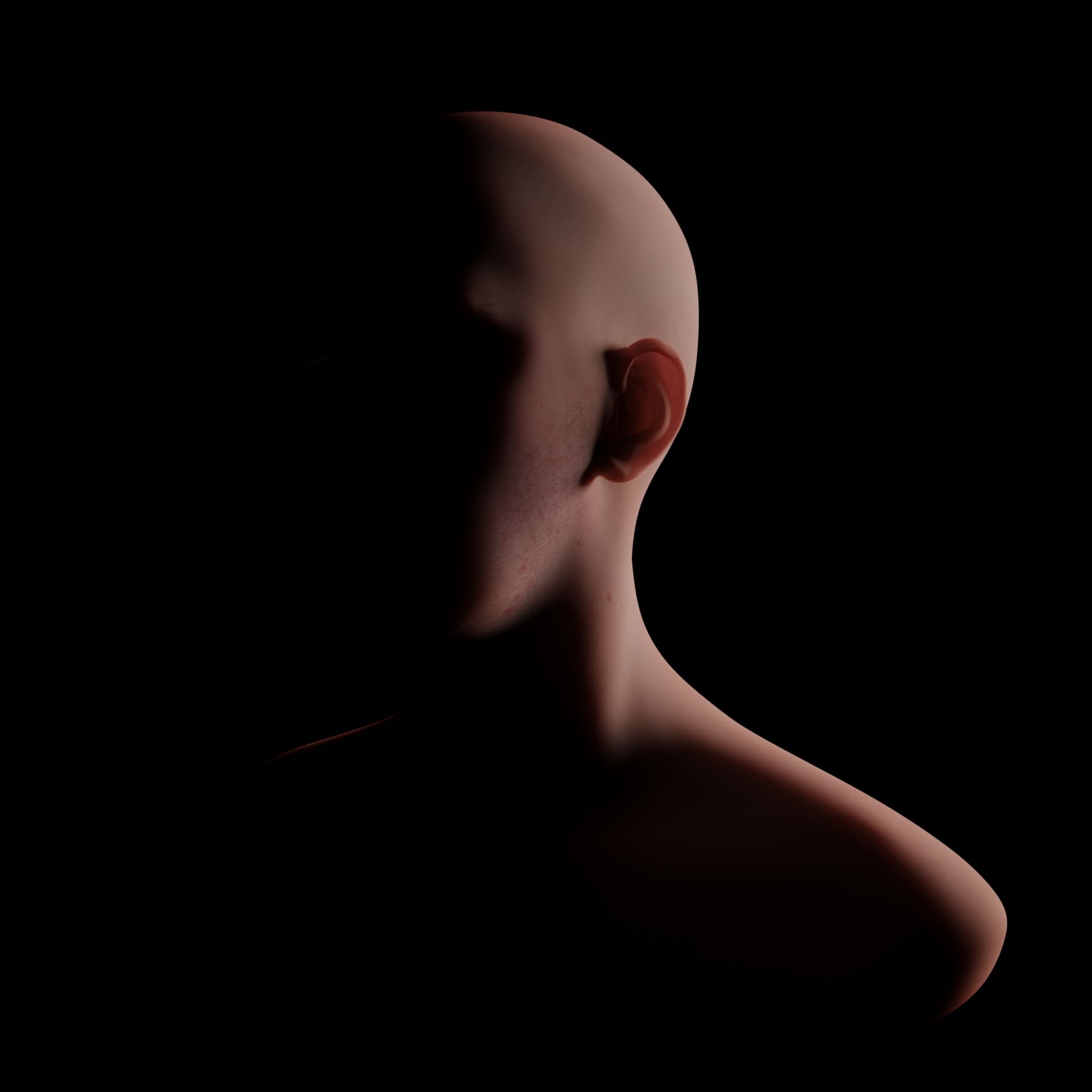}\par
    \vspace{-.3ex}
    \xincludegraphics[width=\linewidth, trim = {32cm 36cm 24cm 20cm}, clip, label = \color{white}(\reflectbox{F}LIP), pos = ne, labelbox = false]{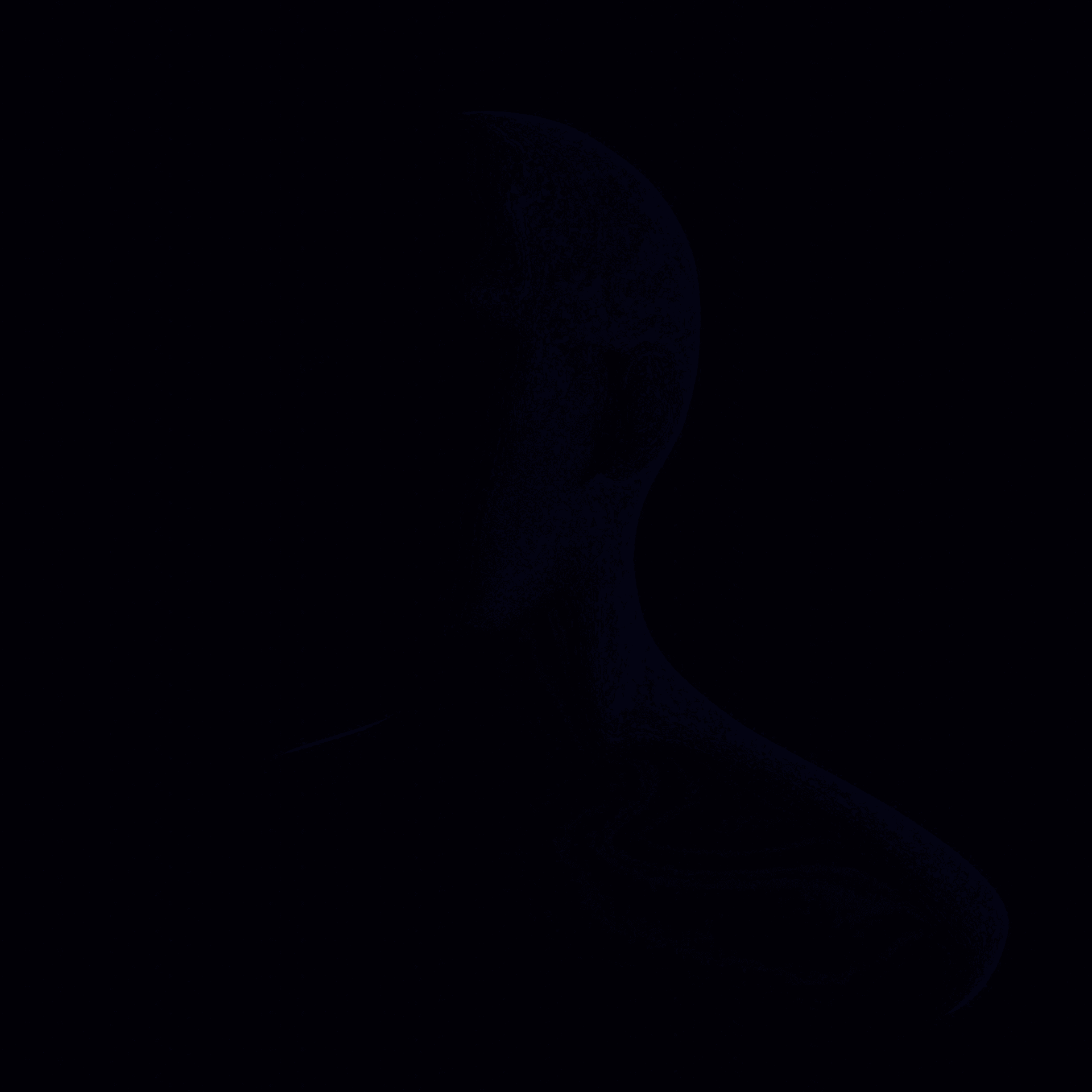}
  \end{minipage}
    \caption{
    The ability of our method to represent the self-shadowing of a translucent object with non-trivial surface geometry. The back-lighting of the head makes the ear cast a shadow while light is also bleeding through the ear. This example also demonstrates that the neural BSSRDF works for a single directional light just as well as for the environment light used in the other examples. The \reflectbox{F}LIP error of the close-up on the ear shows high accuracy as compared with the path traced reference image.} \label{fig:shadow}
\end{figure}

\section{Results}
    \label{results}
    In this section, we demonstrate the ability of our network to faithfully represent and render heterogeneous translucent objects under distant illumination. Figure~\ref{fig:translucency} is a test of our model using objects with different levels of translucency illuminated by an environment. In Figure~\ref{fig:shadow}, we test an example of just a single directional light illuminating an object with non-trivial surface geometry leading to self-shadowing of the translucent object. A performance and accuracy analysis is provided for five different objects in Table~\ref{fig:table_comp_ours}. 
    We employ three objective full image quality metrics: mean-squared error (MSE), structural similarity (SSIM) index~\cite{wang2004image}, and the mean value of \reflectbox{F}LIP~\cite{andersson2020flip}. Our network achieves high fidelity (\reflectbox{F}LIP below 0.1), enabling us to represent the complex appearance of the objects while efficiently storing the data. Additionally, our PRT-based implementation for our translucent object appearance representation achieves real-time rendering frame rates (less than $0.01$ seconds per frame). Our appearance specification module can run at more than 60 samples per pixel per second, meaning that our method supports real-time rendering with one sample per pixel (SPP) and progressive updates.


    The setup and models in the different scenes that we used in our experiments are as follows:
    \begin{itemize}
        \item \textit{Grape} is a model consisting of a scattering volume behind a refractive interface with surface roughness defined by a texture map. The volume contains two seeds that are diffuse with reflectance defined by an image texture.

        \item \textit{Dragon} is the XYZ RGB dragon from the 3D scanning repository of the Stanford Computer Graphics Laboratory (\href{http://graphics.stanford.edu/data/3Dscanrep/}{http://graphics.stanford.edu/data/3Dscan\-rep/}). We use a noise function to define the optical properties inside the model.

        \item \textit{Paperweight} is a cube with a scattering volume. It contains the Stanford bunny (Stanford CG Lab) with solid texture variation of the optical properties of marble from Jensen et al.~\cite{jensen_01}.

        \item \textit{Drink with ice} is a scattering medium in a drinking glass. The model includes two ice cubes inside the liquid.

        \item \textit{Lucy} is another model from the Stanford CG Lab which we augmented with solid texture variation of ketchup and marble materials. The optical properties of these translucent materials are from Jensen et al.~\cite{jensen_01}.
    \end{itemize}

    \begin{figure}
        \includegraphics[trim = 0 0cm 0cm 0cm, clip, width = \linewidth]{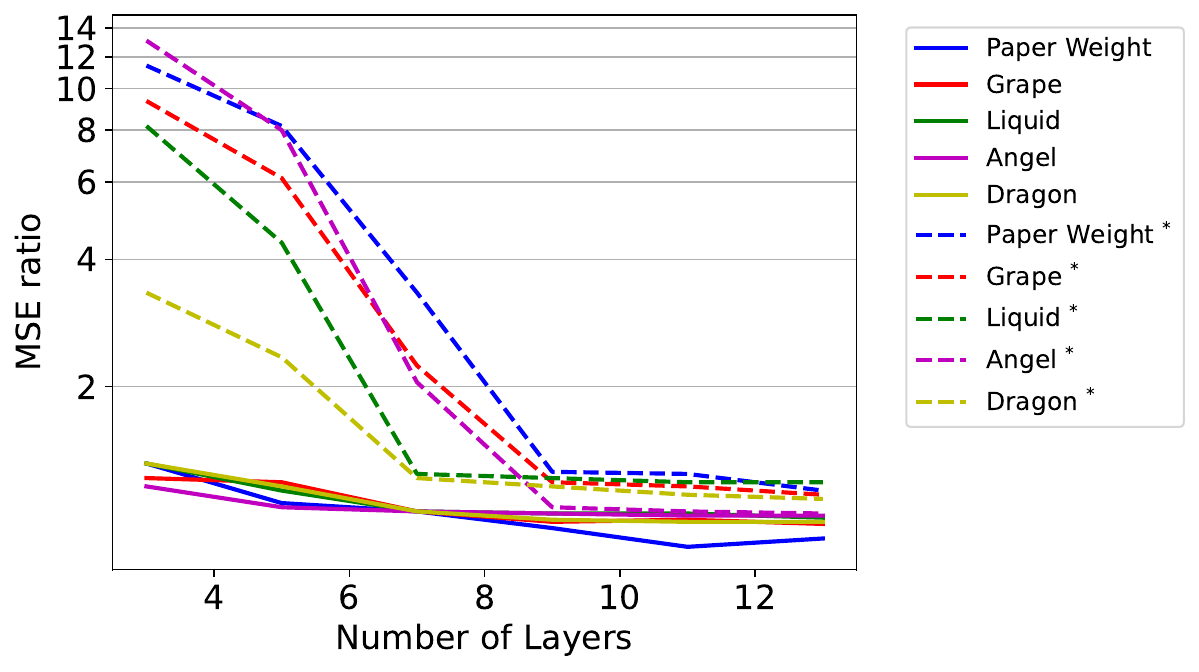}
        \\[-3.5ex]
        \caption{The influence of varying the number of layers in our \emph{appearance specification module} (Figure \ref{network_architecture}b) on the resulting mean squared error (MSE) taken relative to a 7-layers MLP with skip connections. The asterisk $\ast$ marks the same results but without skip connections. This clearly demonstrates the importance of the skip connections and illustrates the trade-off in selecting the number of layers (we chose 5).}
        \label{fig:studies}
    \end{figure}

    \paragraph*{Ablation study.} We conducted an ablation test by evaluating the performance of various network sizes in comparison to a 7-layered network, see Figure~\ref{fig:studies}. Our findings indicate that a 7-layered network is generally sufficient for representing most types of heterogeneous translucent objects. It is worth noting that we incorporate skip connections between alternate layers as our tests have shown that this architectural design choice leads to improved performance with better compression.

    
    \paragraph*{Low storage.} During the design of our architecture for material representation, we considered storage efficiency important. Comparing with other techniques for storing heterogeneous scattering volumes, such as the approach proposed by Kurt et al.~\cite{kurt_13}, our model achieves a more compact representation. However, since we did not have access to the same dataset as Kurt et al., a direct quantitative comparison was not possible. Table~\ref{fig:table_comp_ours} provides evidence of the good storage efficiency of our model. Our storage requirements are lower than the compact factored representations~\cite{peers_06,kurt_13} (for which 4.2 to 39 MB to were spent to represent measured heterogeneous translucent materials).

\begin{figure*}
    \centering    
        \begin{tabularx}{\textwidth}{sbbbb}
        & \large Reference & \large Ours (env IS) & \large Ours (our IS) & \large Ours (SH) \\
    
    \large \rotatebox{90}{Grape} & 
    \xincludegraphics[height=3.75cm, label = \color{white}(a), pos = sw, labelbox = false]{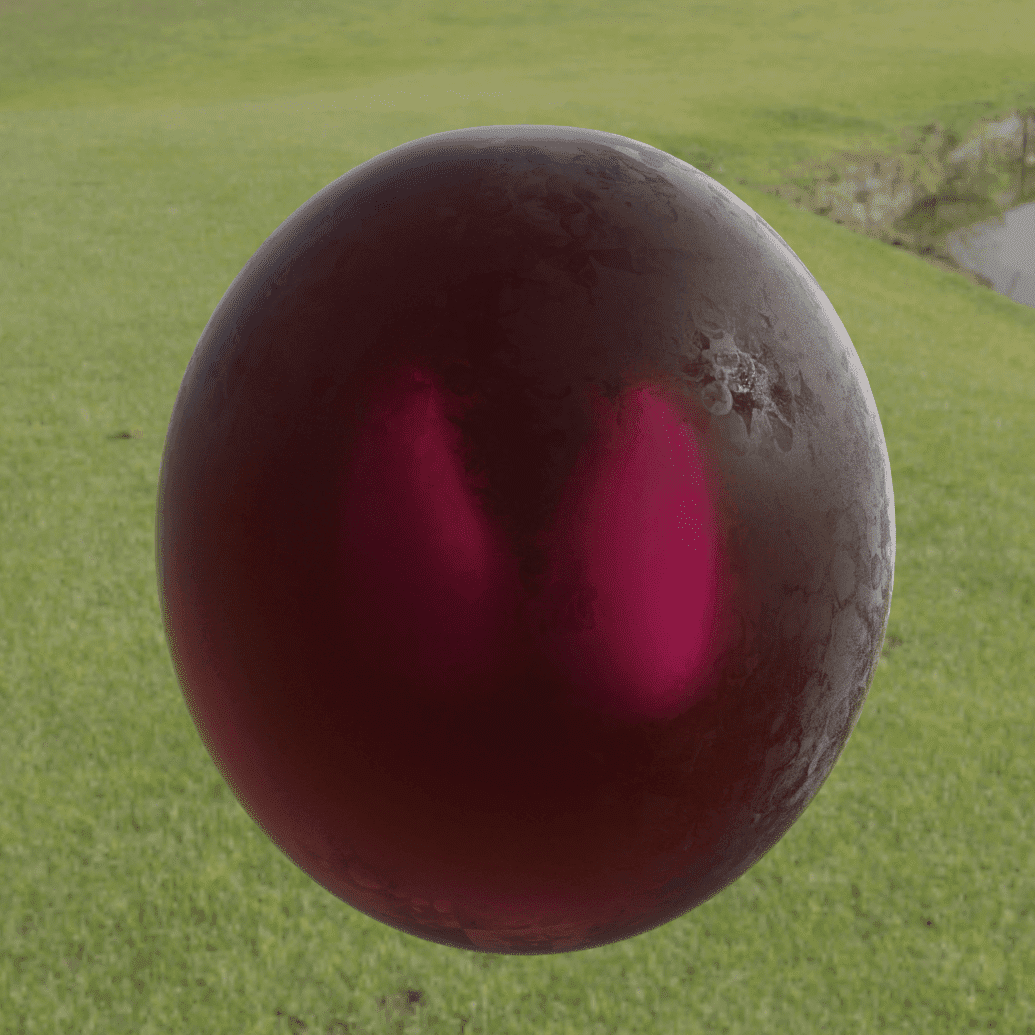}  \label{6a} &
    
    \xincludegraphics[height=3.75cm, label = \color{white}(b), pos = sw, labelbox = false]{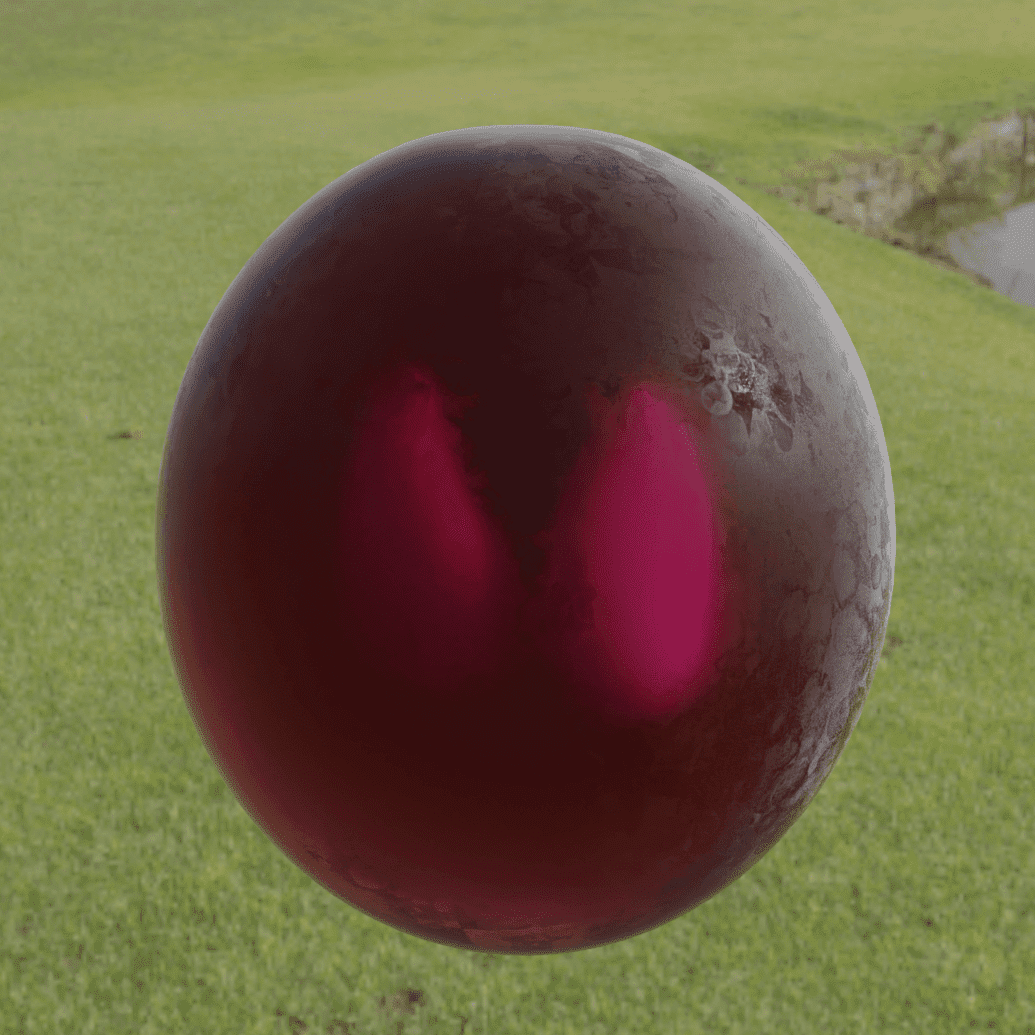}\llap{\includegraphics[height=1cm]{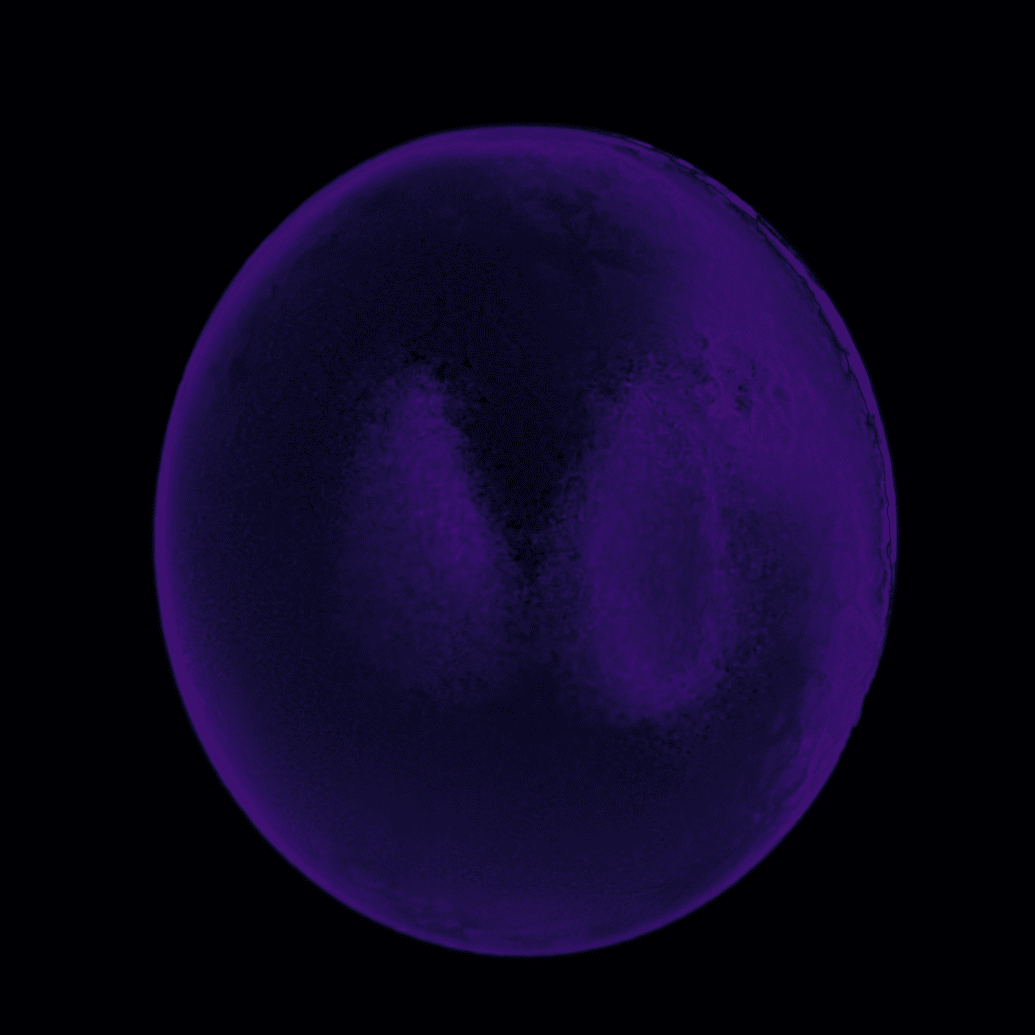}}
     \label{6b}&
    
    \xincludegraphics[height=3.75cm, label = \color{white}(c), pos = sw, labelbox = false]{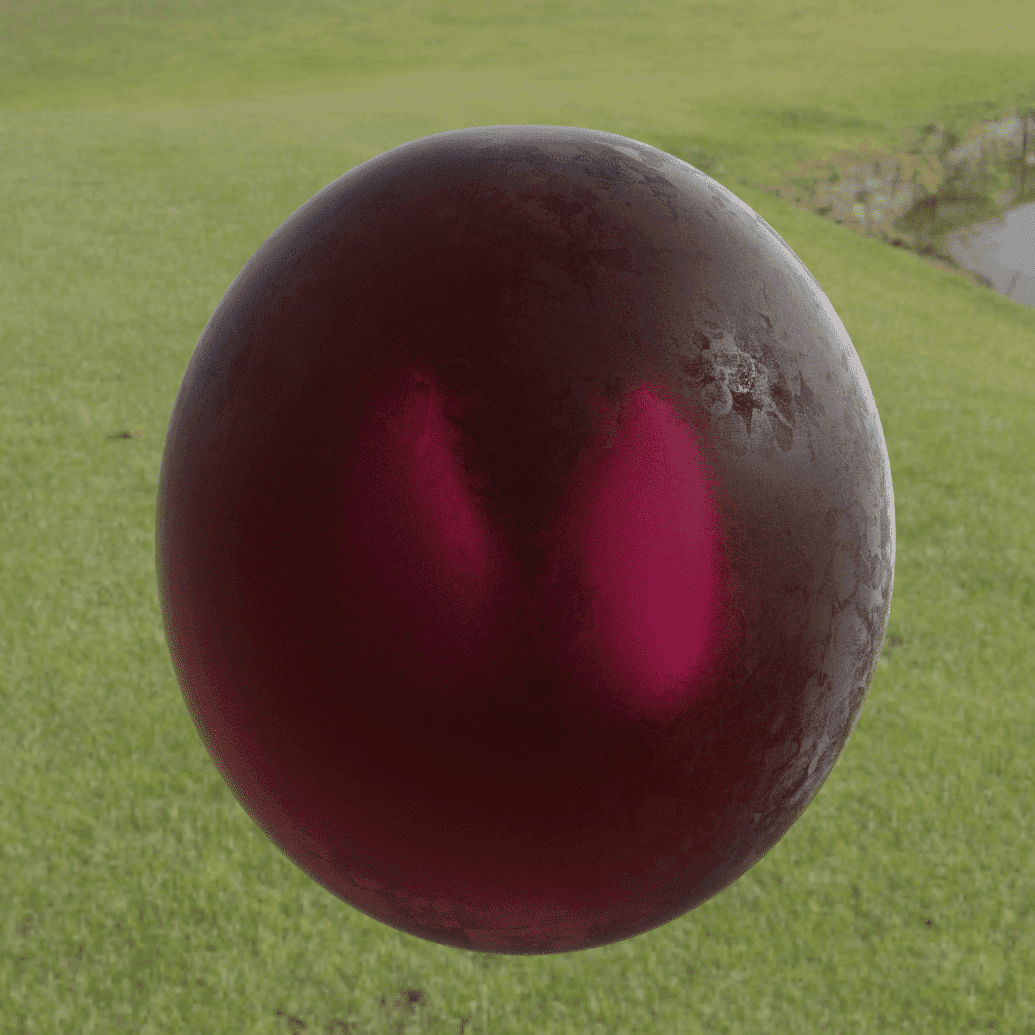}\llap{\includegraphics[height=1cm]{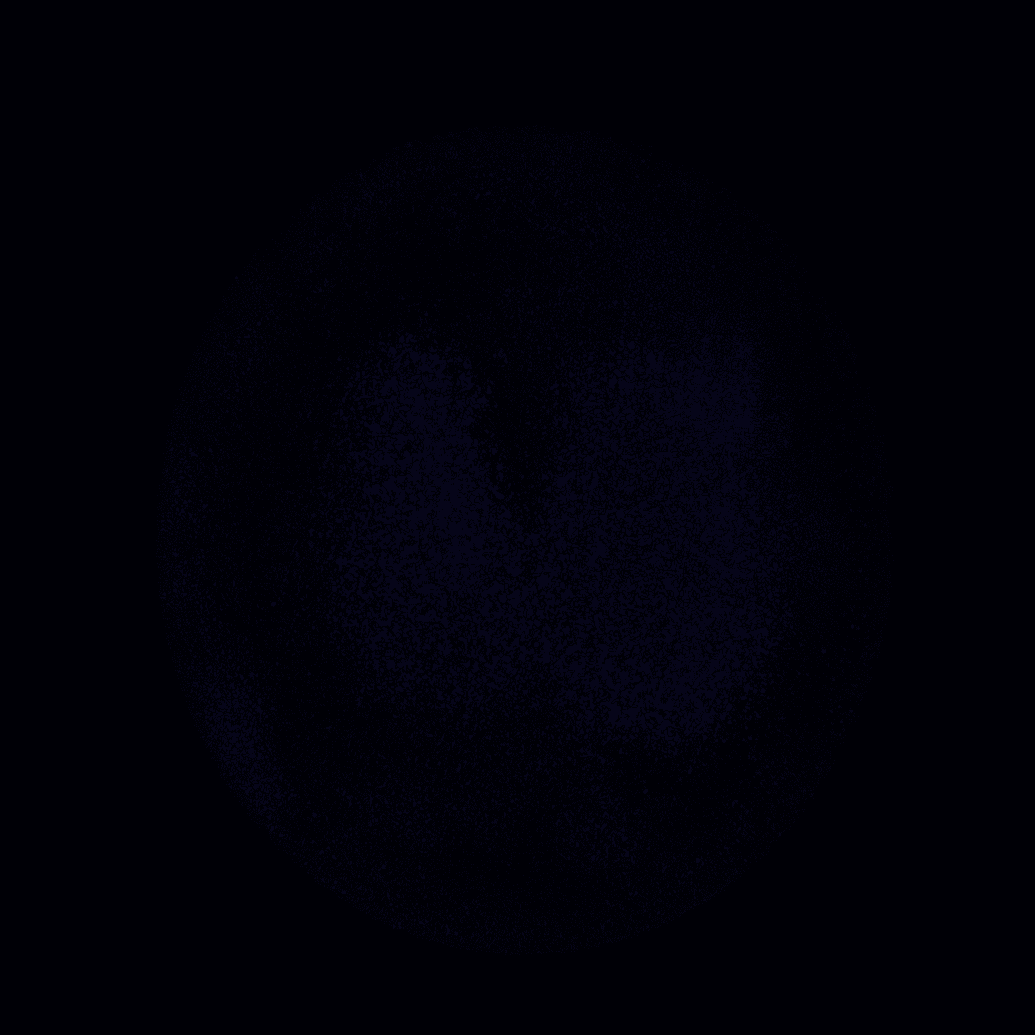}} \label{6c}&
    
    \xincludegraphics[height=3.75cm, label = \color{white}(d), pos = sw, labelbox = false]{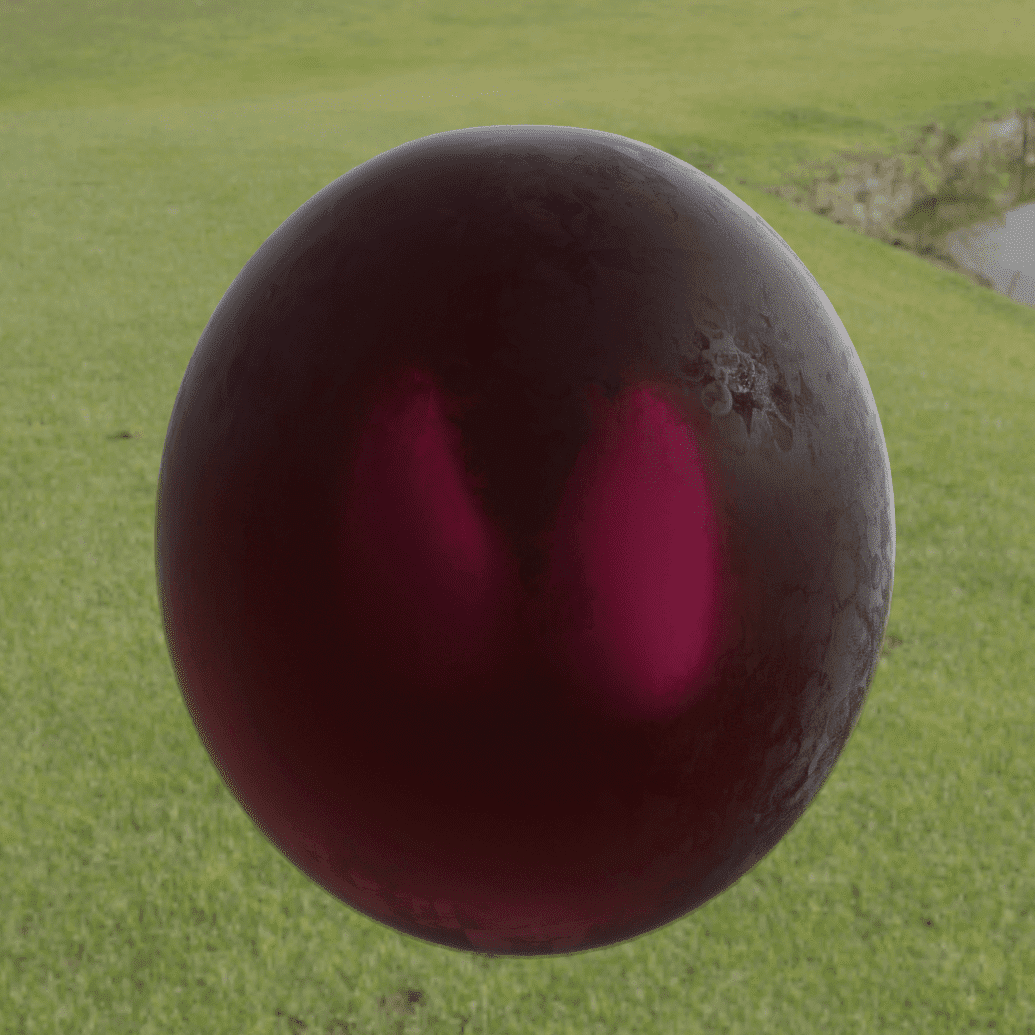}\llap{\includegraphics[height=1cm]{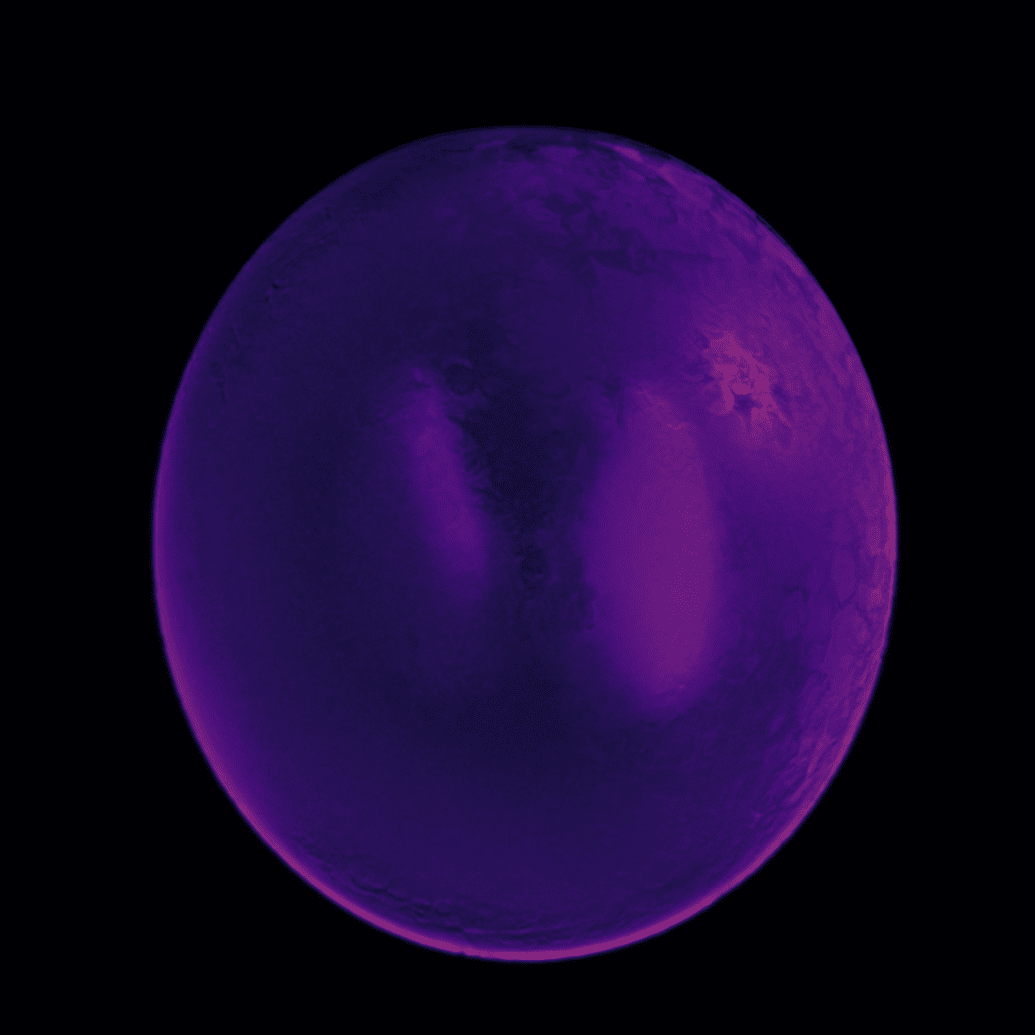}} \label{6d}\\[-2ex]

    \large \rotatebox{90}{Dragon} & 
    \xincludegraphics[height=3.75cm, label = \color{white}(e), pos = sw, labelbox = false]{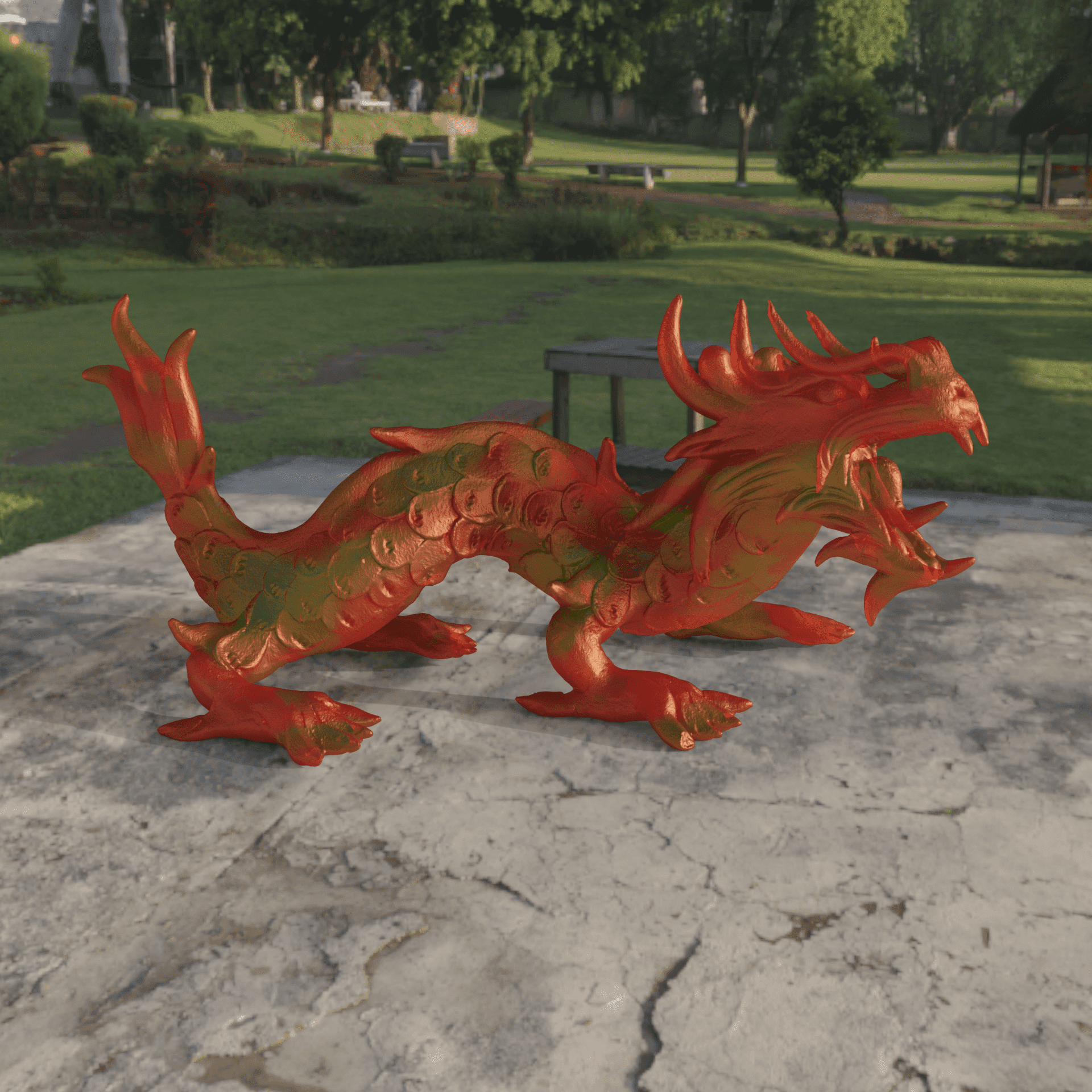}  \label{6e}&
    
    \xincludegraphics[height=3.75cm, label = \color{white}(f), pos = sw, labelbox = false]{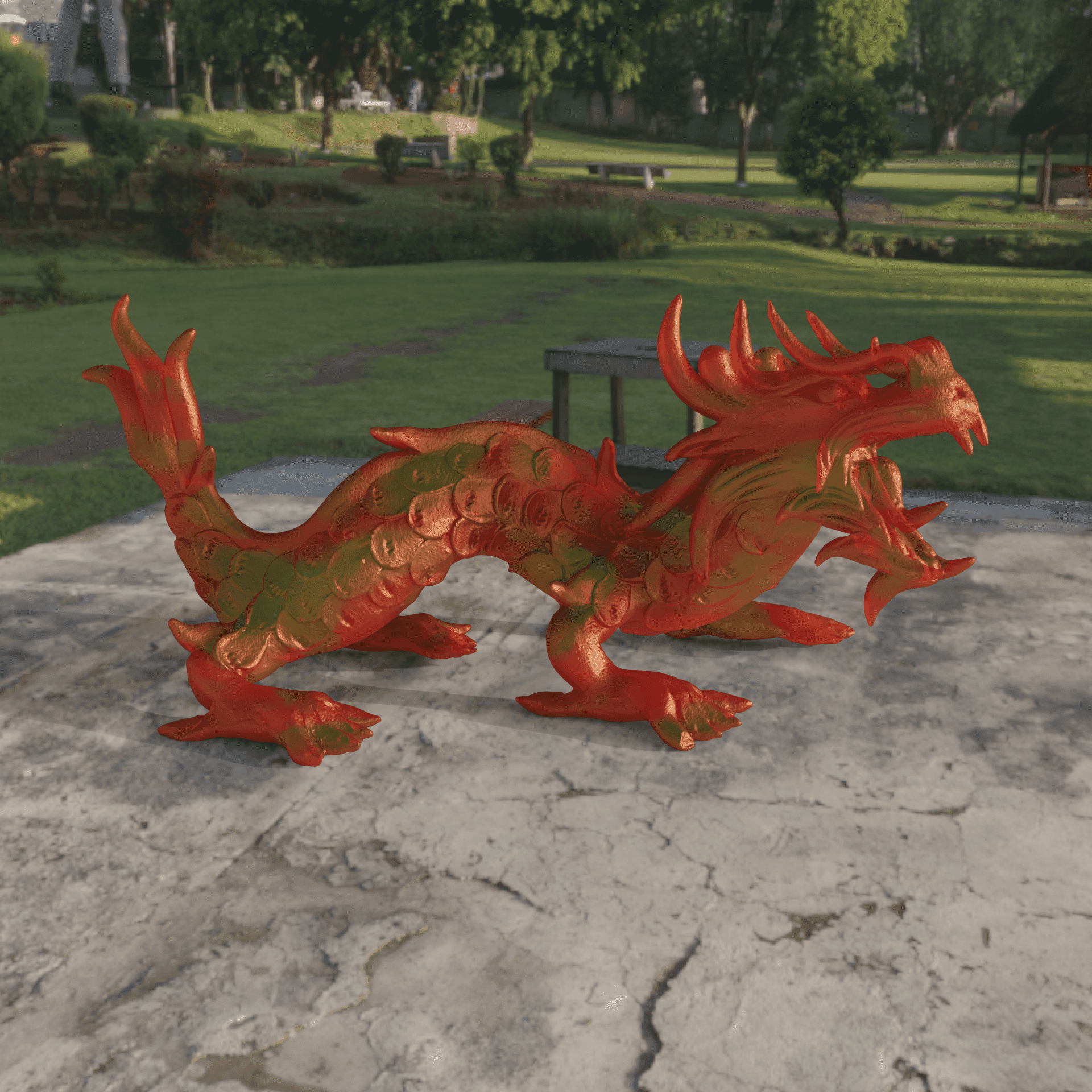}\llap{\includegraphics[height=1cm]{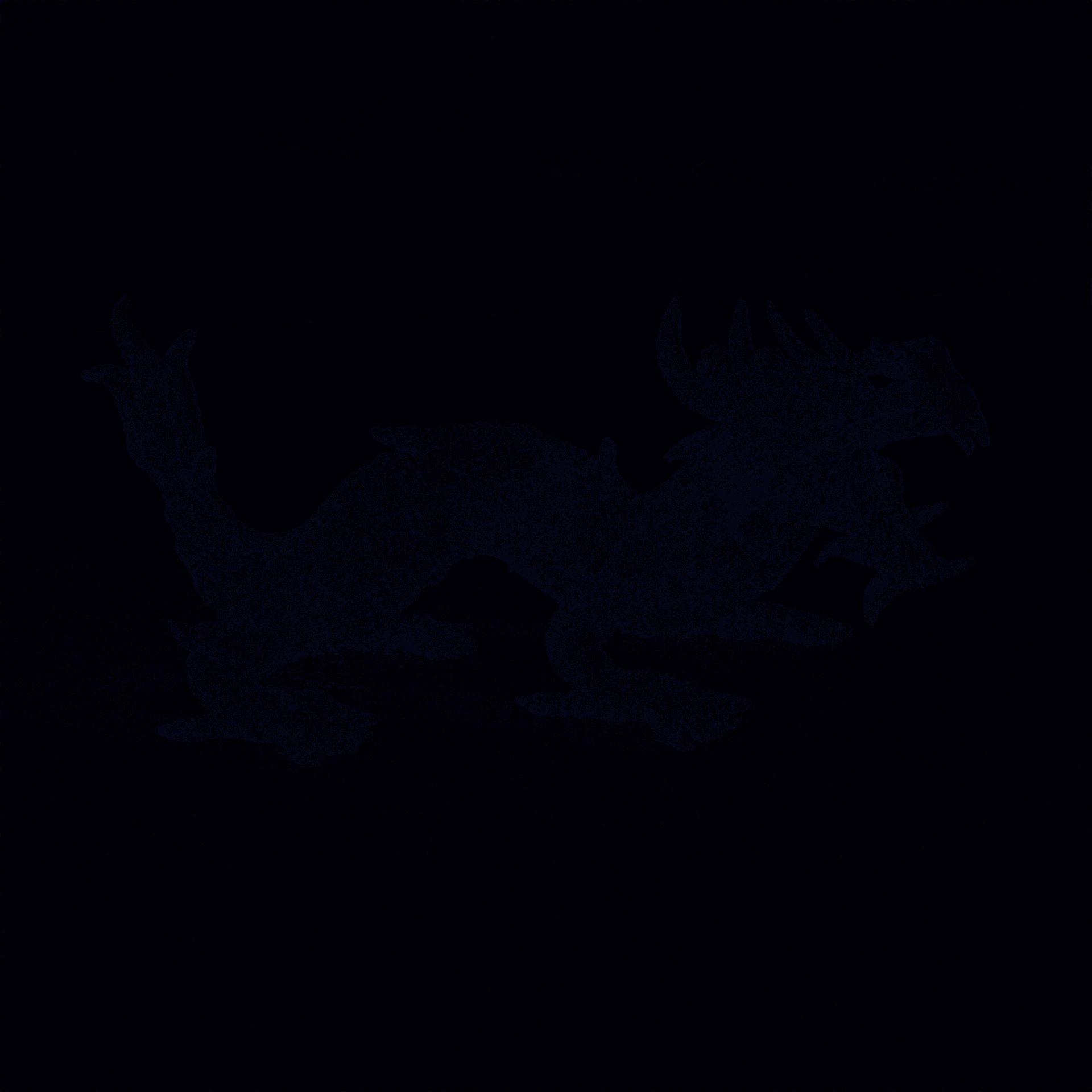}} \label{6f}&
    
    \xincludegraphics[height=3.75cm, label = \color{white}(g), pos = sw, labelbox = false]{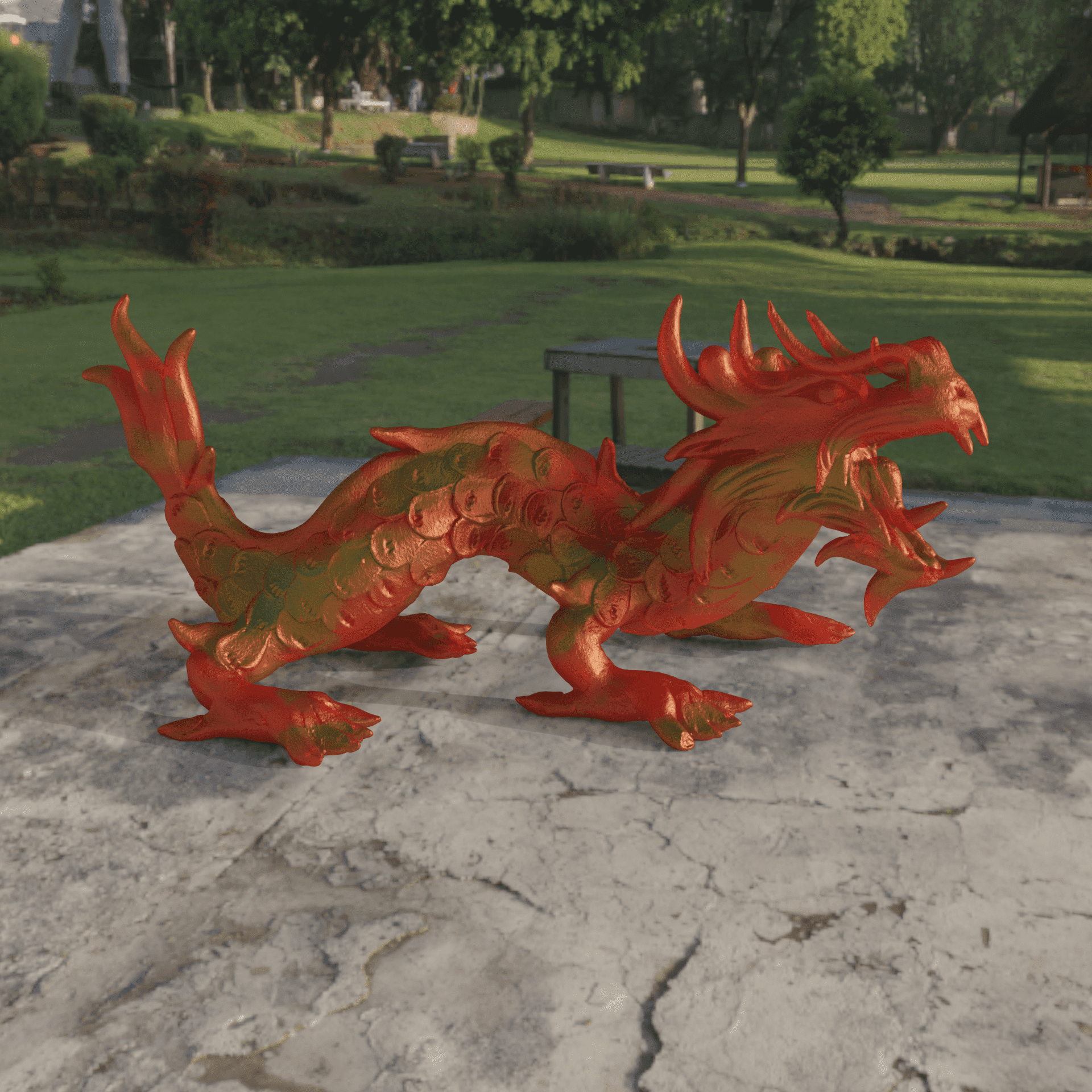}\llap{\includegraphics[height=1cm]{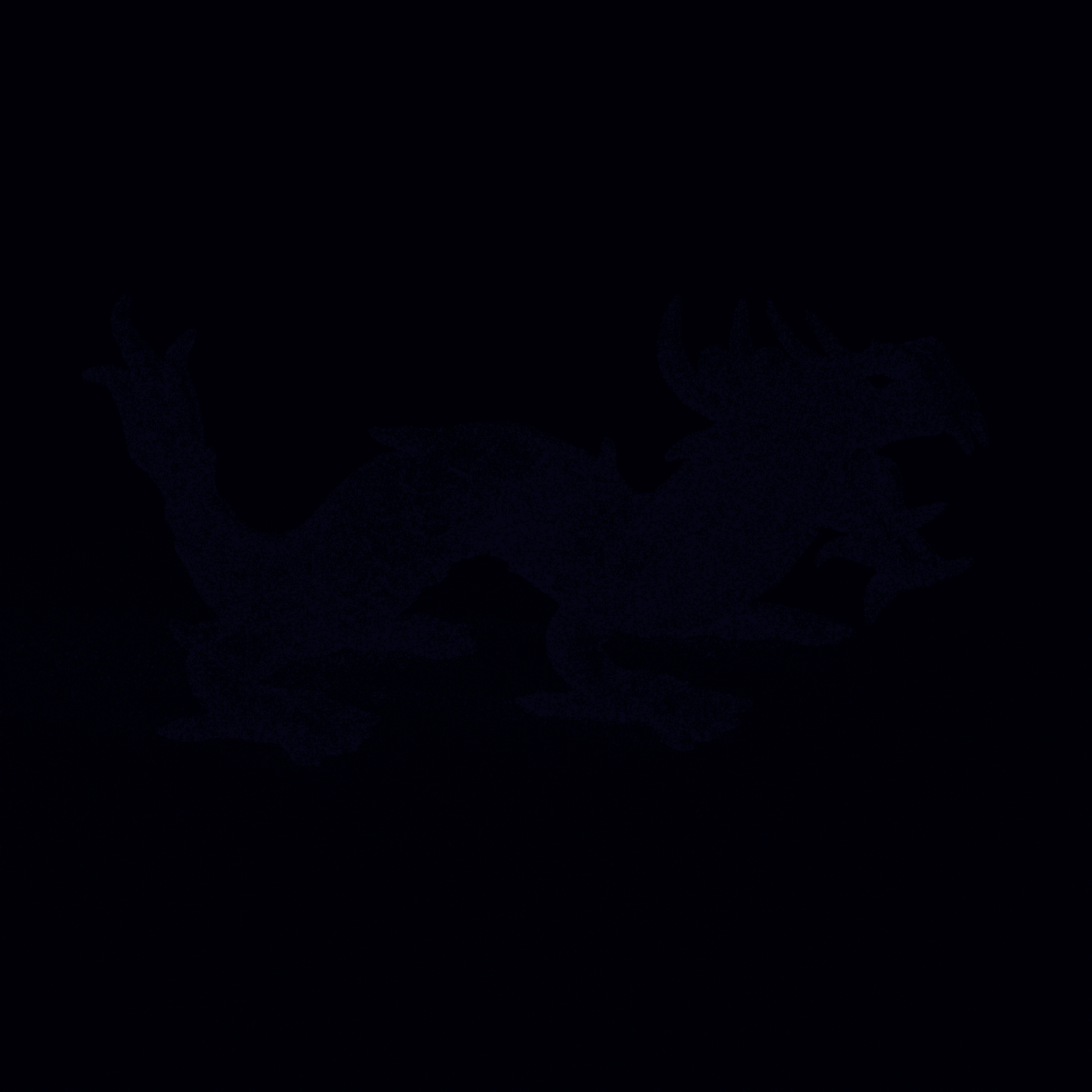}} \label{6g}&
    
    \xincludegraphics[height=3.75cm, label = \color{white}(h), pos = sw, labelbox = false]{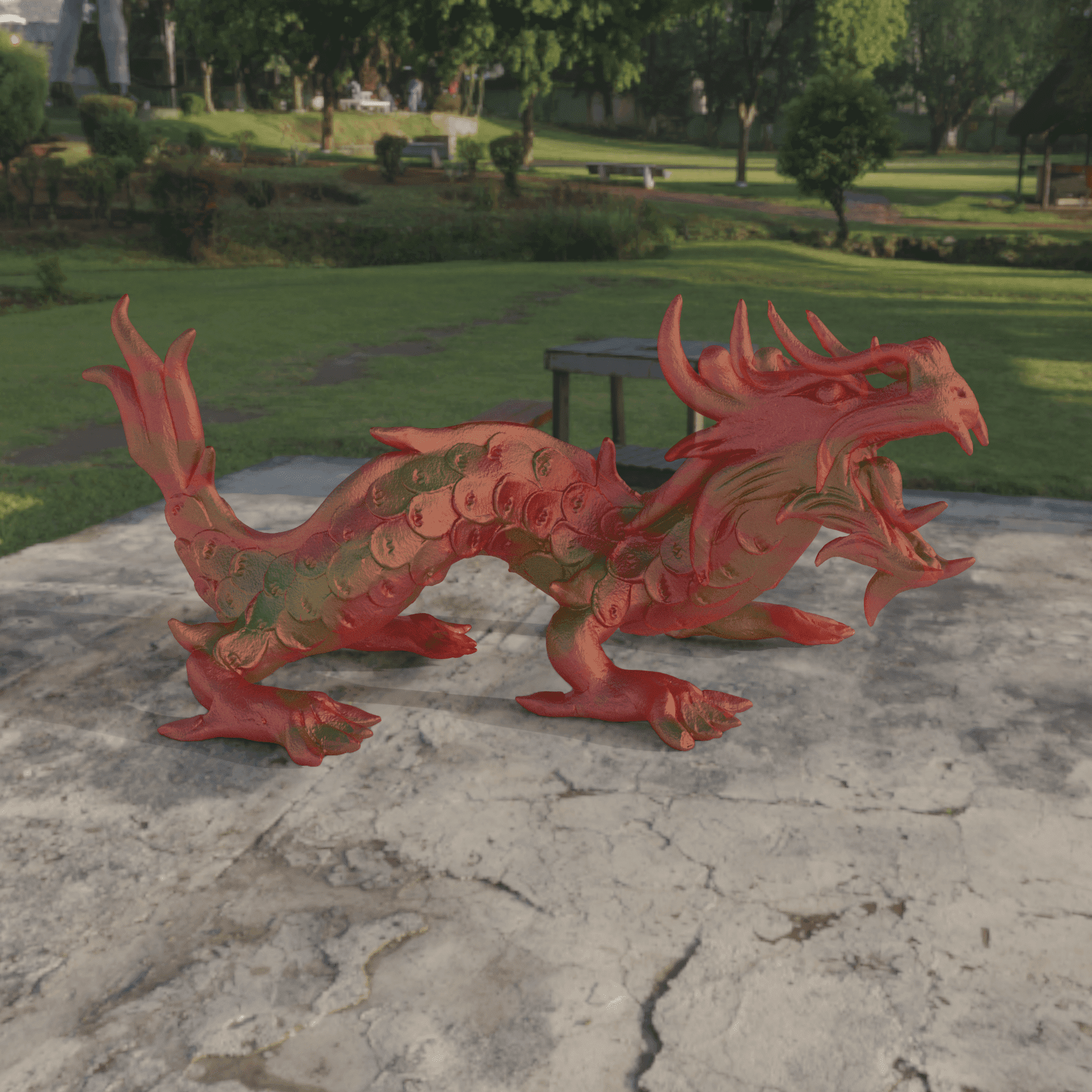}\llap{\includegraphics[height=1cm]{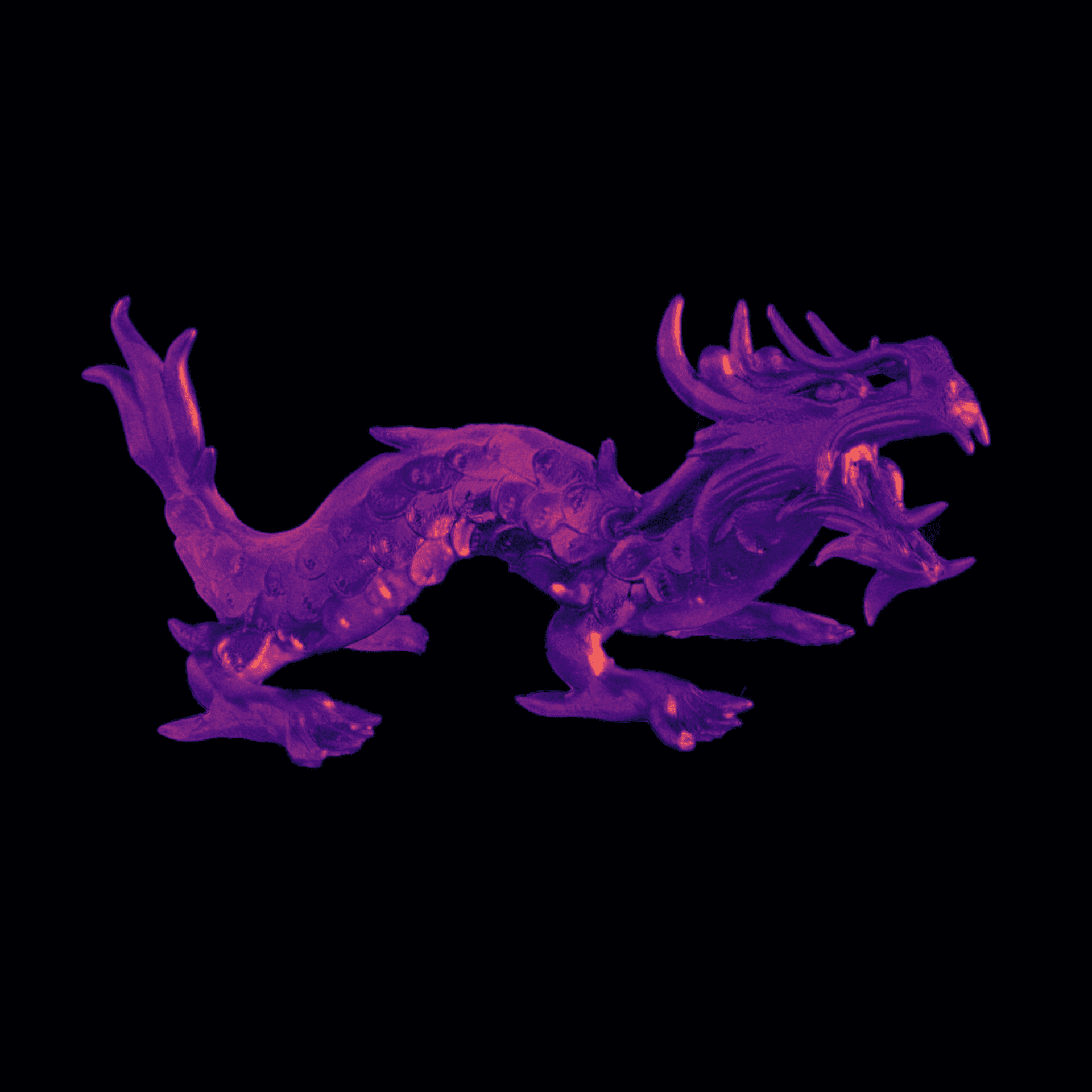}} \label{6h}\\[-2ex]

    \large \rotatebox{90}{Paperweight} & 
    \xincludegraphics[height=3.75cm, label = \color{white}(i), pos = sw, labelbox = false]{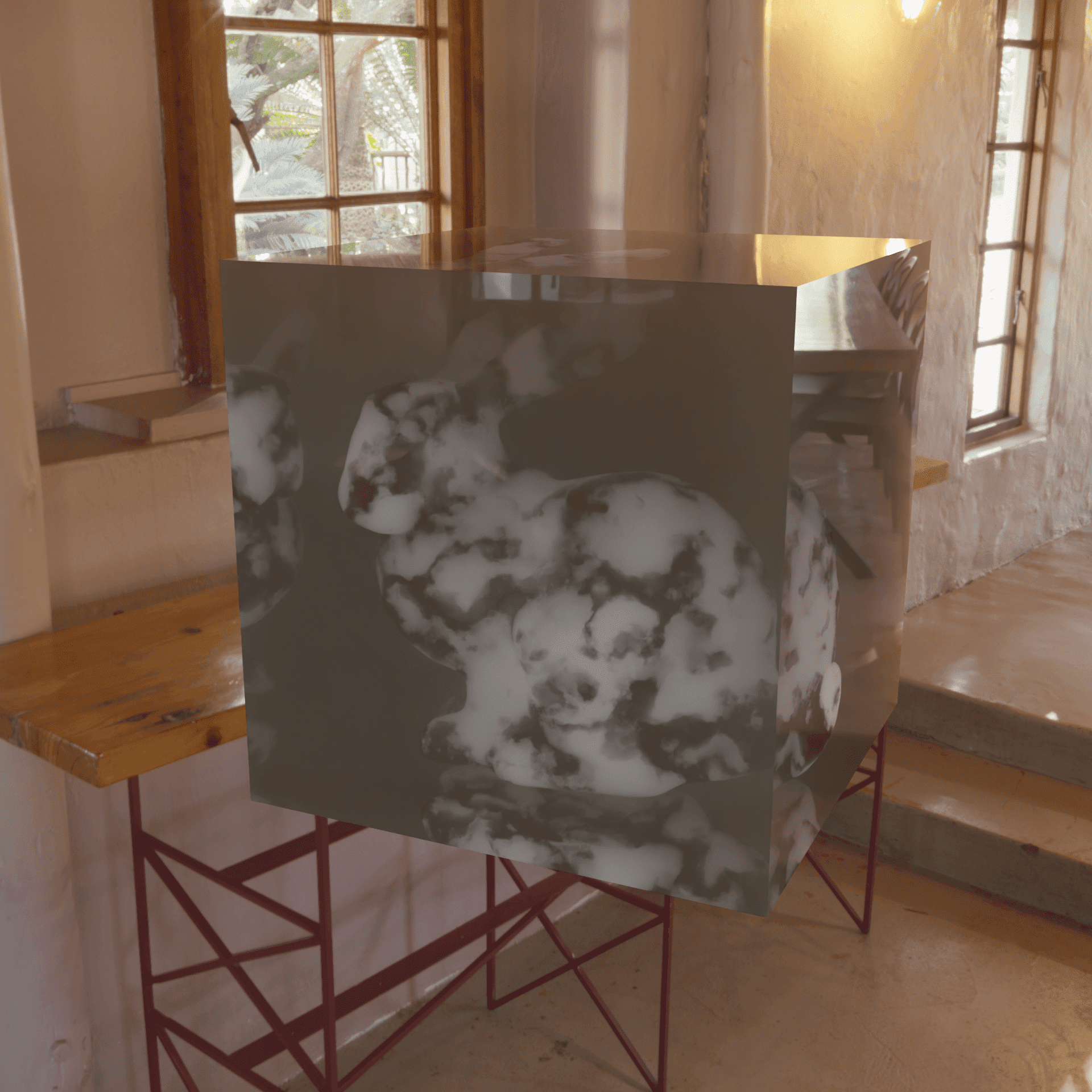}  \label{6i}&
    
    \xincludegraphics[height=3.75cm, label = \color{white}(j), pos = sw, labelbox = false]{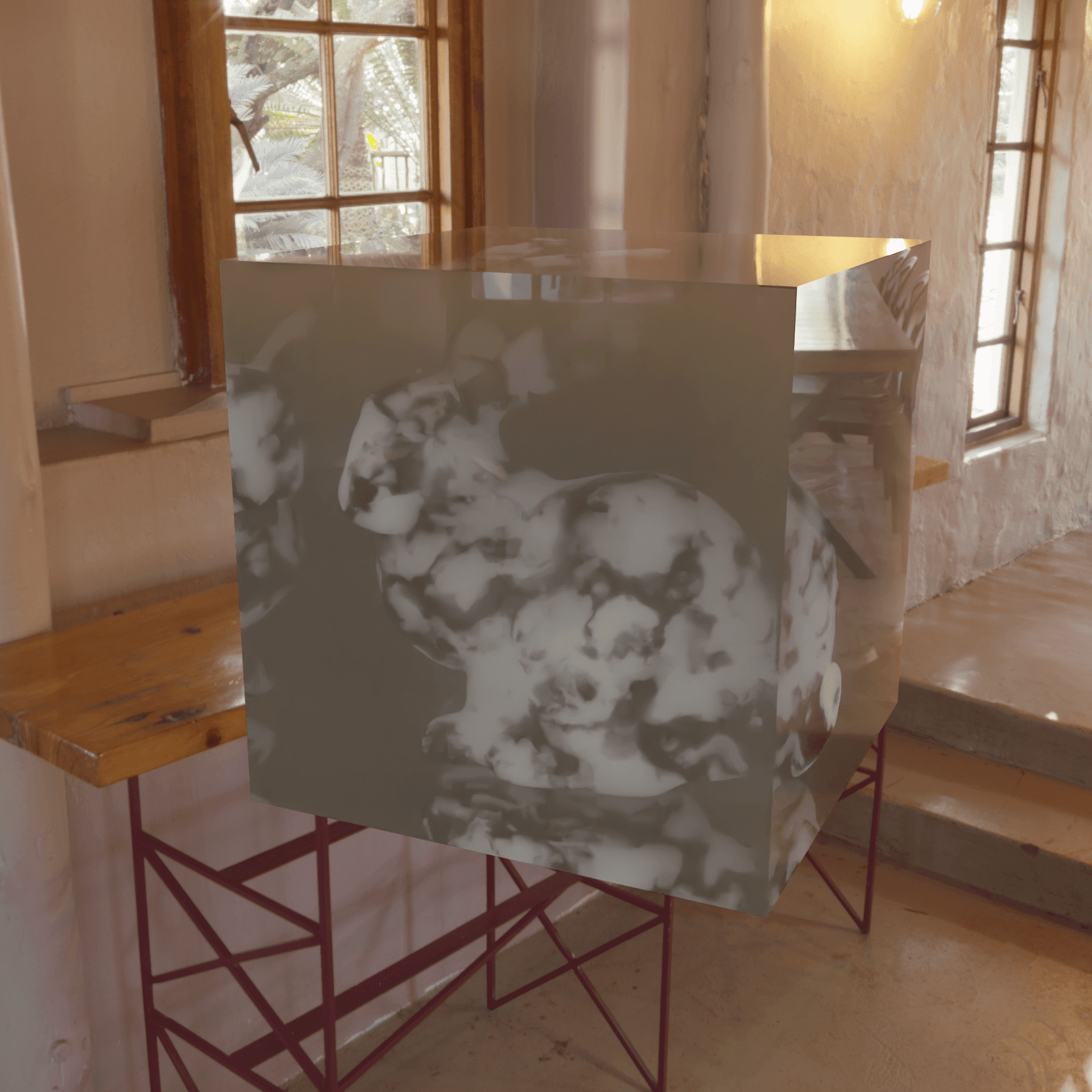}\llap{\includegraphics[height=1cm]{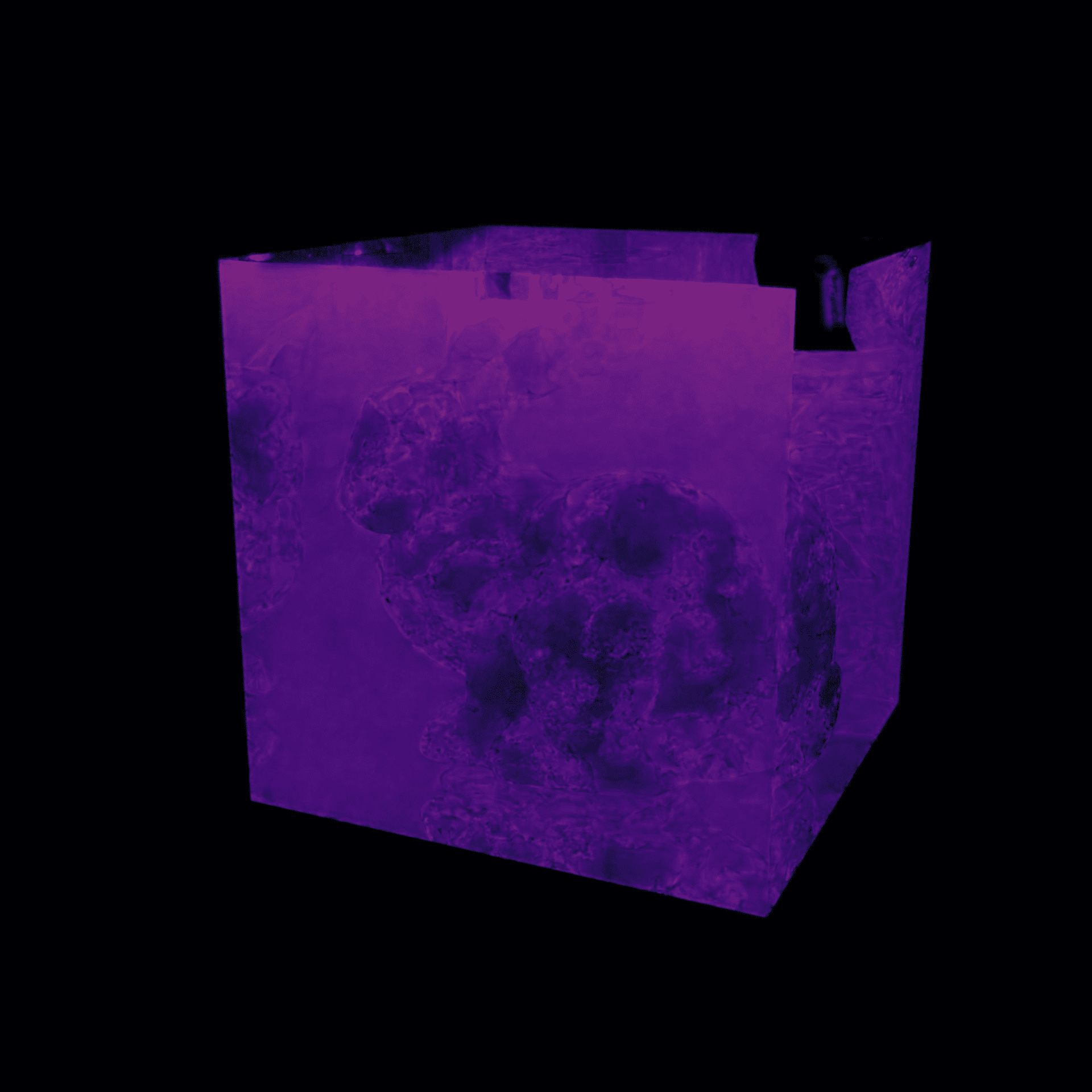}} \label{6j}&
    
    \xincludegraphics[height=3.75cm, label = \color{white}(k), pos = sw, labelbox = false]{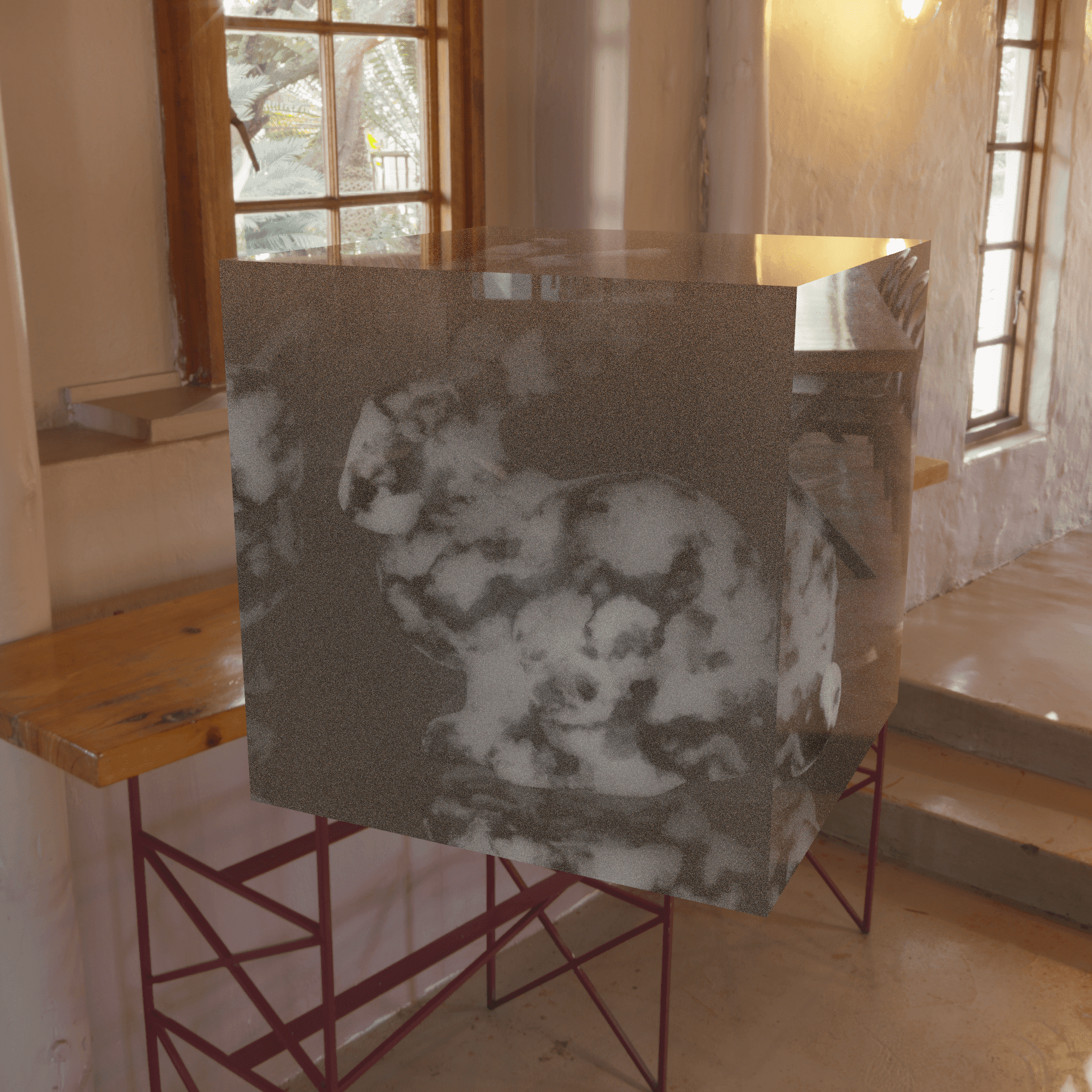}\llap{\includegraphics[height=1cm]{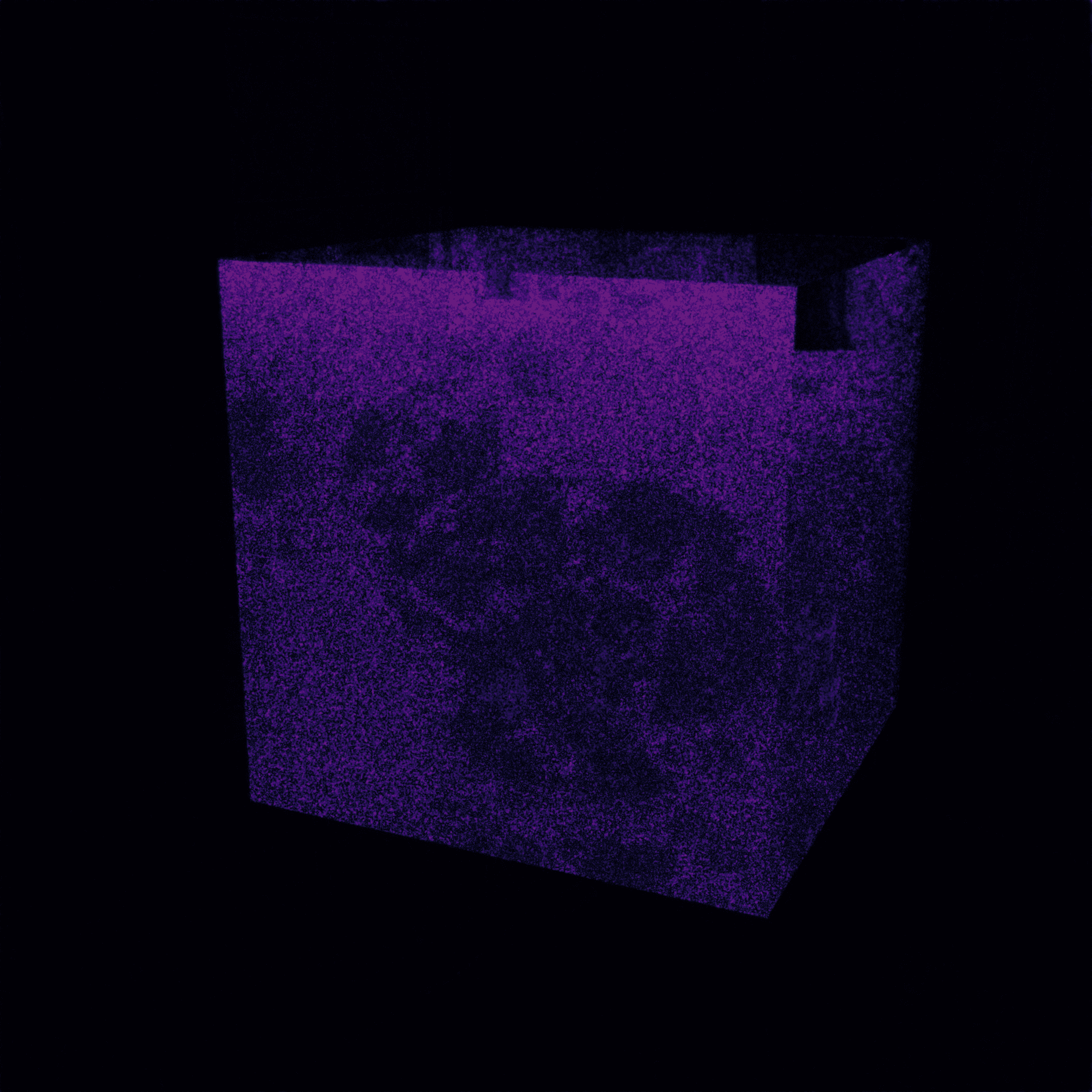}} \label{6k}&
    
    \xincludegraphics[height=3.75cm, label = \color{white}(l), pos = sw, labelbox = false]{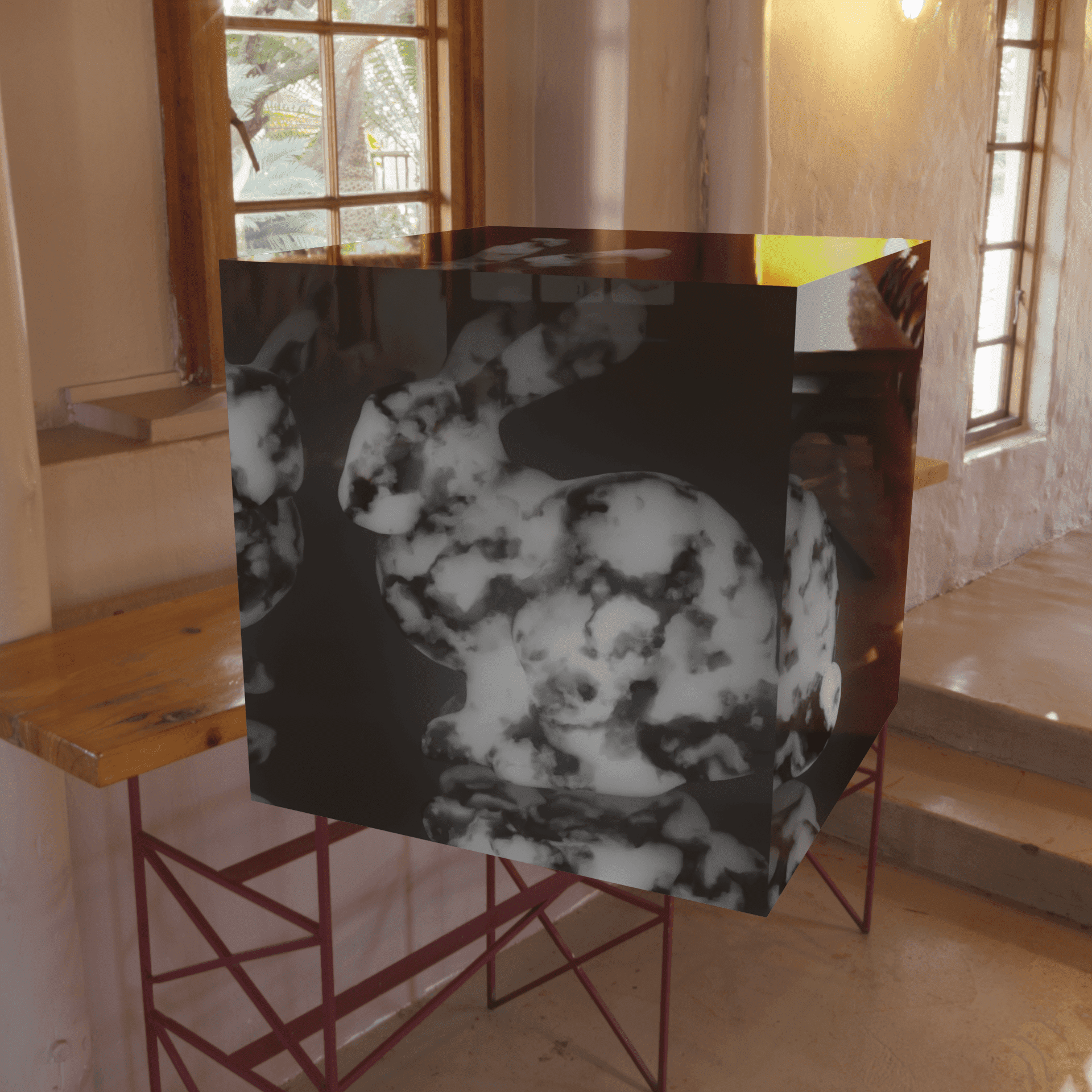}\llap{\includegraphics[height=1cm]{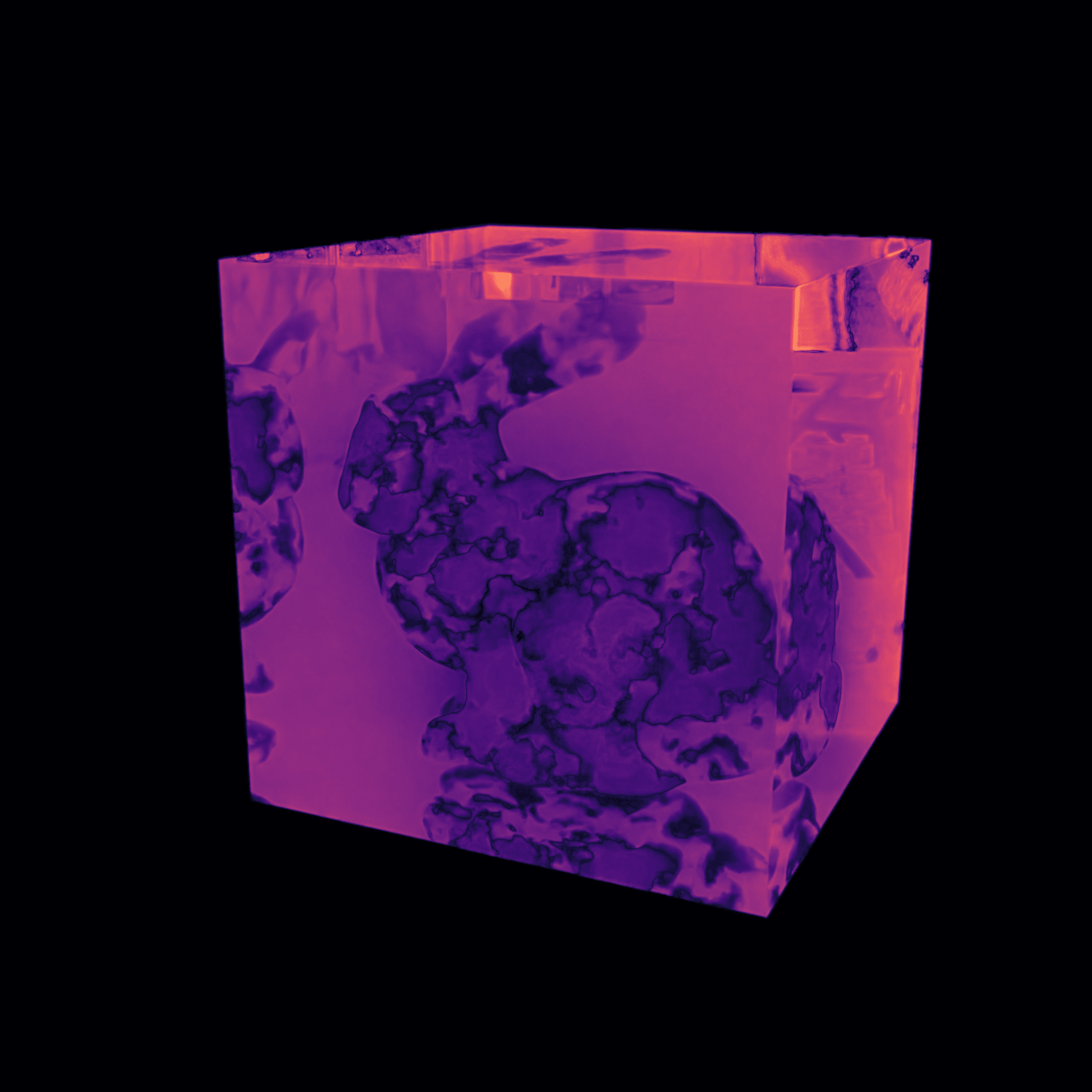}} \label{6l}\\[-2ex]

    \large \rotatebox{90}{Drink with ice} & 
    \xincludegraphics[height=3.75cm, label = \color{white}(m), pos = sw, labelbox = false]{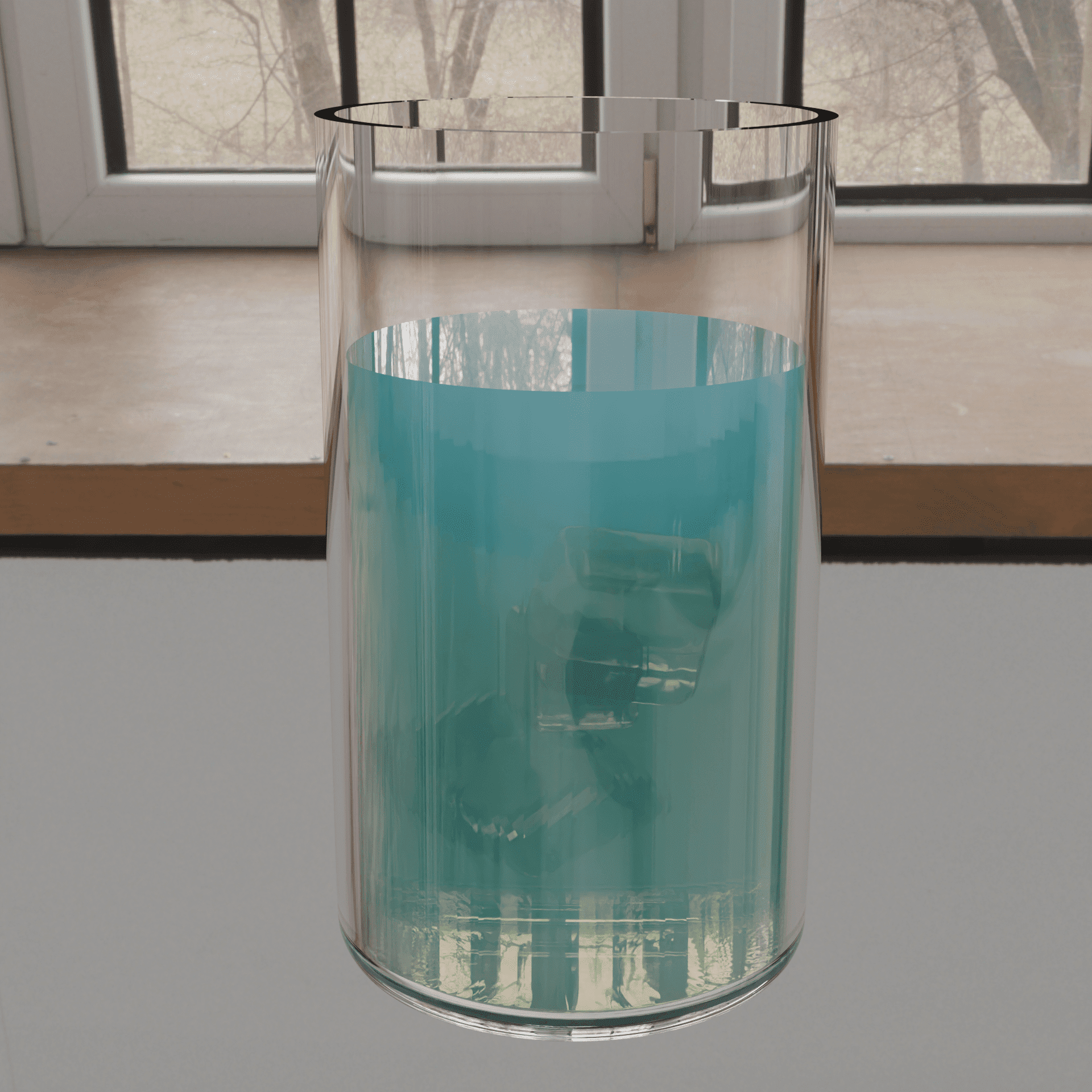}  \label{6m}&
    
    \xincludegraphics[height=3.75cm, label = \color{white}(n), pos = sw, labelbox = false]{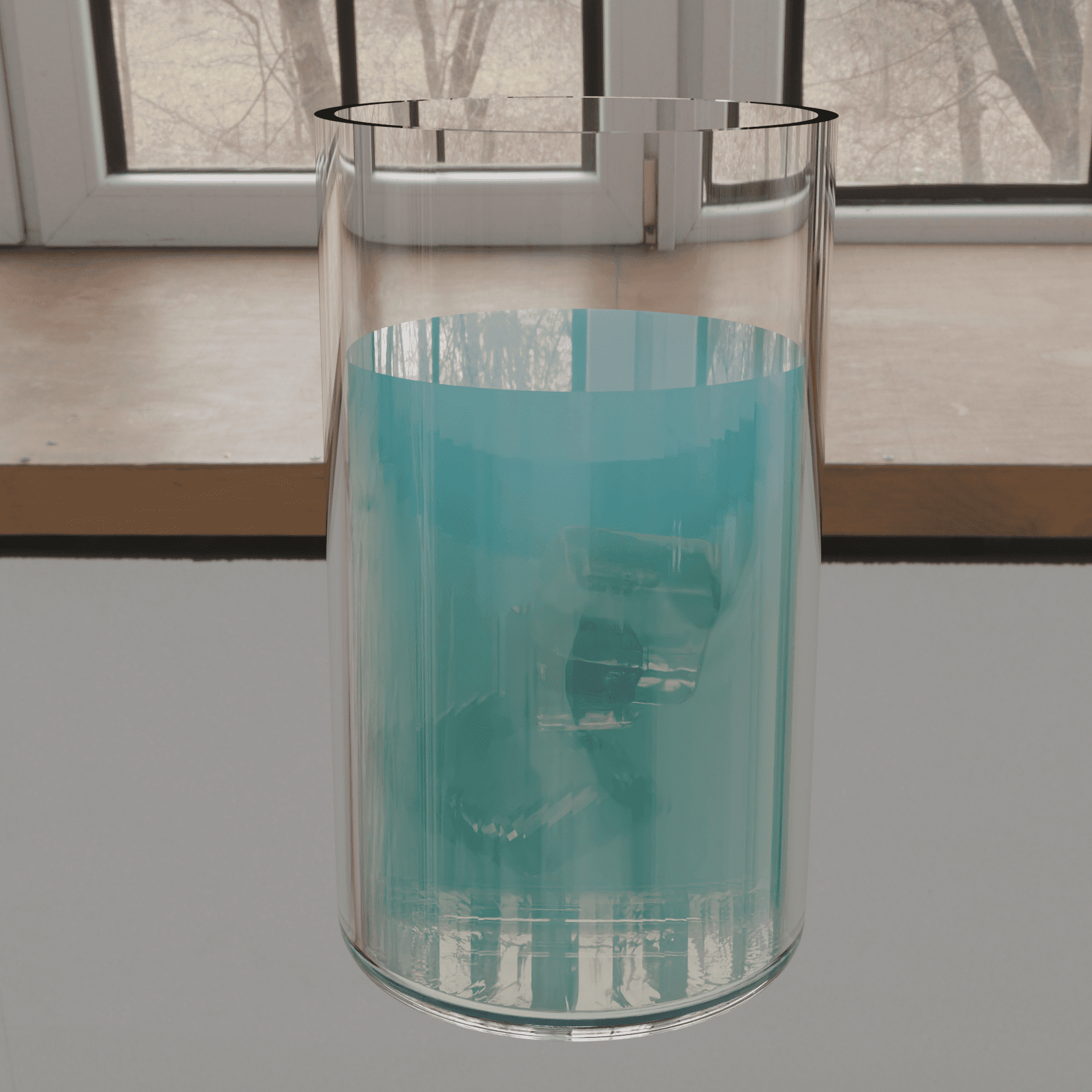}\llap{\includegraphics[height=1cm]{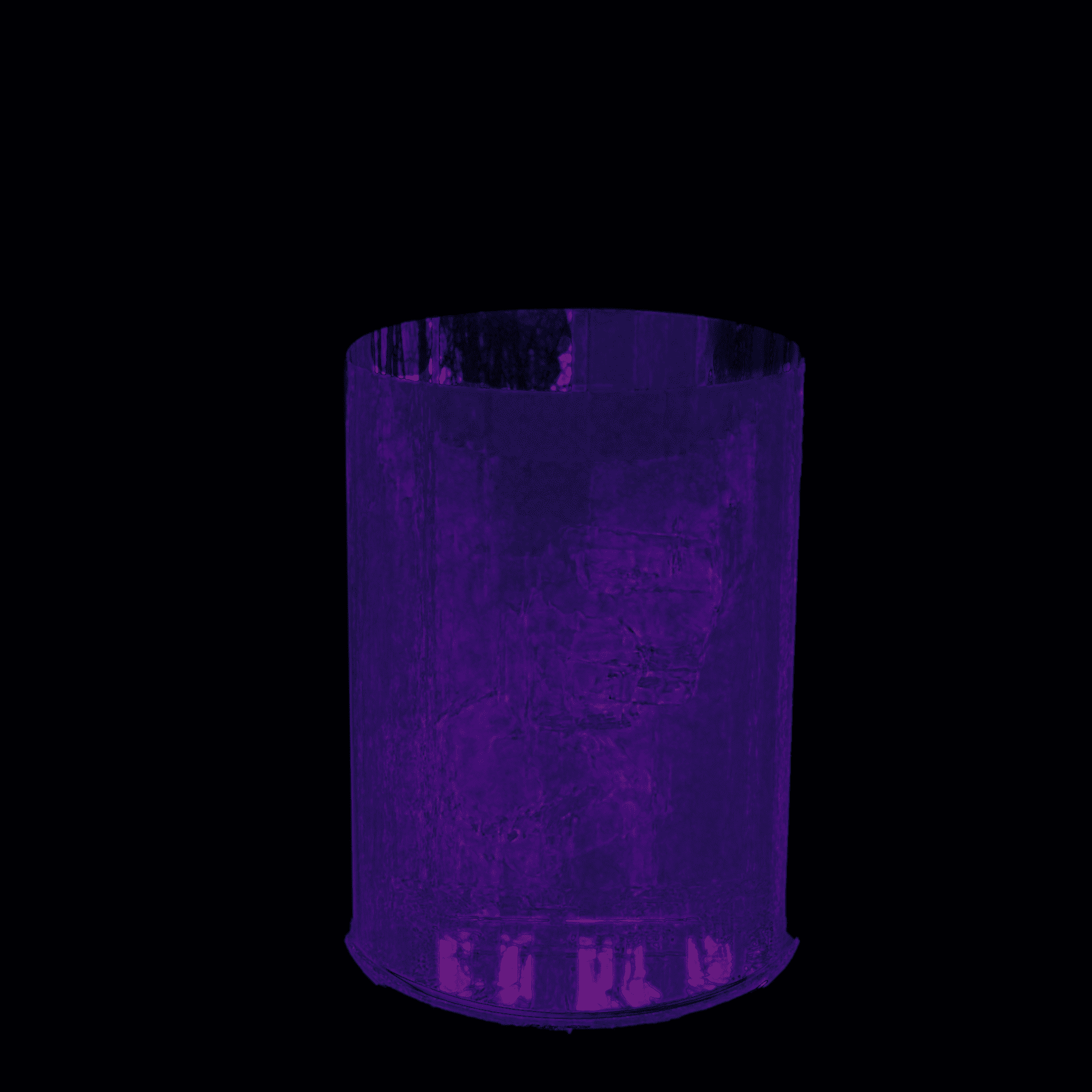}} \label{6n}&
    
    \xincludegraphics[height=3.75cm, label = \color{white}(o), pos = sw, labelbox = false]{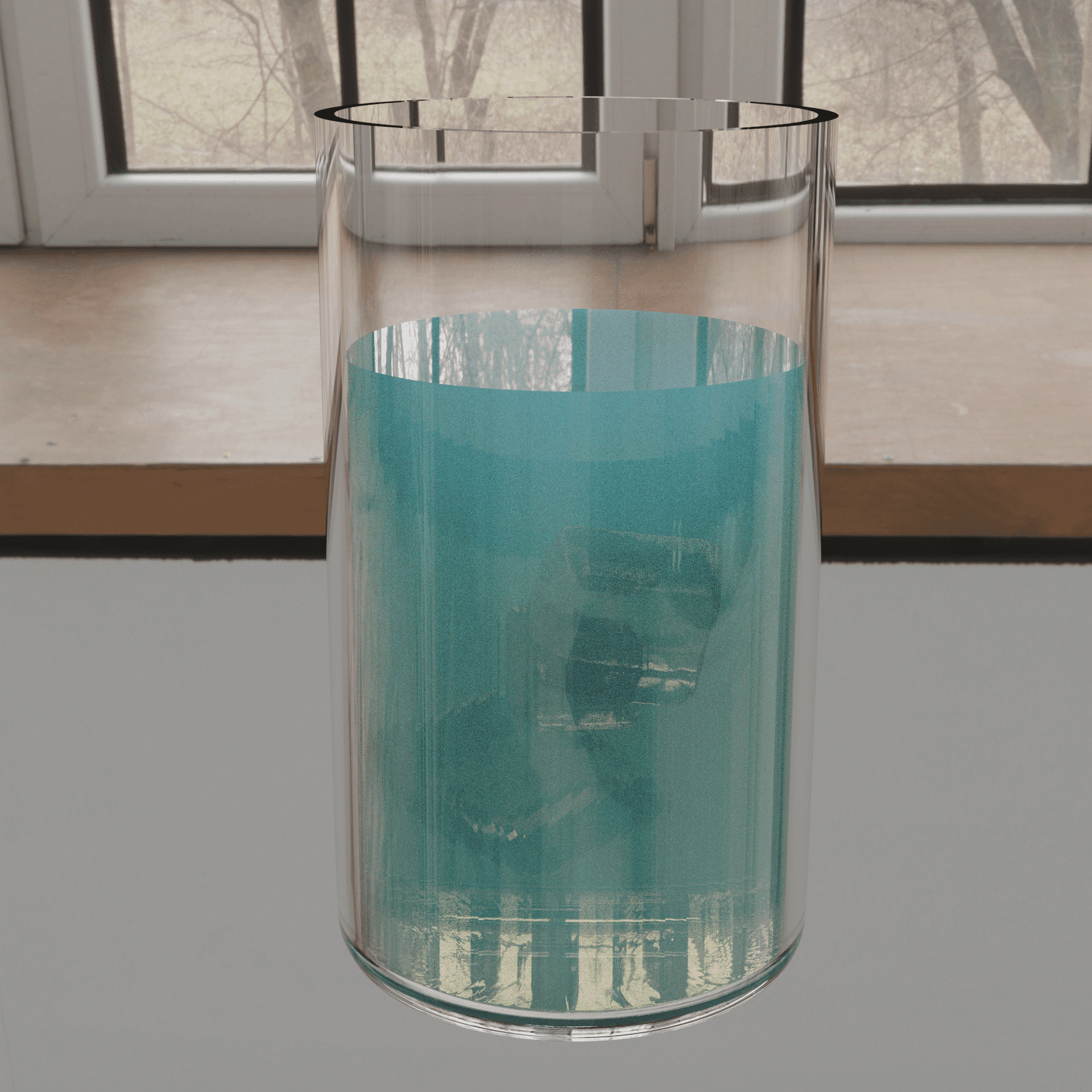}\llap{\includegraphics[height=1cm]{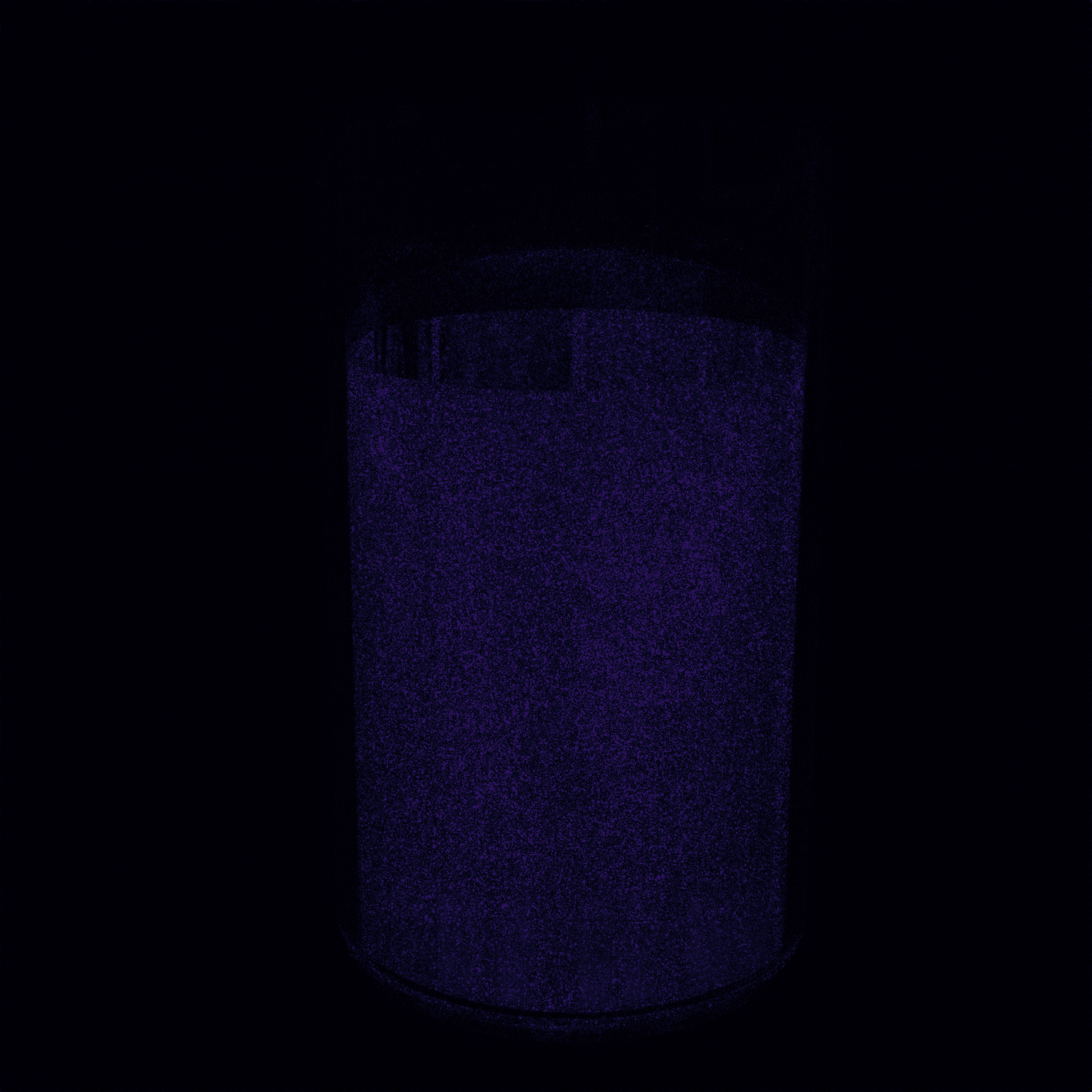}} \label{6o}&
    
    \xincludegraphics[height=3.75cm, label = \color{white}(p), pos = sw, labelbox = false]{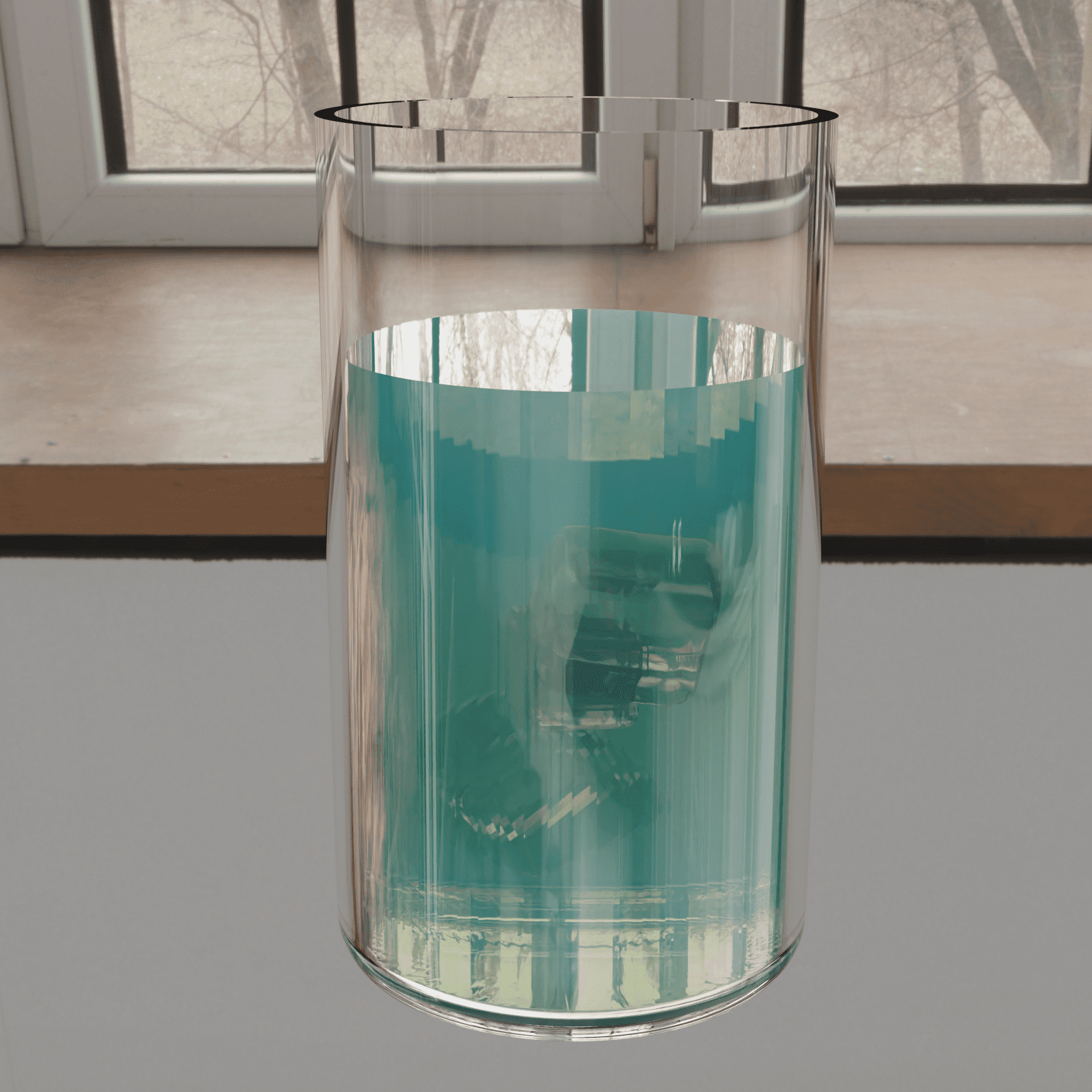}\llap{\includegraphics[height=1cm]{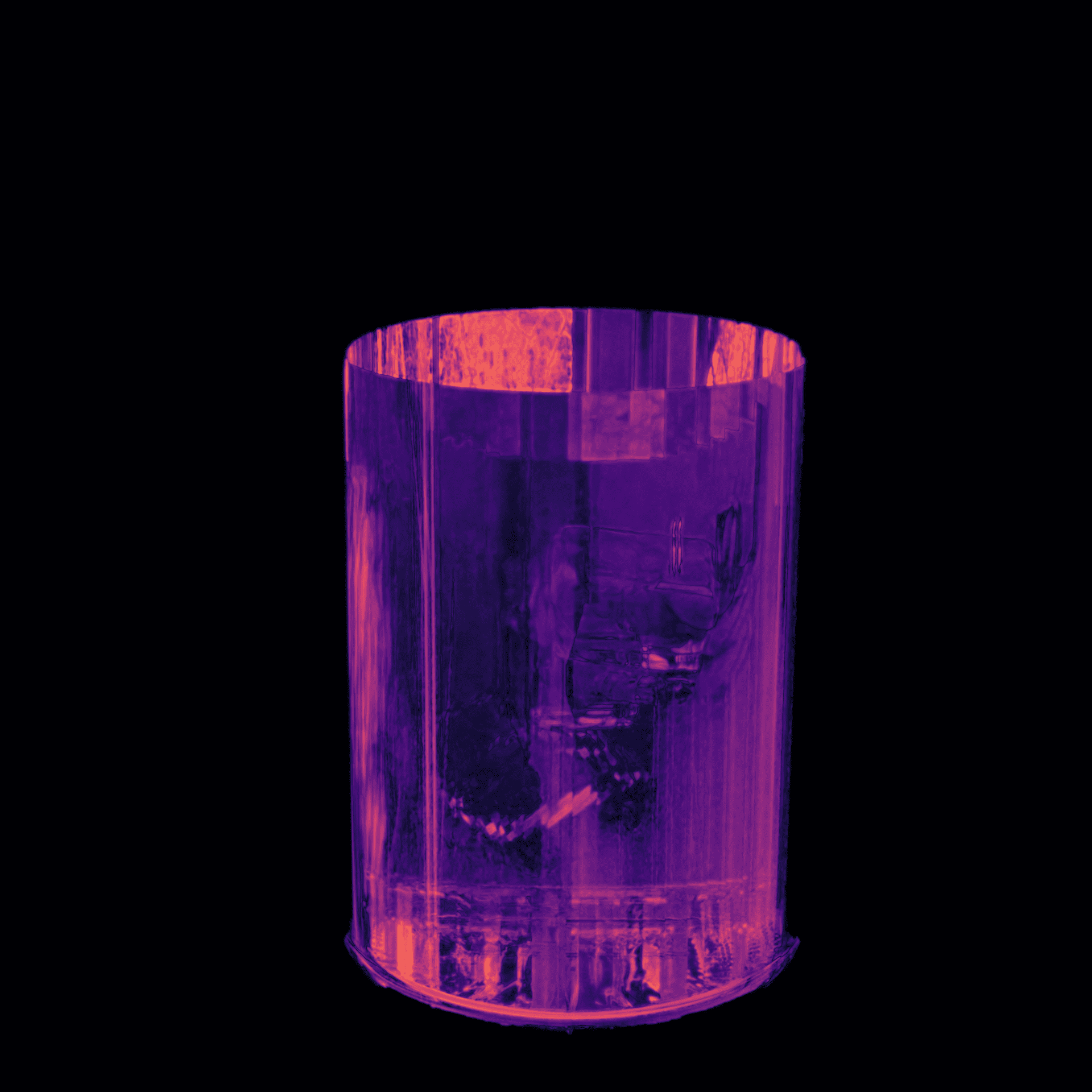}} \label{6p}\\[-2ex]

    \large \rotatebox{90}{Lucy} & 
    \xincludegraphics[height=3.75cm, label = \color{white}(q), pos = sw, labelbox = false]{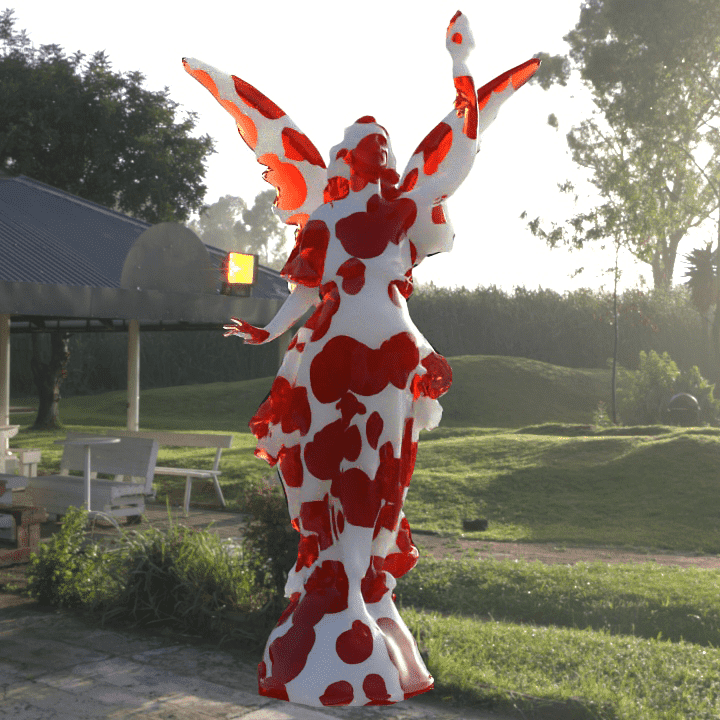}  \label{6q}&
    
    \xincludegraphics[height=3.75cm, label = \color{white}(r), pos = sw, labelbox = false]{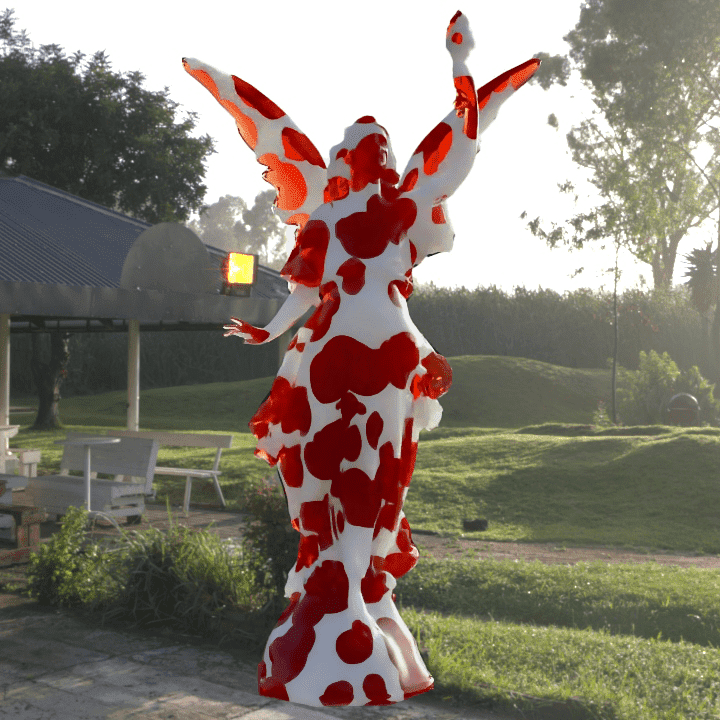}\llap{\includegraphics[height=1cm]{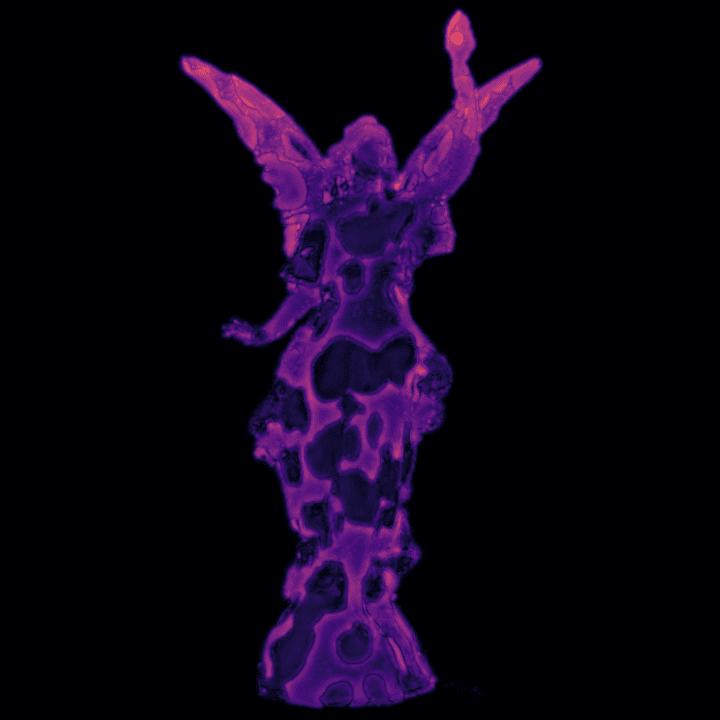}} \label{6r}&
    
    \xincludegraphics[height=3.75cm, label = \color{white}(s), pos = sw, labelbox = false]{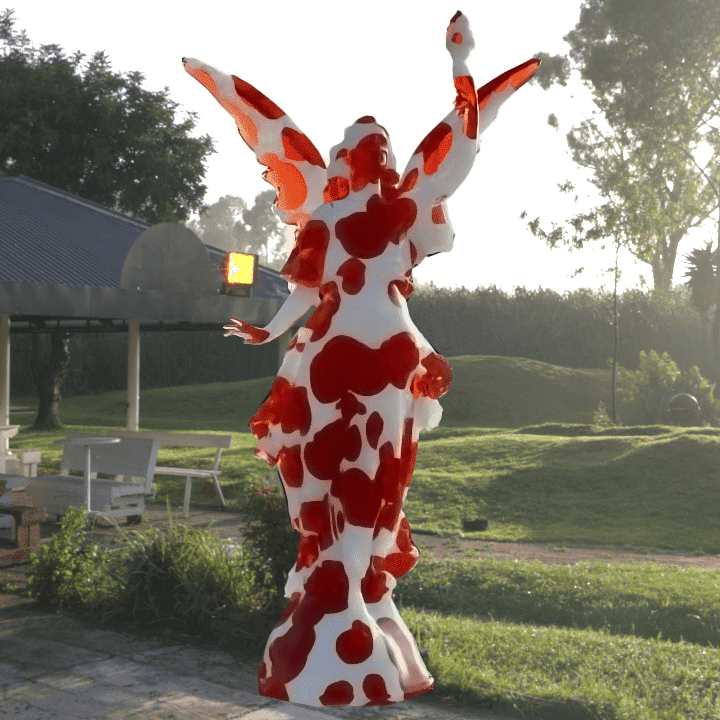}\llap{\includegraphics[height=1cm]{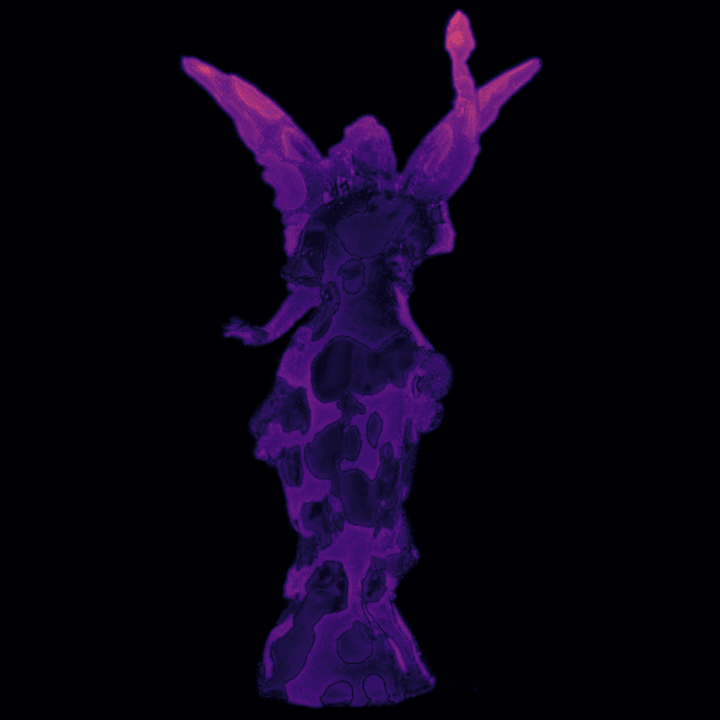}} \label{6s}&
    
    \xincludegraphics[height=3.75cm, label = \color{white}(t), pos = sw, labelbox = false]{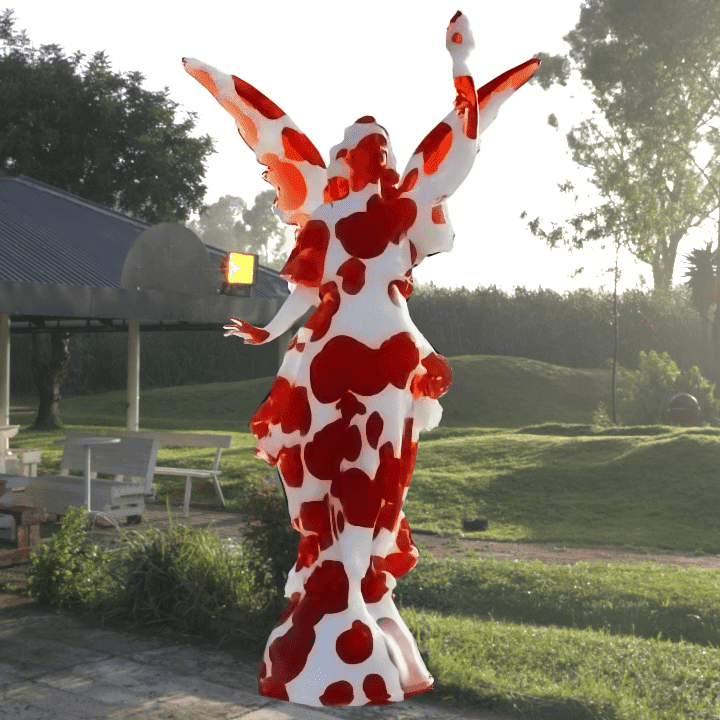}\llap{\includegraphics[height=1cm]{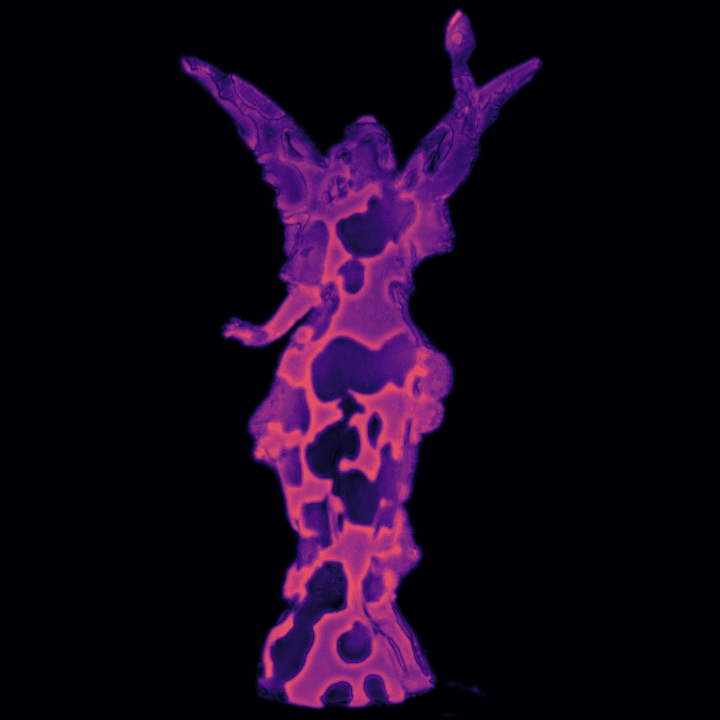}} \label{6t}\\[-4ex]

    \end{tabularx}
  \caption{The ability of different versions of our model to represent path traced reference images. We show traditional importance sampling of the environment map~\cite{pharr2023physically} (env IS) compared against our importance sampling technique described in Sec.~\ref{importance_sampling} (our IS) and the spherical harmonic representation of our model for real-time rendering described in Sec.~\ref{PRT} (SH). Each image is rendered at 256 samples per pixel (SPP). The bottom right corners of the images show the \reflectbox{F}lip error metric.} 
  \label{fig:maincomparison_ours} 
\end{figure*}

    \paragraph*{Quality of representations.} To provide the reader with an opportunity to qualitatively assess the ability of our method to represent object appearance, Figure \ref{fig:maincomparison_ours} provides a comparison of our different types of BSSRDF representations with path traced reference images. We specifically compare use of our appearance specification module with environment importance sampling~\cite{pharr2023physically} as well as use of our own importance sampling module. The scenes are the same as in the quantitative analysis in Table~\ref{fig:table_comp_ours}.

    When using environment importance sampling, we may sample directions that are not very relevant to our appearance representation. In contrast, our importance sampling module puts more focus on our appearance specification module resulting in reduced variance at equal sample counts. To ensure a fair comparison, we render the images at 256 samples per pixel with and without our importance sampling module. If an object is placed in a high dynamic range (HDR) environment with large variation of intensities, or in a scene with an idealized directional source, our importance sampling would not necessarily be better. In general, multiple importance sampling would be desirable, and a network can be used to find multiple importance sampling weights~\cite{is_neuralbrdf}.

    Additionally, we compare with our implementation of the spherical harmonic (SH) representation of our neural BSSRDF to demonstrate a notable performance improvement at the expense of accuracy. While the \reflectbox{F}LIP error rates is higher in the SH representation compared to our appearance specification module, the objects seem to maintain an acceptable visual fidelity. For a visual comparison, we refer the reader to the supplementary video which demonstrates interactive rendering of our network representation and real-time rendering of our spherical harmonic representation.
    
    

    \begin{table}
    \centering
    \caption{Image quality metrics for NeRFactor~\cite{zhangnerf_21}, NeuMip~\cite{kuznetsov2021neumip}, and the network version of our method when comparing with a path traced reference image. The values were computed for rendered images of resolution $1920^2$ (MSE, SSIM1, \reflectbox{F}LIP). The SSIM metric prefers blur to noise. If we downsample the images to $480^2$, the SSIM values change significantly (SSIM2).}
    \label{fig:table_comp}
    \begin{tabular}{@{\:}p{0.18\linewidth}@{\hspace{2ex}}c@{\hspace{2ex}}c@{\hspace{2ex}}c@{\hspace{2ex}}c@{\:}} 
    &&&&\\[-4.5ex]
    \hline
    \rule{0pt}{2.5ex}\textbf{Experiment} & \textbf{Metric} & \textbf{NeRFactor} & \textbf{NeuMip} & \textbf{Ours(Net)}\\[0.5ex] 
    \hline
    \multirow{3}{=}{Grape (outdoors)} & \rule{0pt}{2.5ex}MSE$\downarrow$ & $3.34\cdot10^{-3}$ & $1.82\cdot10^{-4}$ & $\mathbf{9.28\cdot10^{-5}}$\\ 
        & SSIM1$\uparrow$ & $8.86\cdot10^{-1}$ & $\mathbf{9.82\cdot10^{-1}}$ & $9.49\cdot10^{-1}$\\
        & SSIM2$\uparrow$ & $8.89\cdot10^{-1}$ & $9.90\cdot10^{-1}$ & $\mathbf{9.98\cdot10^{-1}}$\\
        & \reflectbox{F}LIP$\downarrow$ & $1.32\cdot10^{-1}$ & $5.12\cdot10^{-2}$ & $\mathbf{1.74\cdot10^{-2}}$ \\[0.5ex]
    \hline
     
    \multirow{3}{=}{Grape (indoors)} & \rule{0pt}{2.5ex}MSE$\downarrow$ & $2.84\cdot10^{-3}$ & $9.81\cdot10^{-4}$ & $\mathbf{3.11\cdot10^{-4}}$\\ 
        & SSIM1$\uparrow$ & $9.12\cdot10^{-1}$ & $\mathbf{9.79\cdot10^{-1}}$ & $8.70\cdot10^{-1}$ \\
        & SSIM2$\uparrow$ & $9.08\cdot10^{-1}$ & $9.83\cdot10^{-1}$ & $\mathbf{9.94\cdot10^{-1}}$ \\
        & \reflectbox{F}LIP$\downarrow$ & $1.37\cdot10^{-1}$ & $1.60\cdot10^{-1}$ & $\mathbf{3.01\cdot10^{-2}}$ \\[0.5ex]
    \hline
     
    \multirow{3}{=}{Paperweight (outdoors)} & \rule{0pt}{2.5ex}MSE$\downarrow$ & $9.31\cdot10^{-3}$ & $7.40\cdot10^{-4}$ & $\mathbf{6.91\cdot10^{-4}}$ \\ 
        & SSIM1$\uparrow$ & $9.19\cdot10^{-1}$ & $\mathbf{9.73\cdot10^{-1}}$ & $7.57\cdot10^{-1}$\\
        & SSIM2$\uparrow$ & $8.03\cdot10^{-1}$ & $9.43\cdot10^{-1}$ & $\mathbf{9.74\cdot10^{-1}}$\\
        & \reflectbox{F}LIP$\downarrow$ & $1.63\cdot10^{-1}$ & $6.08\cdot10^{-2}$ & $\mathbf{2.79\cdot10^{-2}}$\\[0.5ex]
    \hline
      
    \multirow{3}{=}{Paperweight (indoors)} & \rule{0pt}{2.5ex}MSE$\downarrow$ & $1.01\cdot10^{-2}$ & $1.95\cdot10^{-3}$ & $\mathbf{1.15\cdot10^{-3}}$\\ 
        & SSIM1$\uparrow$ & $9.14\cdot10^{-1}$ & $\mathbf{9.60\cdot10^{-1}}$ & $7.54\cdot10^{-1}$\\
        & SSIM2$\uparrow$ & $8.07\cdot10^{-1}$ & $9.14\cdot10^{-1}$ & $\mathbf{9.71\cdot10^{-1}}$\\
        & \reflectbox{F}LIP$\downarrow$ & $1.70\cdot10^{-1}$ & $8.90\cdot10^{-2}$ & $\mathbf{4.91\cdot10^{-2}}$\\[0.5ex]
    \hline

    \end{tabular}
    \end{table}
    
    \begin{figure*}[]

\begin{tabularx}{\textwidth}{sbbbb}

    & \large Reference & \large NeRFactor & \large NeuMip & \large Ours\\
    \rotatebox{90}{Grape (outdoors)} & 
    \includegraphics[height=3.75cm]{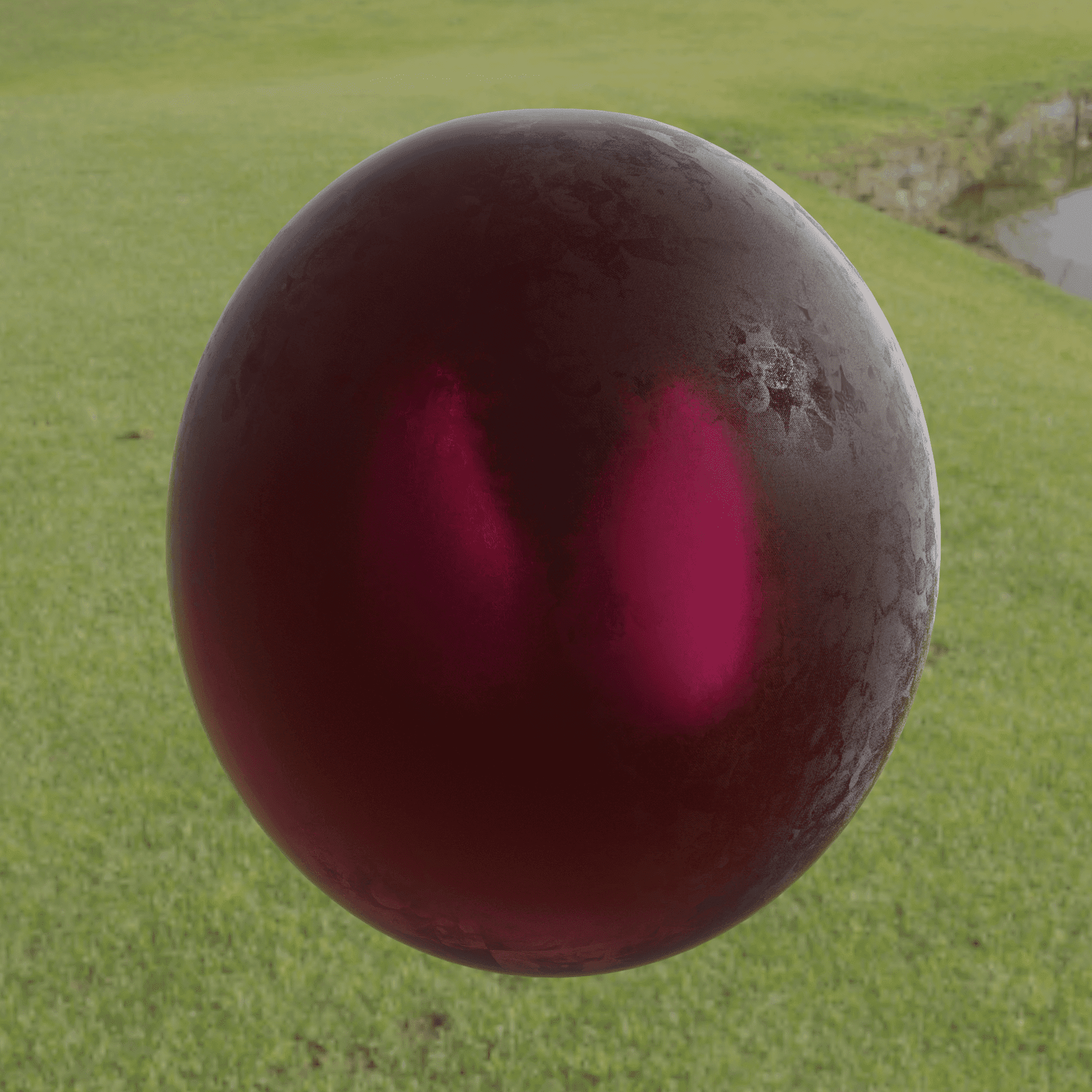} \subcaption{} \label{compa}&

    \includegraphics[height=3.75cm]{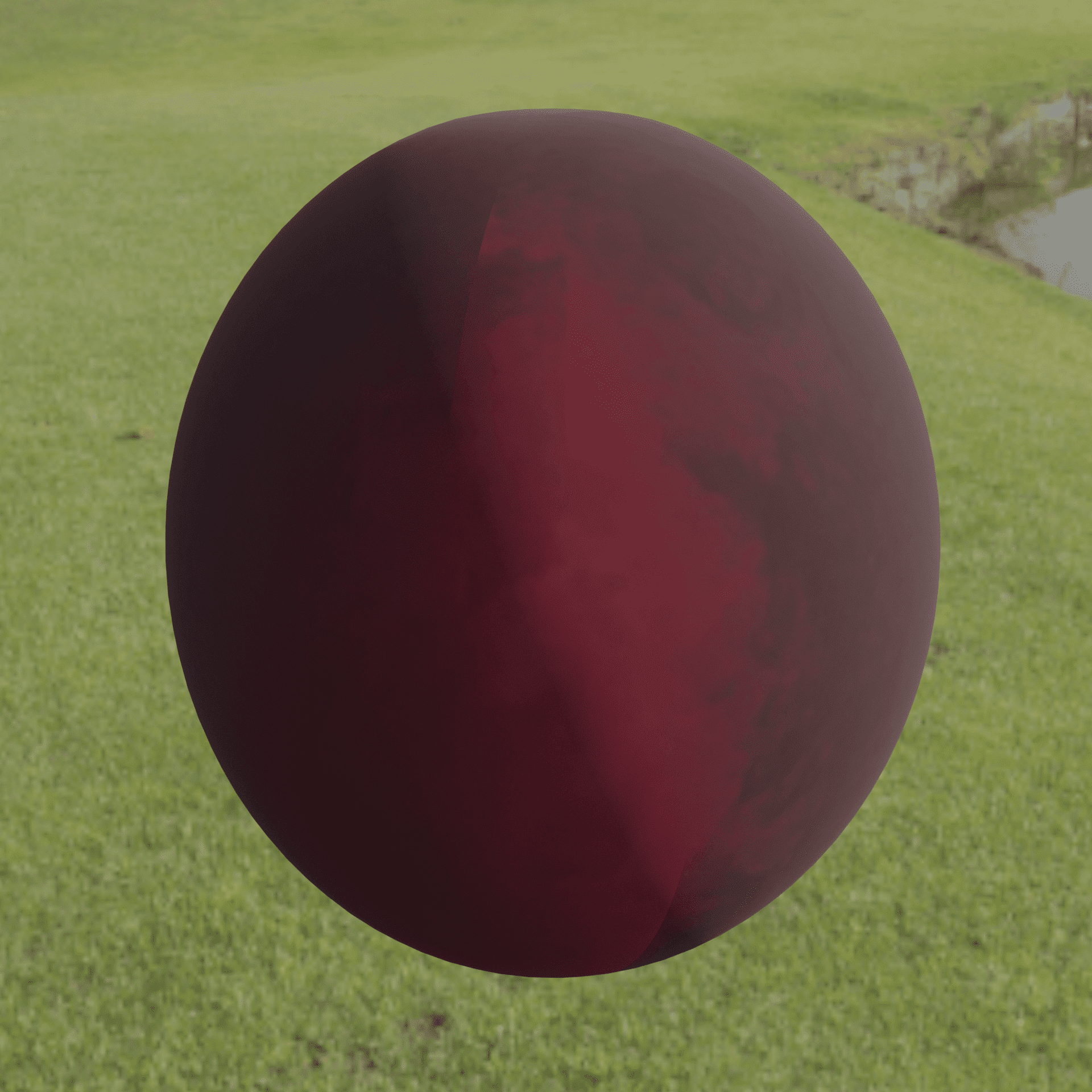}\llap{\includegraphics[height=1cm]{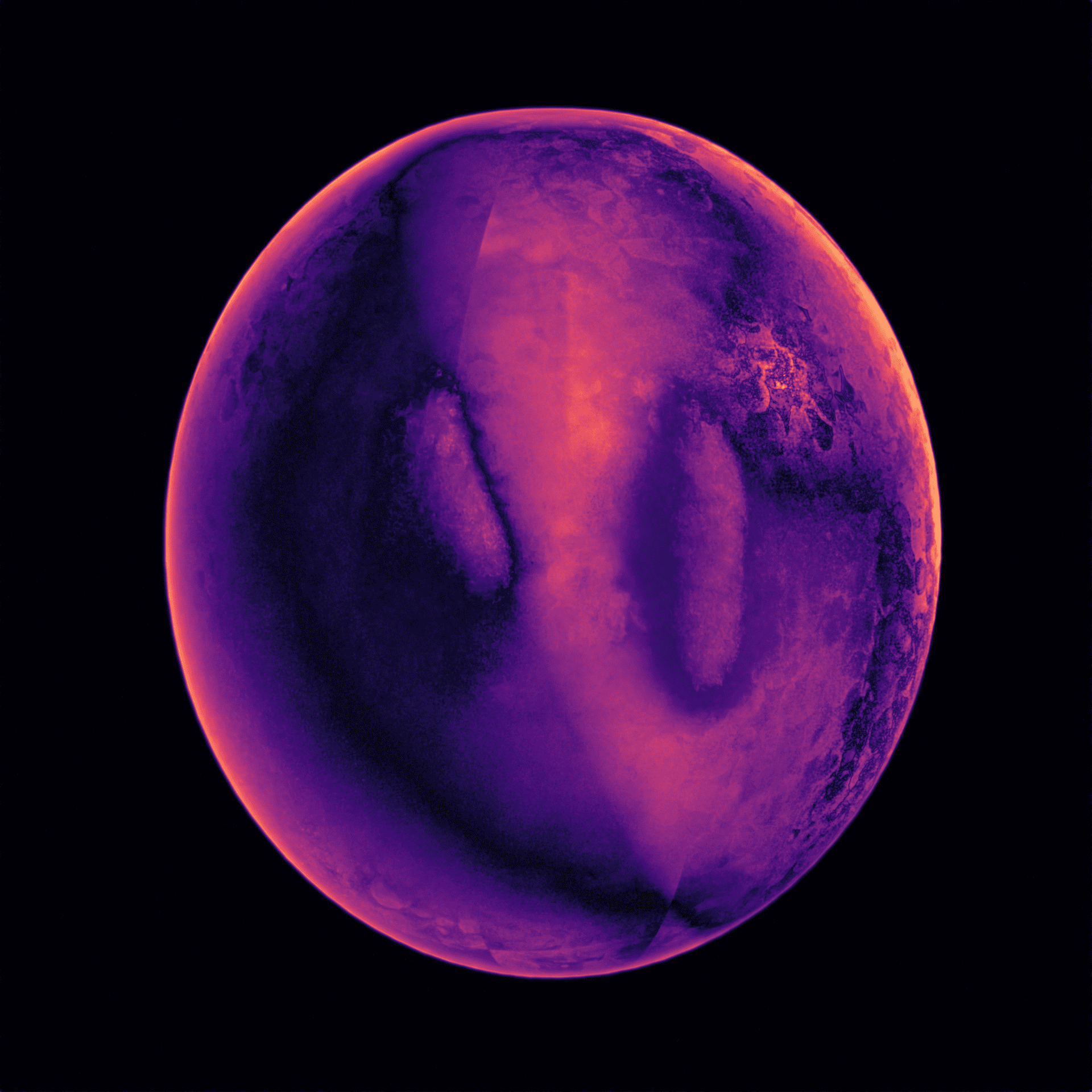}} \subcaption{} \label{compb}&
    
    \includegraphics[height=3.75cm]{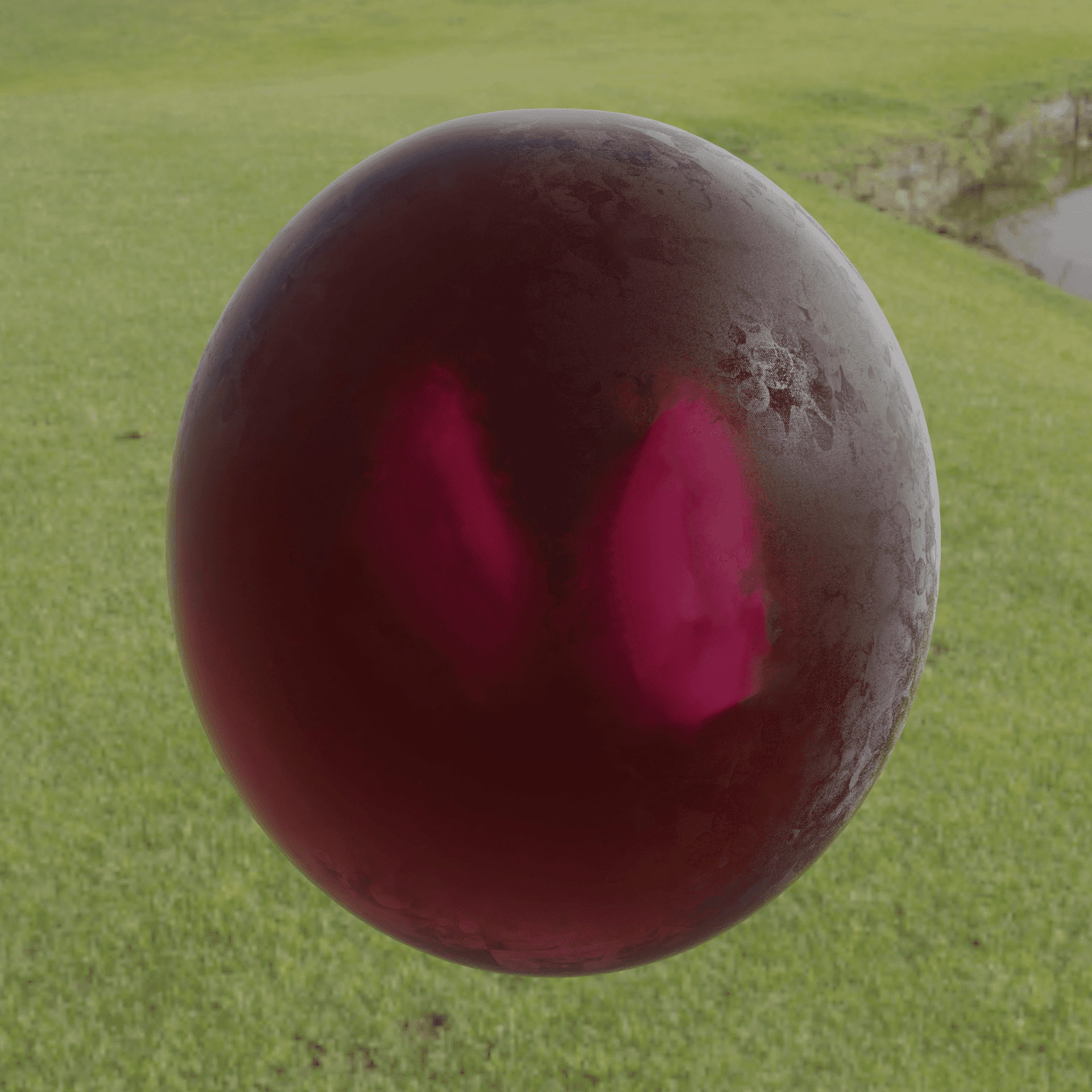}\llap{\includegraphics[height=1cm]{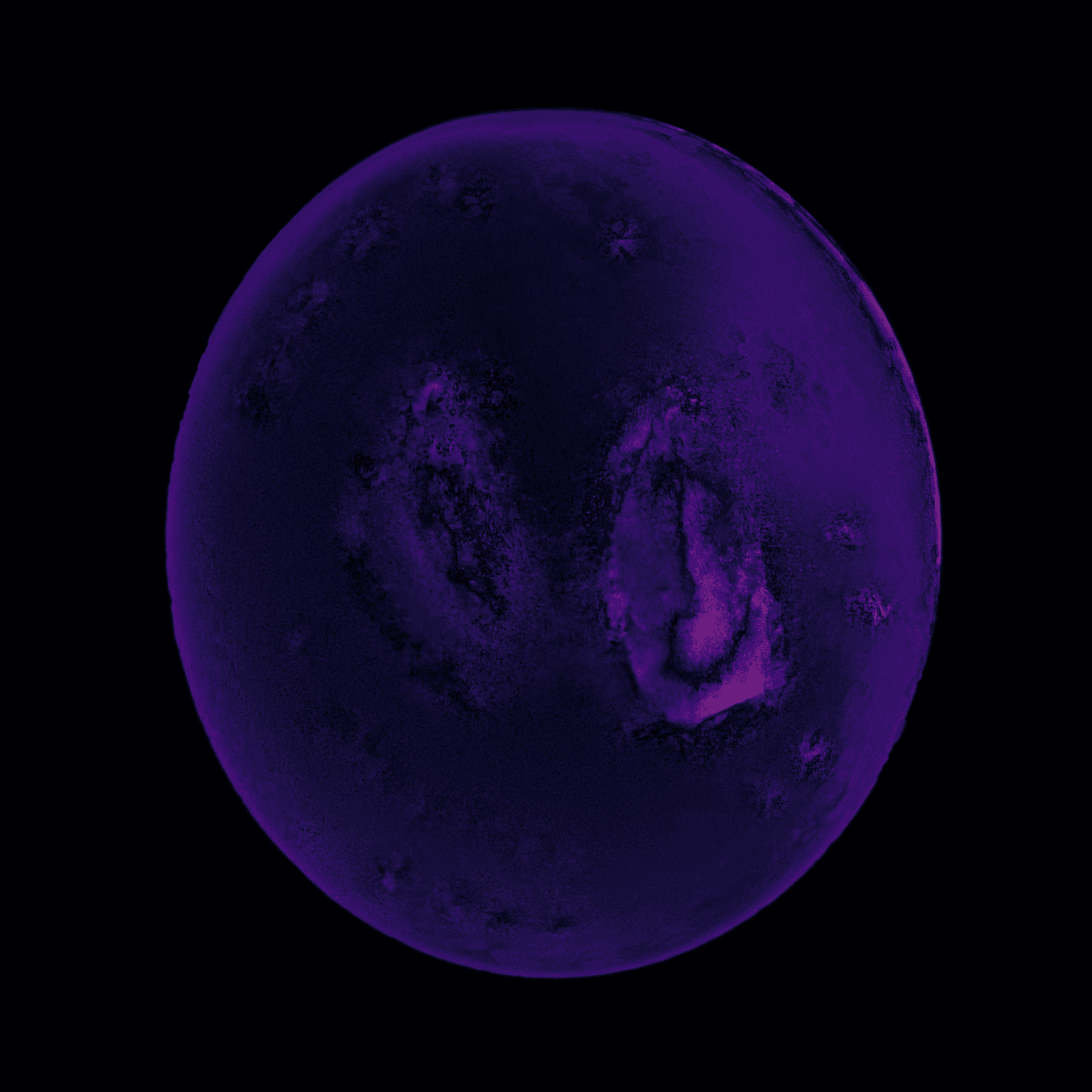}} \subcaption{} \label{compc}&
    
    \includegraphics[height=3.75cm]{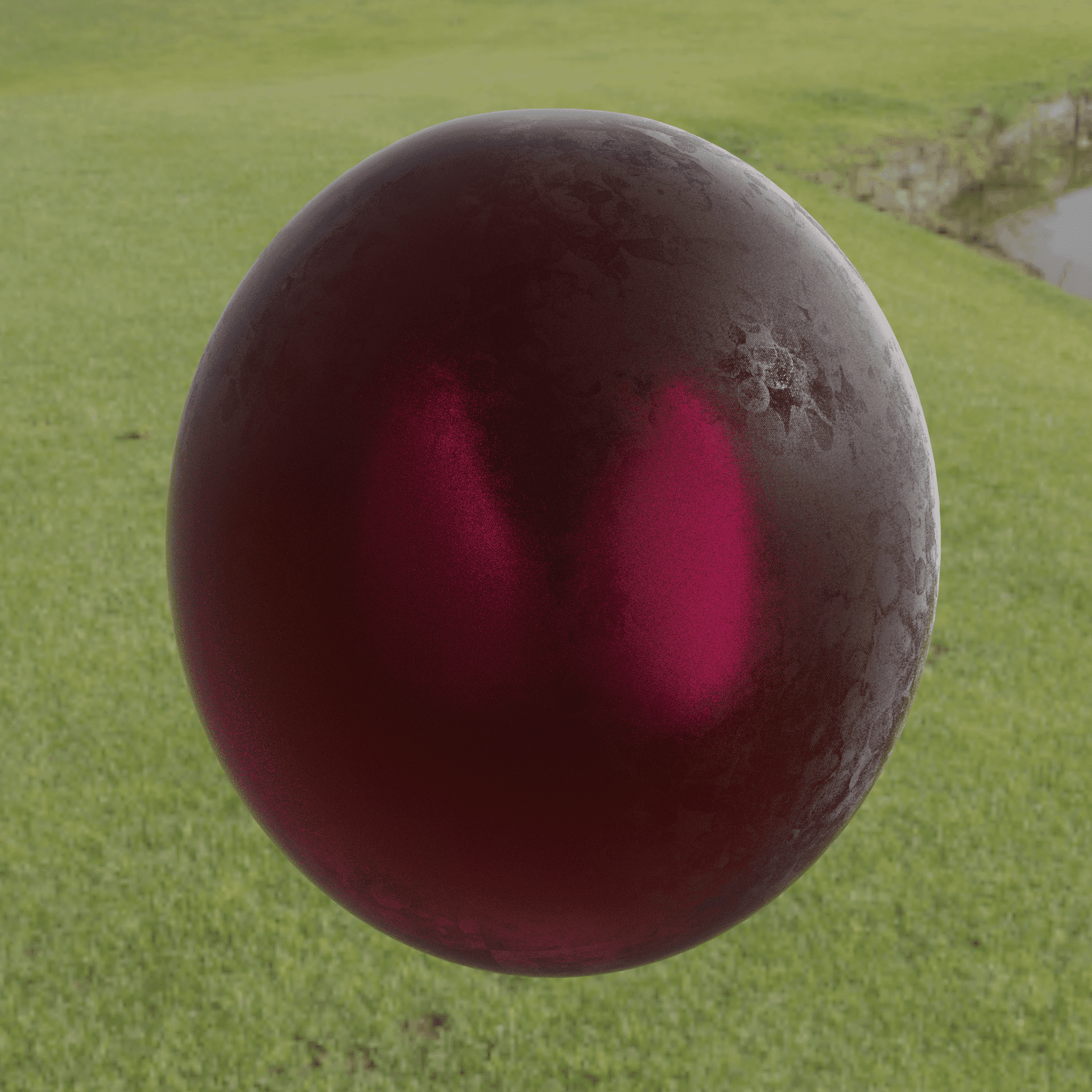}\llap{\includegraphics[height=1cm]{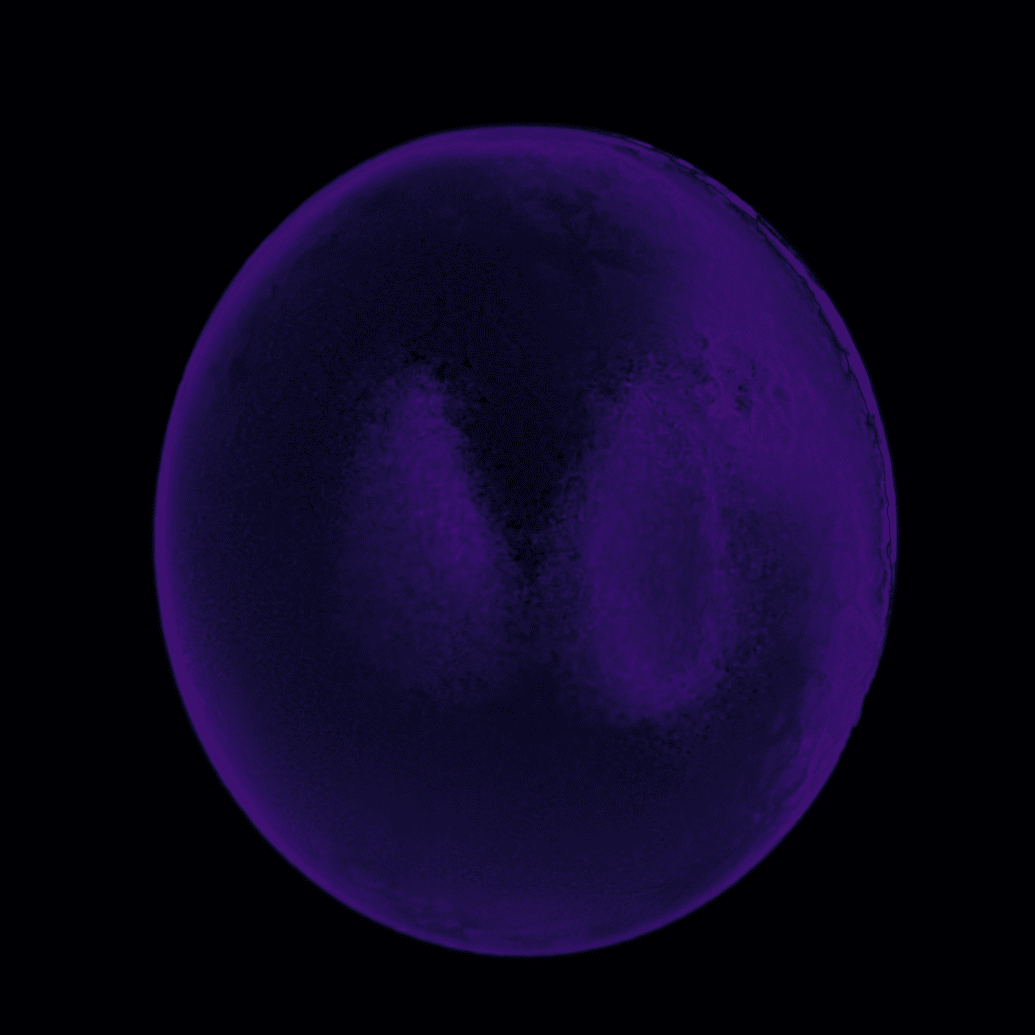}} \subcaption{} \label{compd}\\[-1ex]
    
    \rotatebox{90}{Grape (indoors)} & 
    \includegraphics[height=3.75cm]{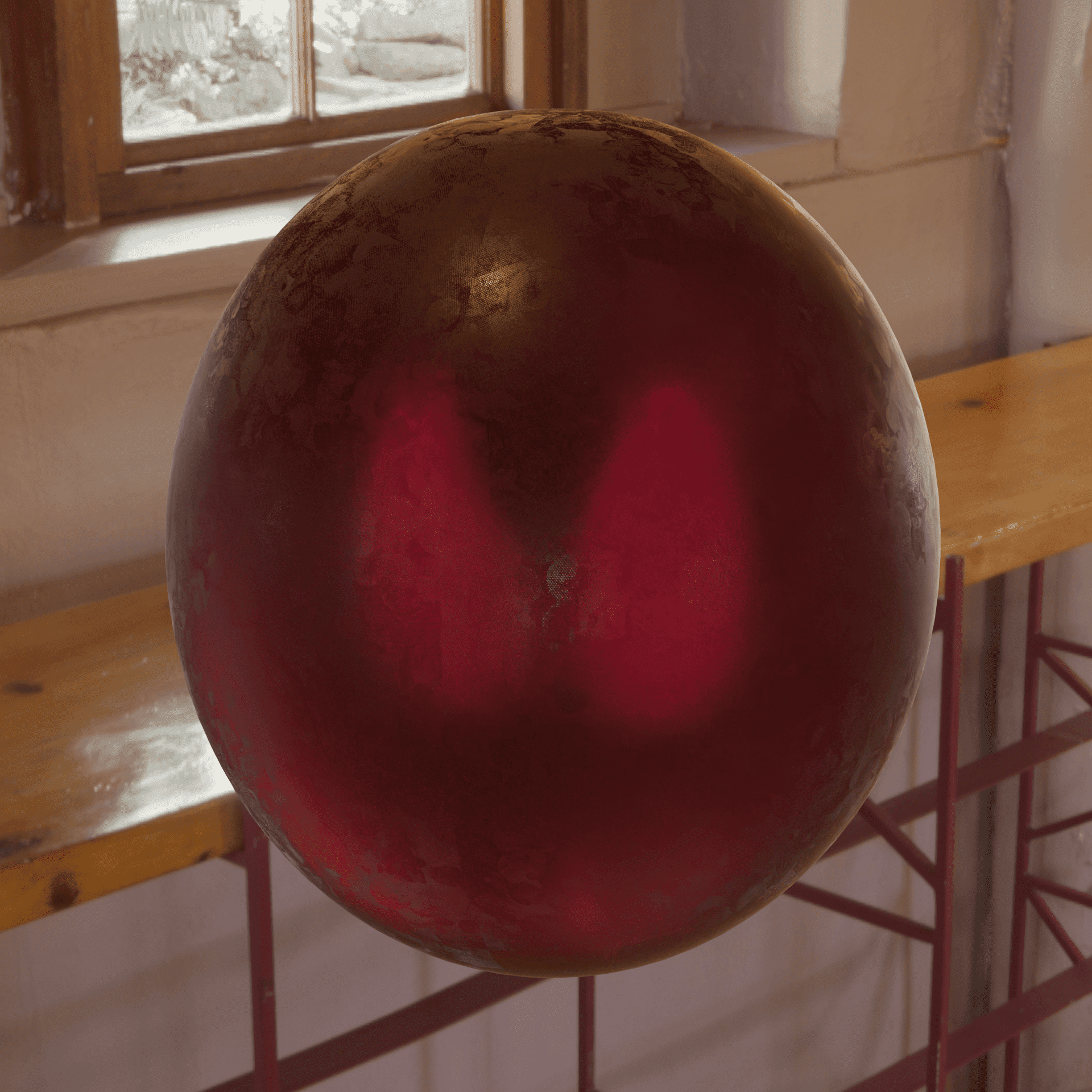} \subcaption{} \label{compe}&
    
    \includegraphics[height=3.75cm]{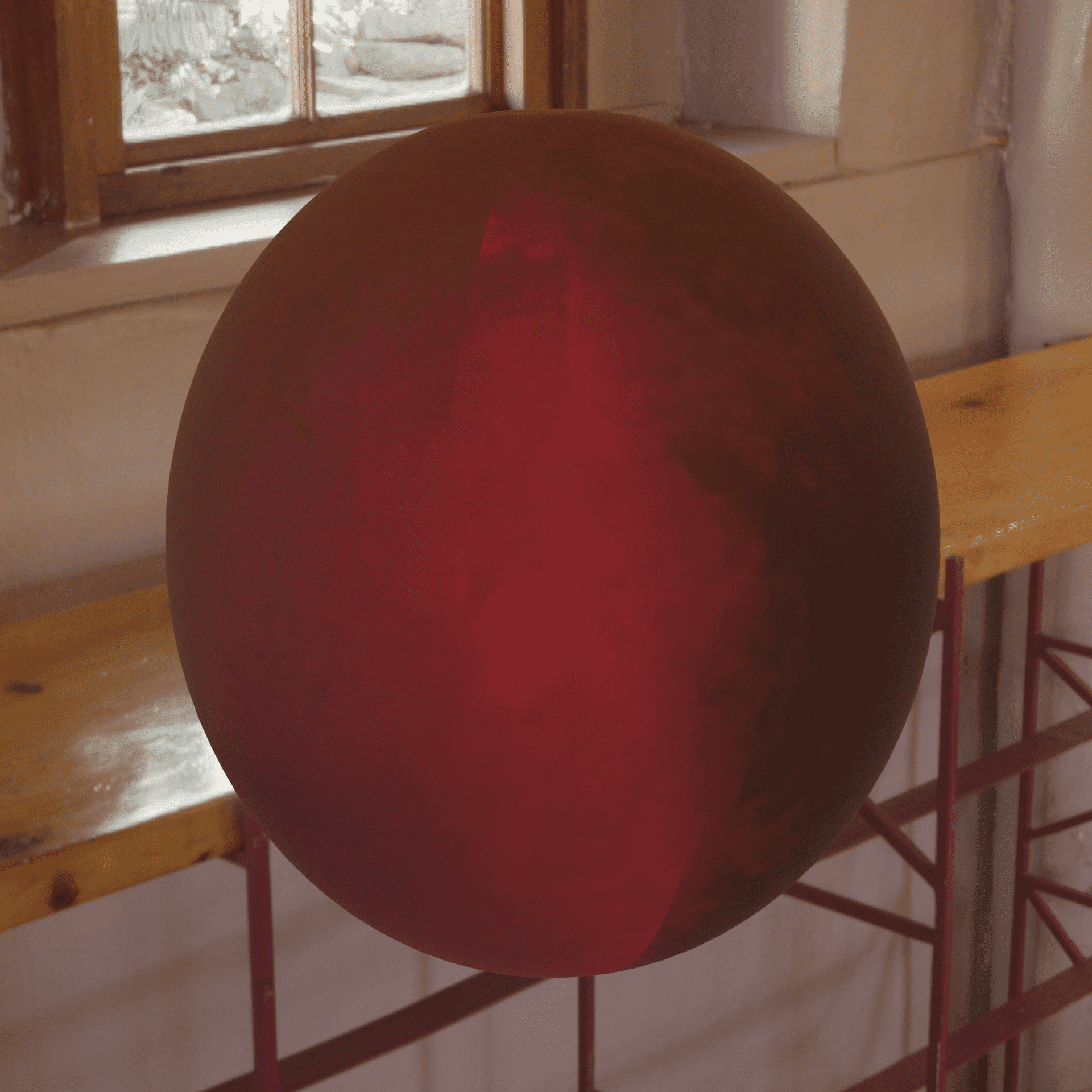}\llap{\includegraphics[height=1cm]{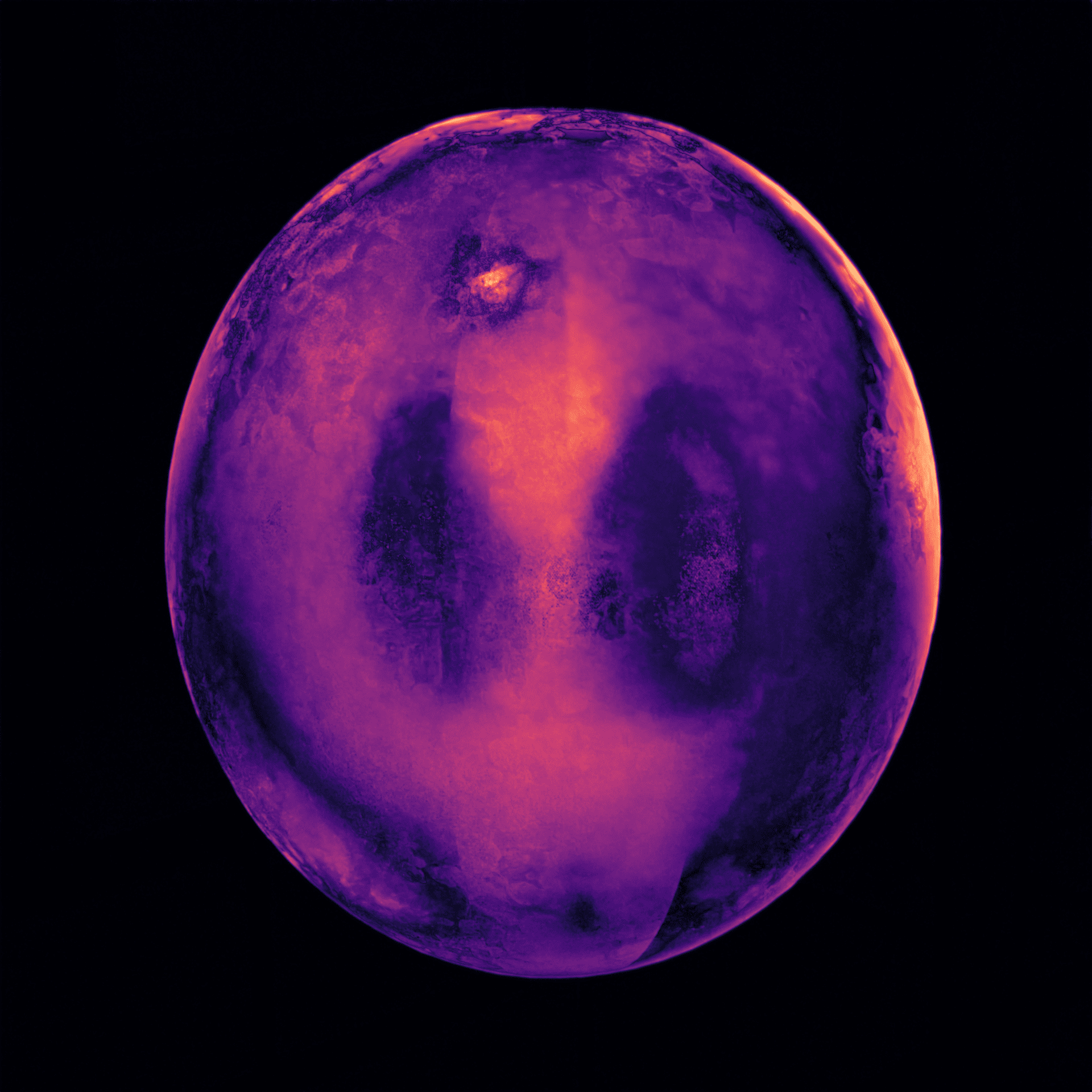}} \subcaption{} \label{compf}&
    
    \includegraphics[height=3.75cm]{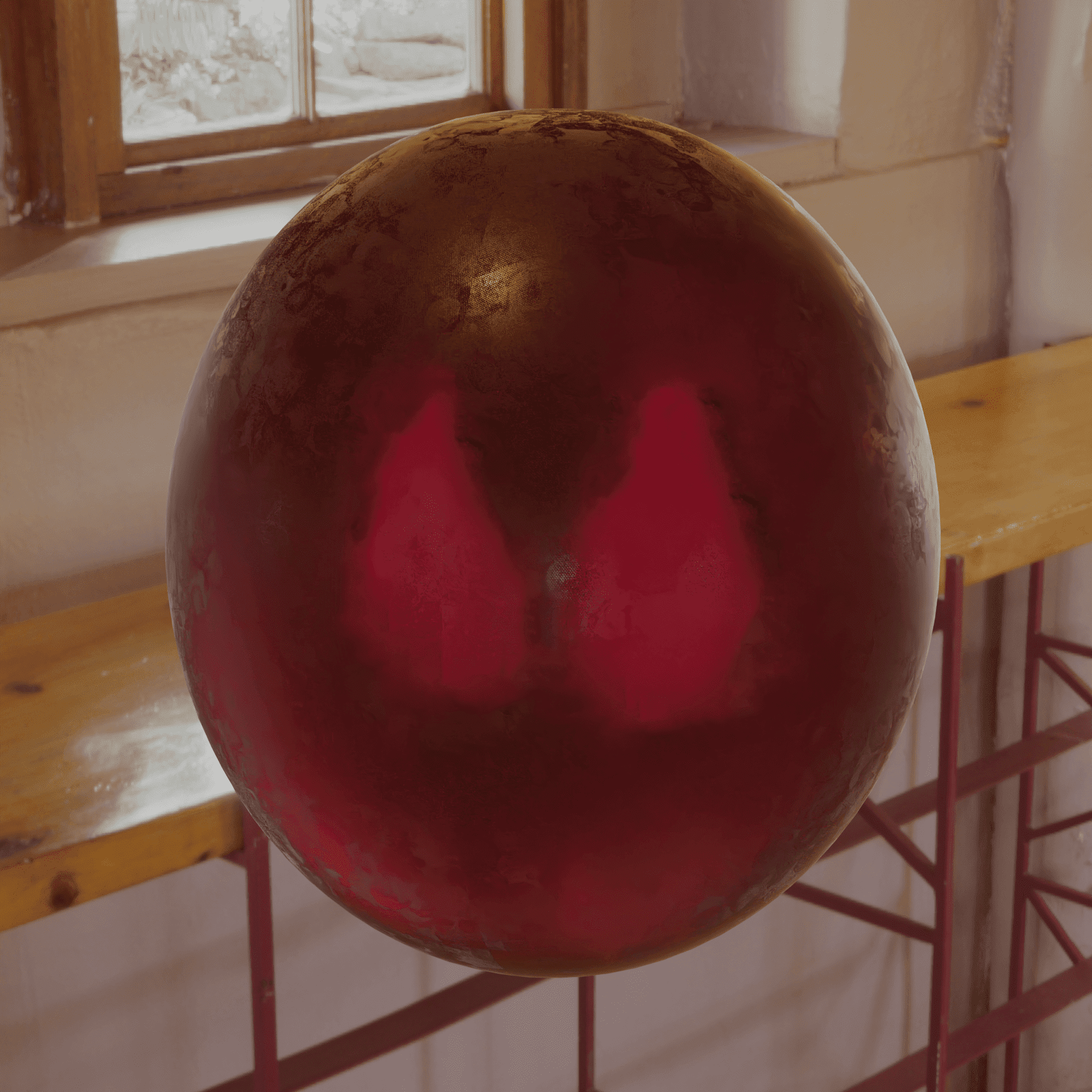}\llap{\includegraphics[height=1cm]{figures/comparsions/grape/lg/flip.grape.grape_neumip.67ppd.ldr_2.png}} \subcaption{} \label{compg}&
    
    \includegraphics[height=3.75cm]{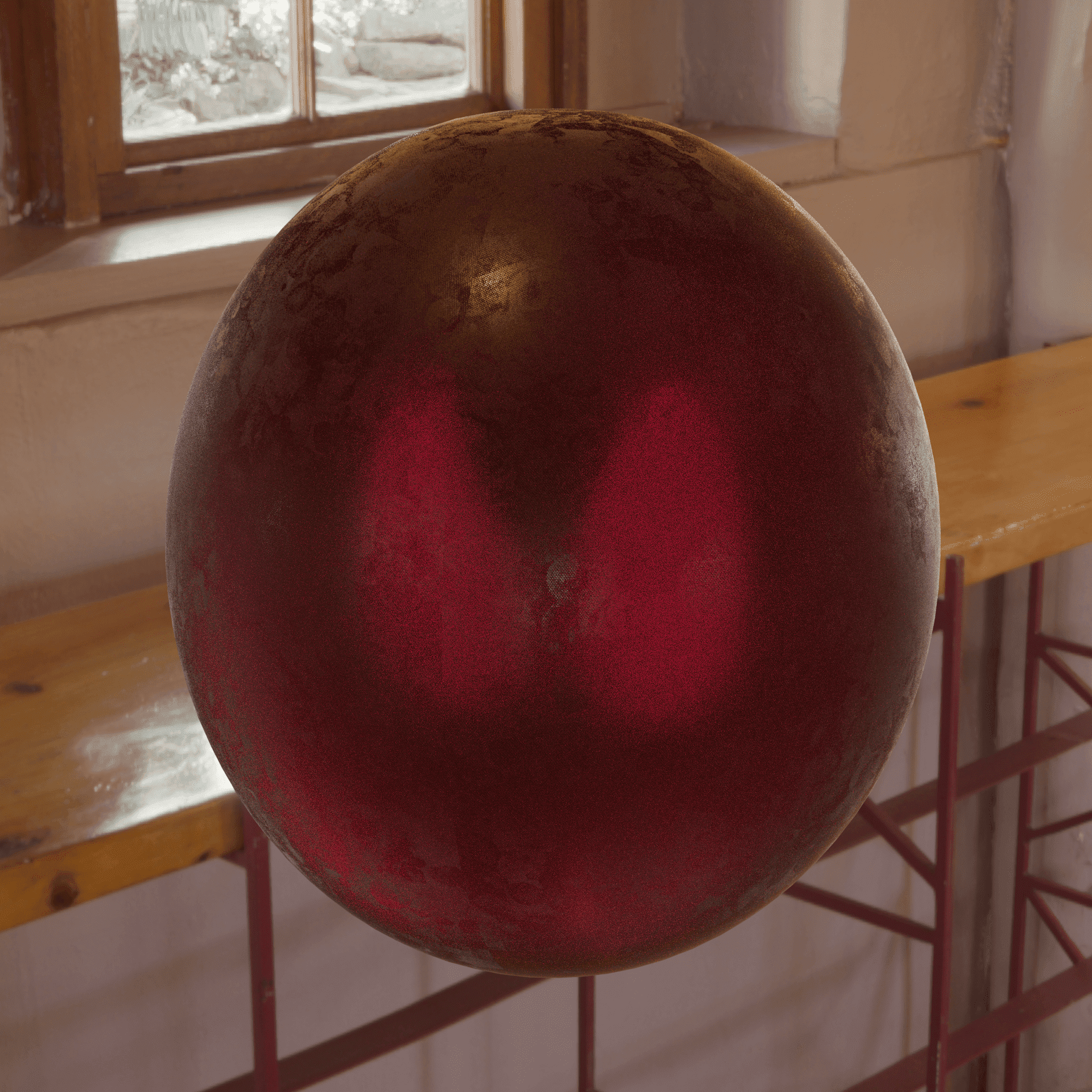}\llap{\includegraphics[height=1cm]{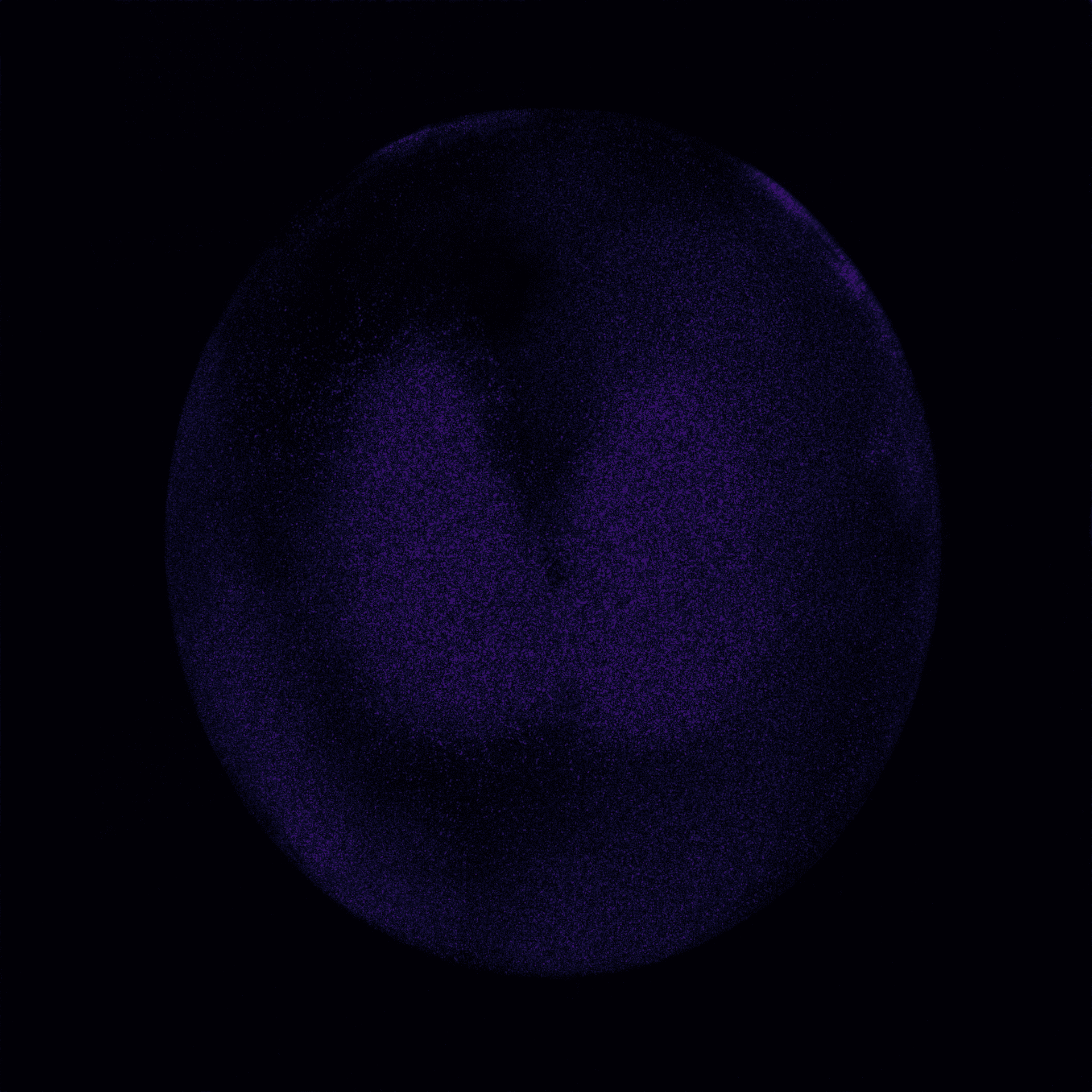}} \subcaption{} \label{comph}\\[-1ex]

    \rotatebox{90}{Paperweight (outdoors)} & 
    \includegraphics[height=3.75cm]{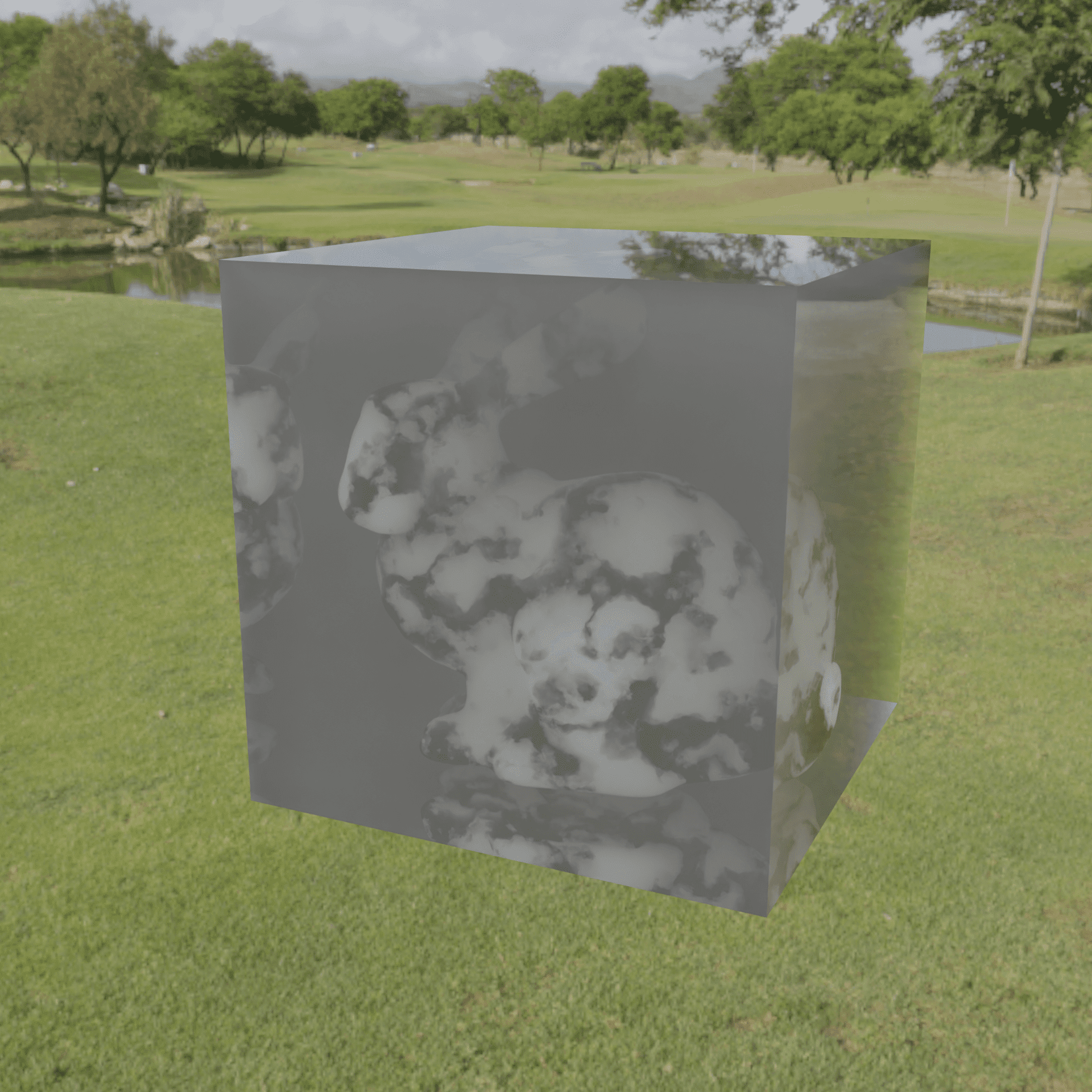} \subcaption{} \label{compi}&
    
    \includegraphics[height=3.75cm]{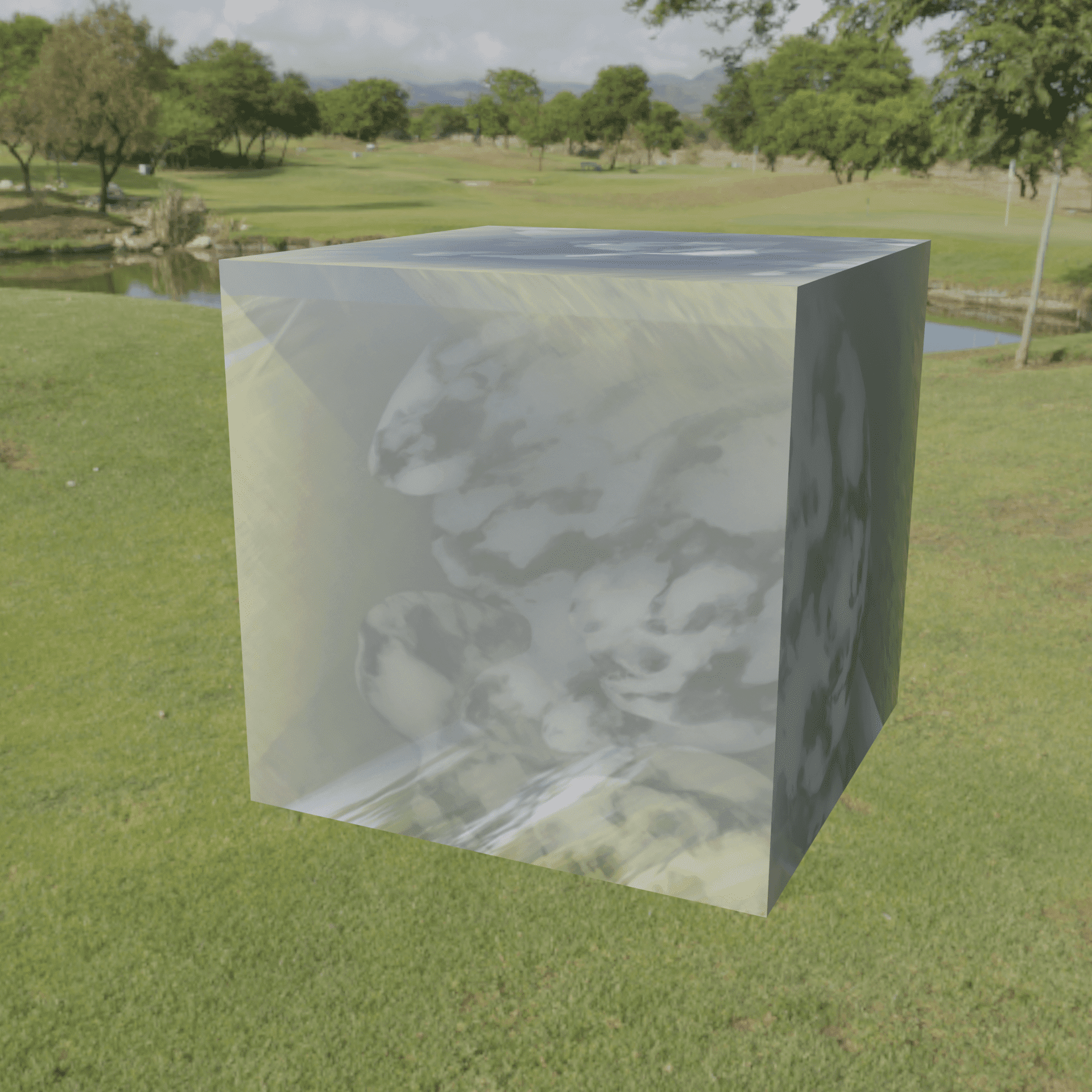}\llap{\includegraphics[height=1cm]{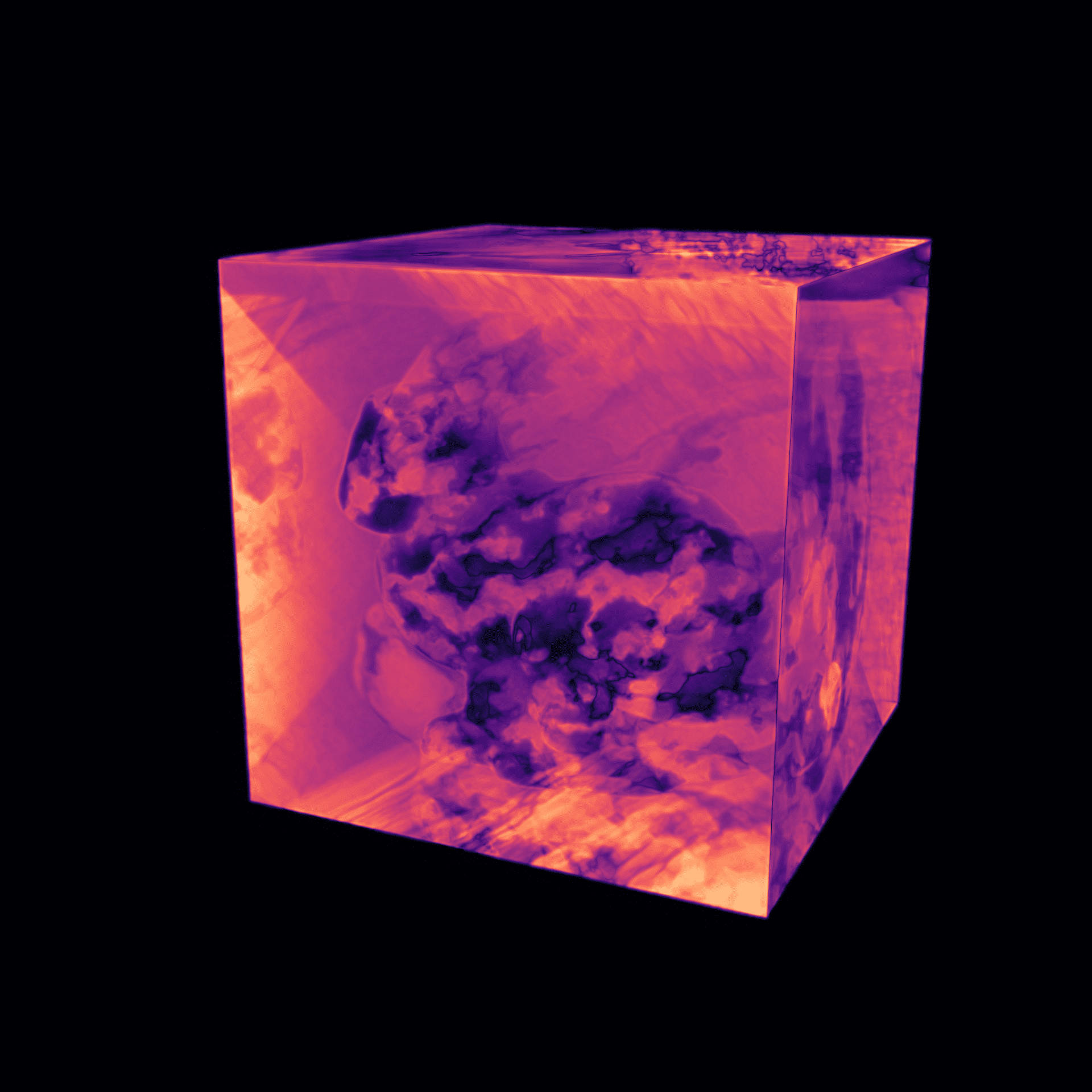}} \subcaption{} \label{compj}&
    
    \includegraphics[height=3.75cm]{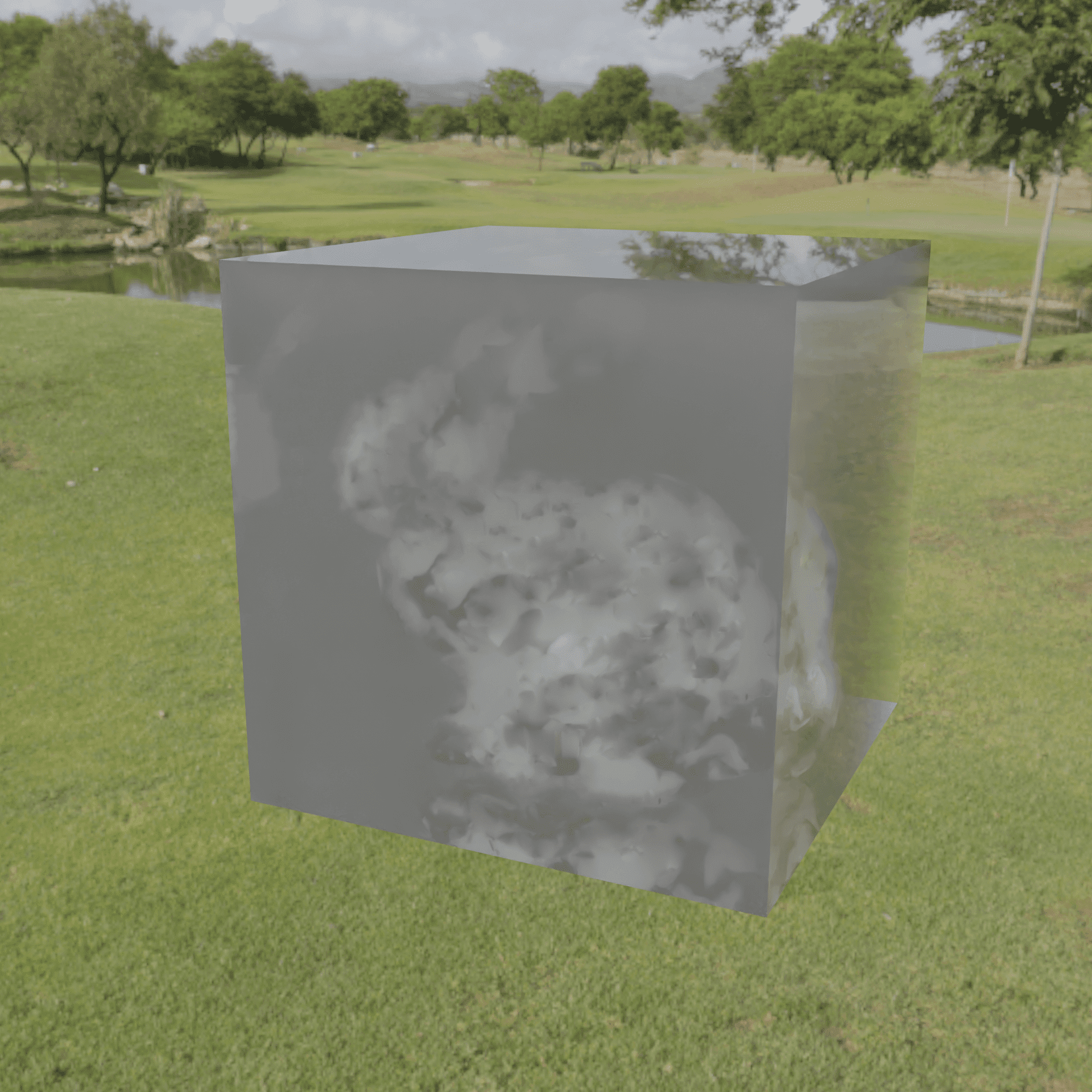}\llap{\includegraphics[height=1cm]{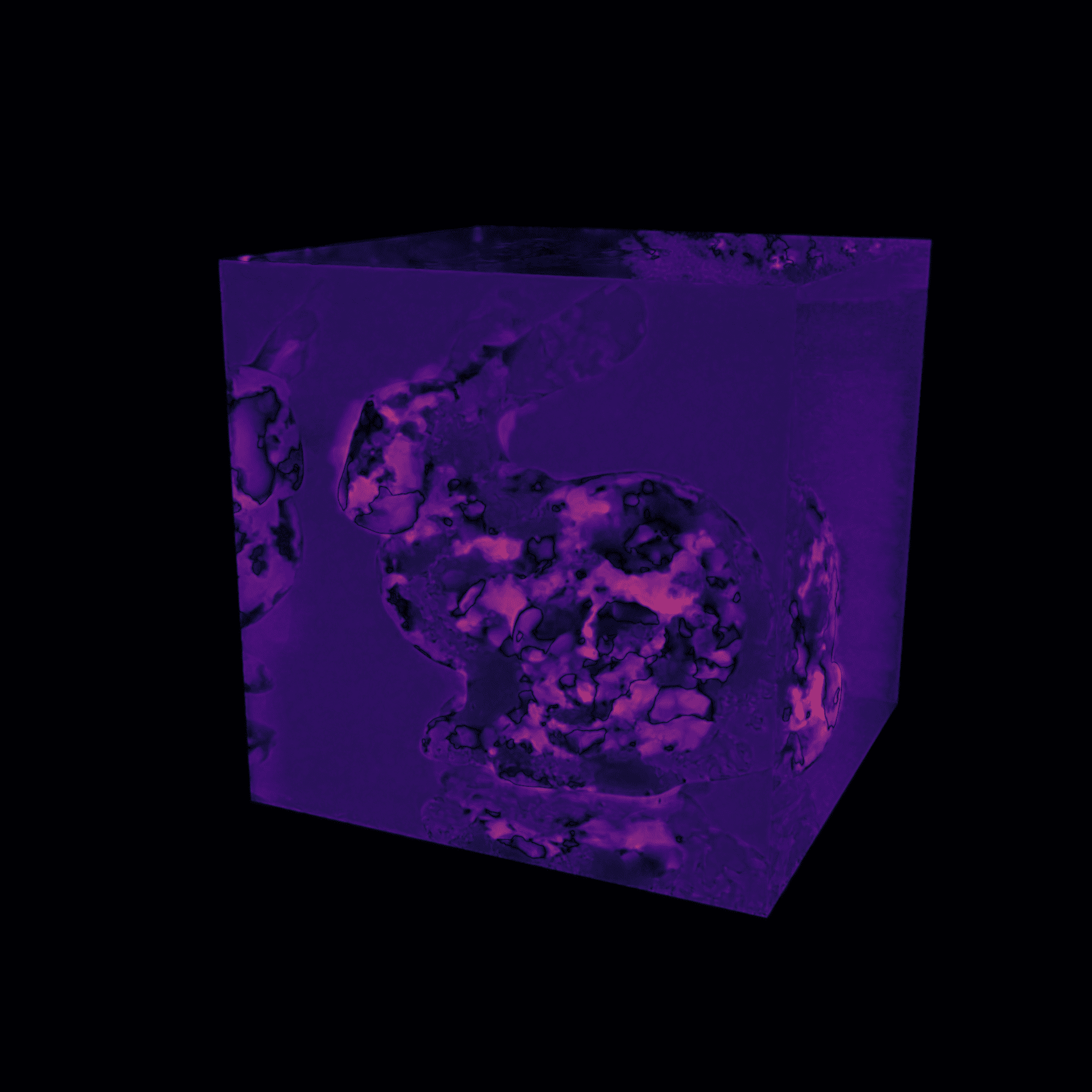}} \subcaption{} \label{compk}&
    
    \includegraphics[height=3.75cm]{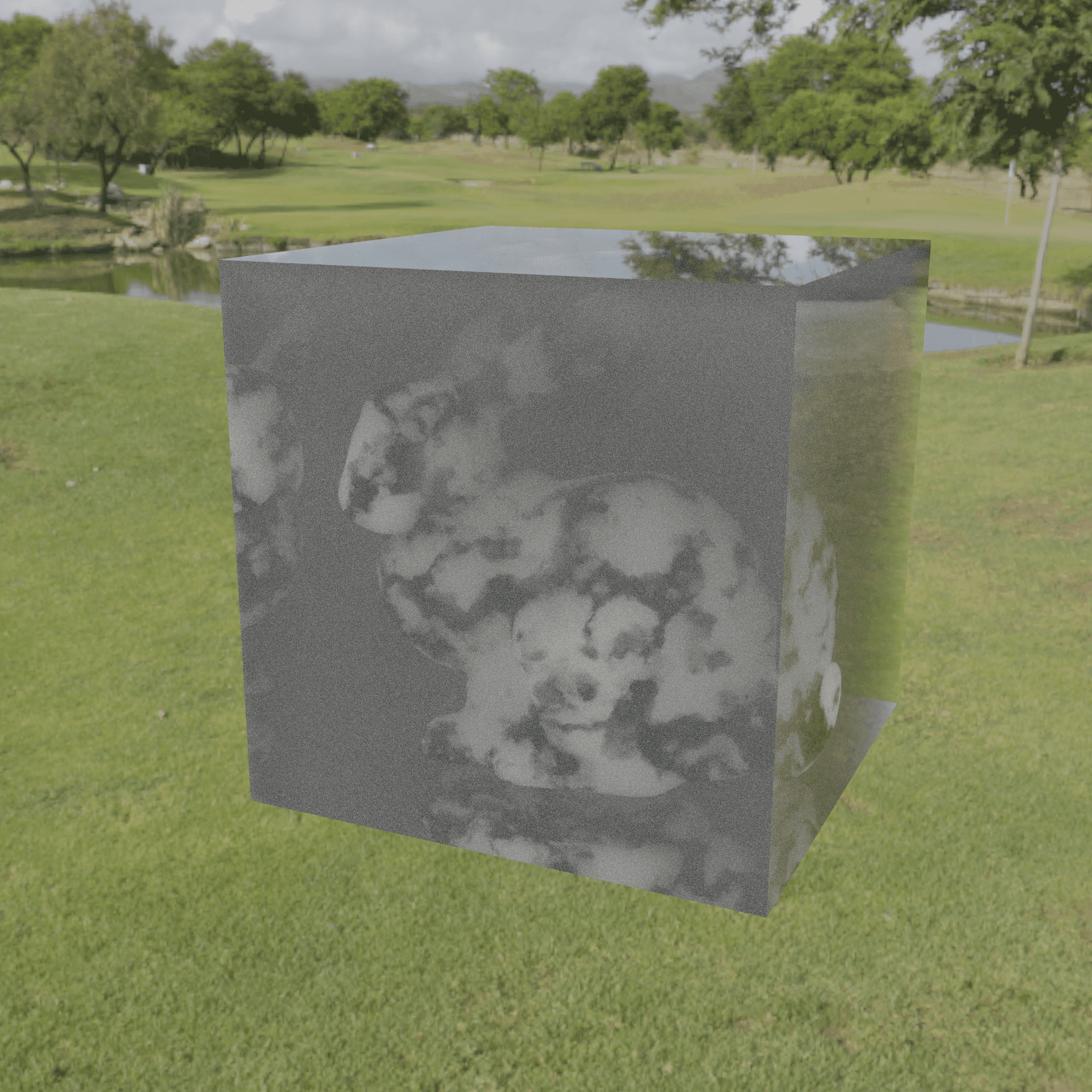}\llap{\includegraphics[height=1cm]{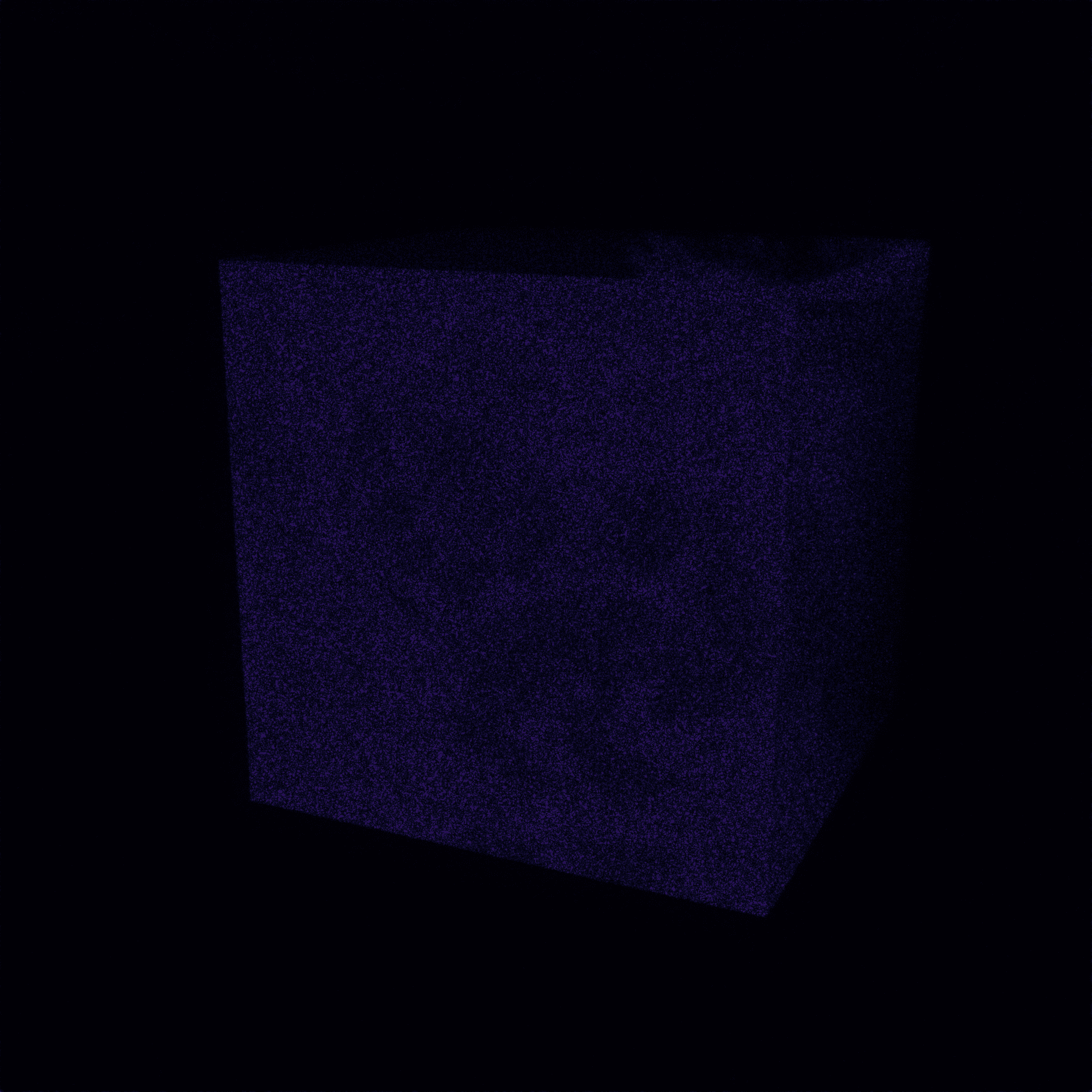}} \subcaption{} \label{compl}\\[-1ex]

    \rotatebox{90}{Paperweight (indoors)} & 
    \includegraphics[height=3.75cm]{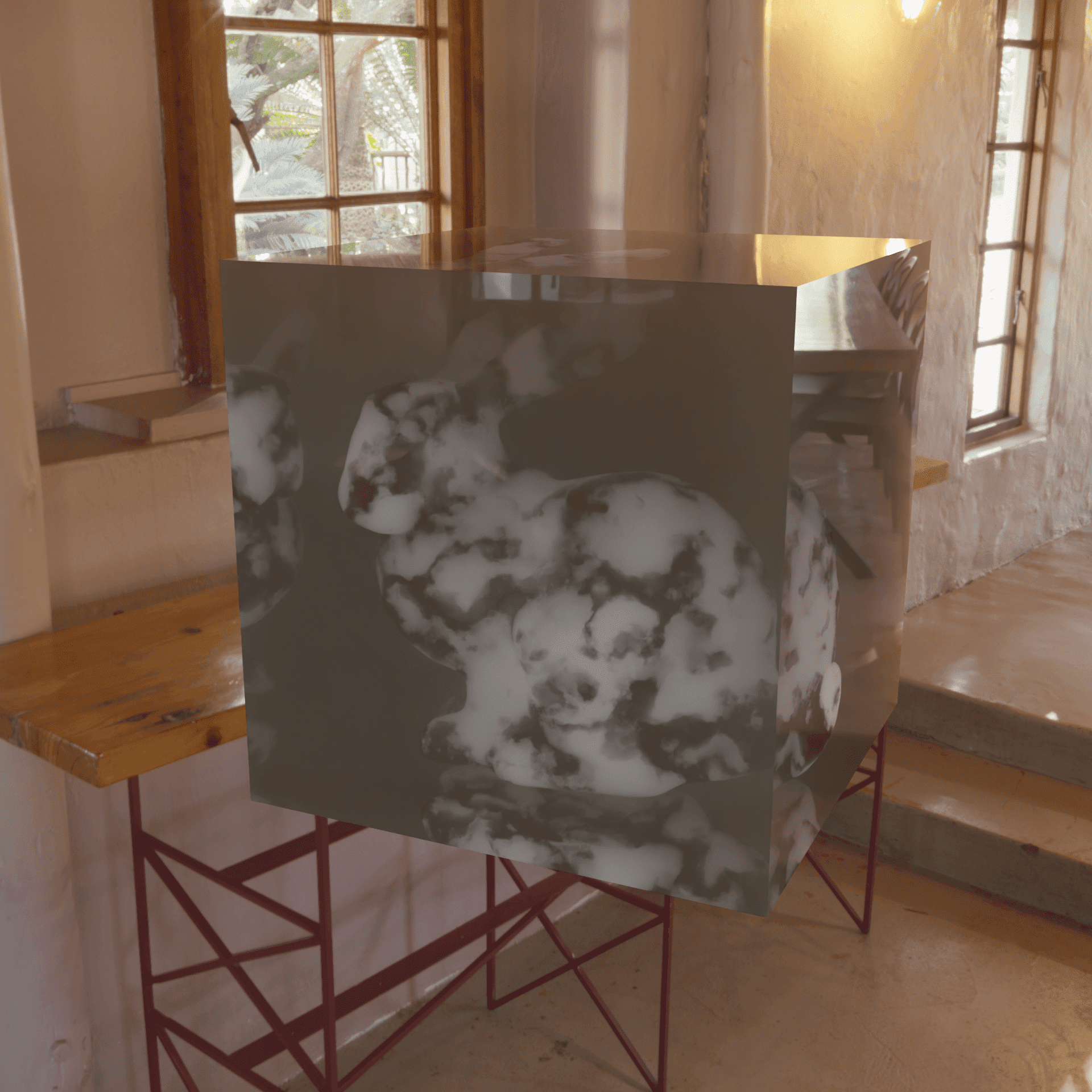} \subcaption{} \label{compm}&
    
    \includegraphics[height=3.75cm]{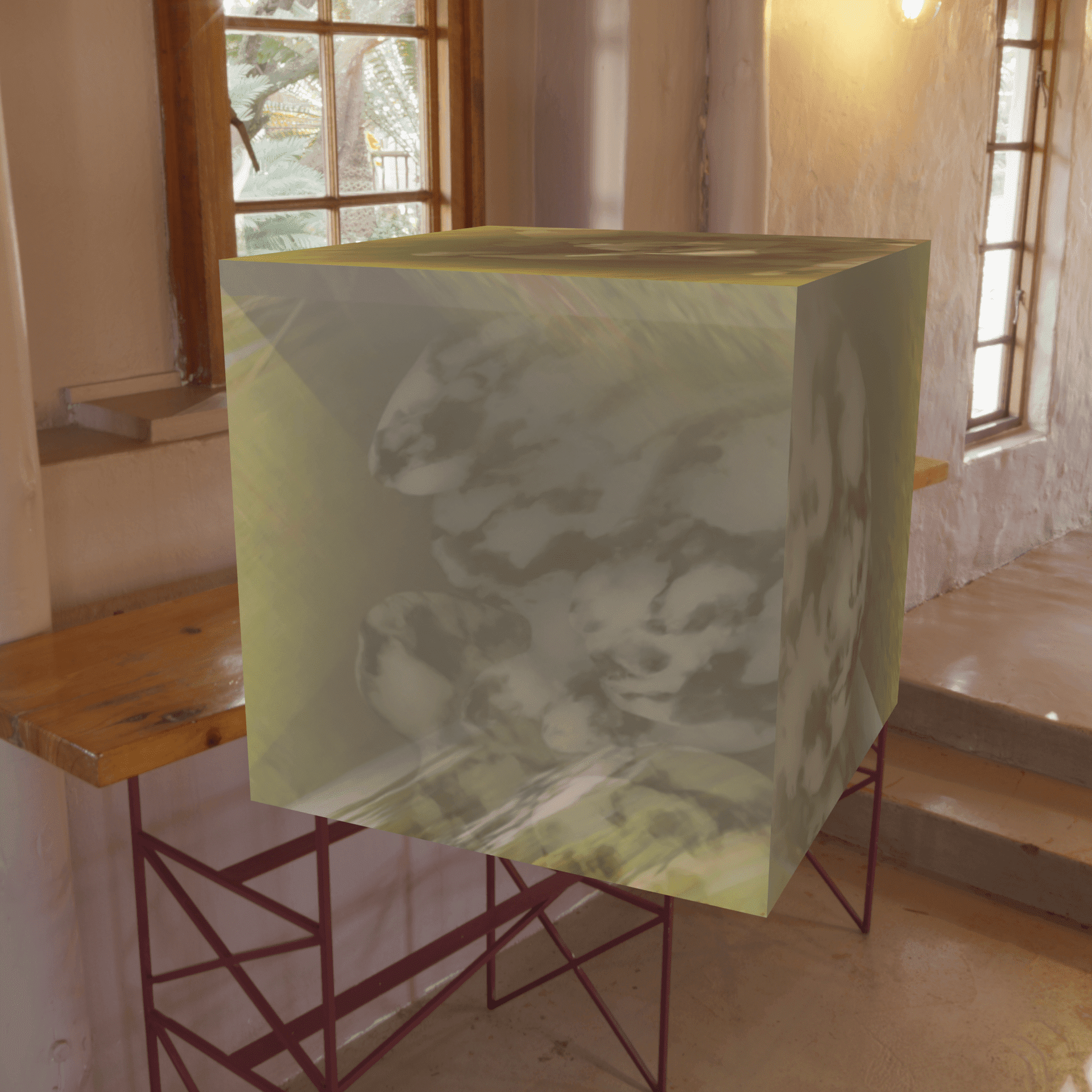}\llap{\includegraphics[height=1cm]{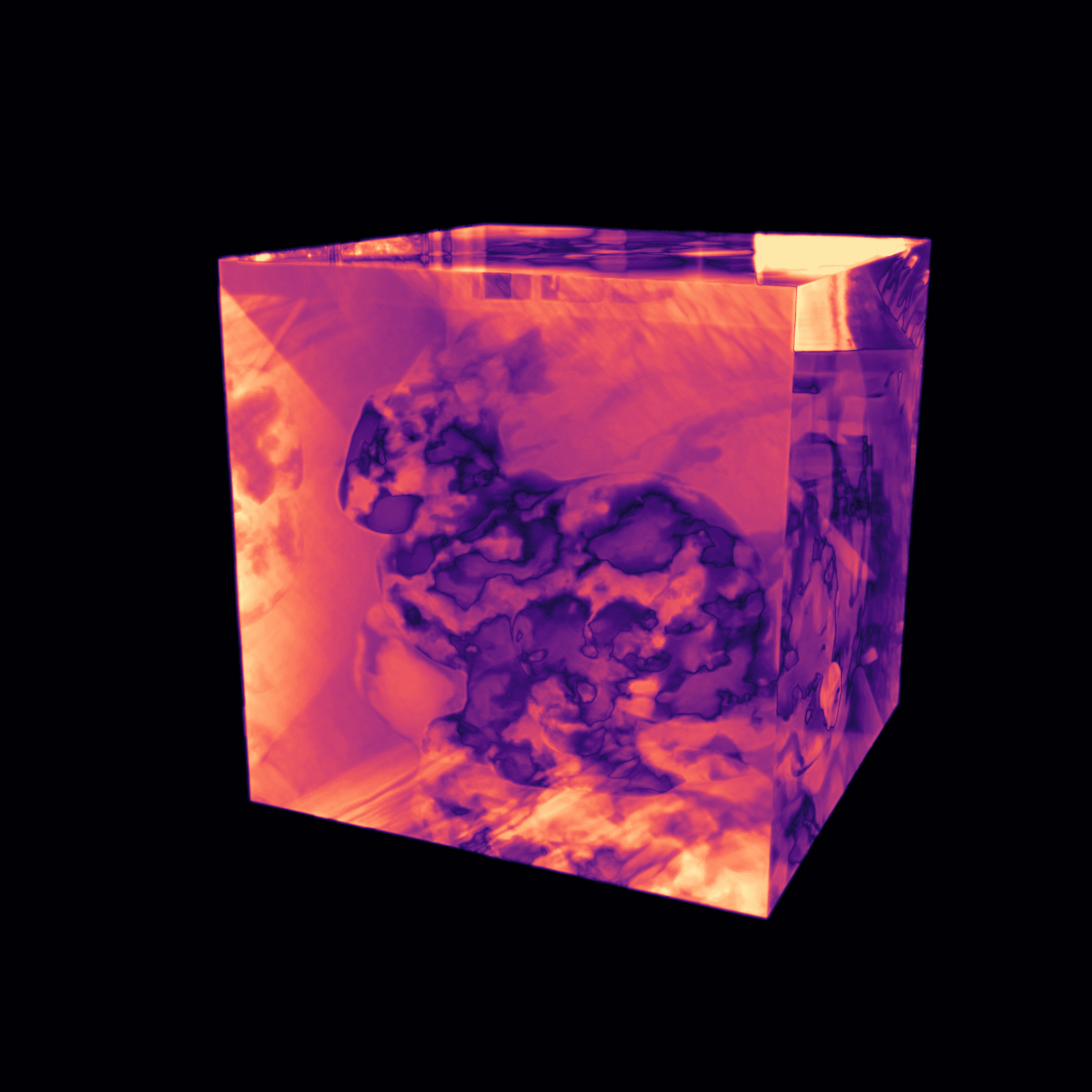}} \subcaption{} \label{compn}&
    
    \includegraphics[height=3.75cm]{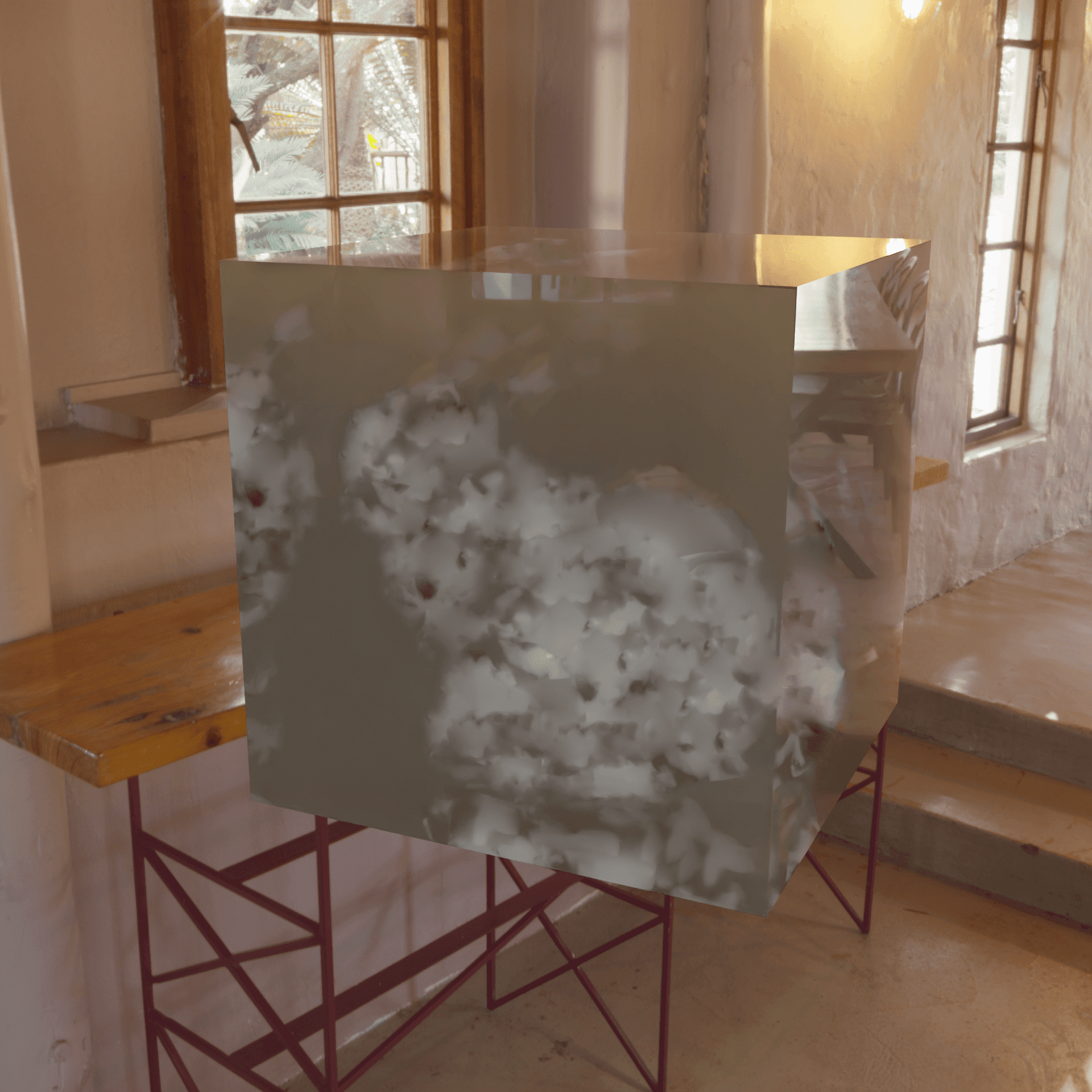}\llap{\includegraphics[height=1cm]{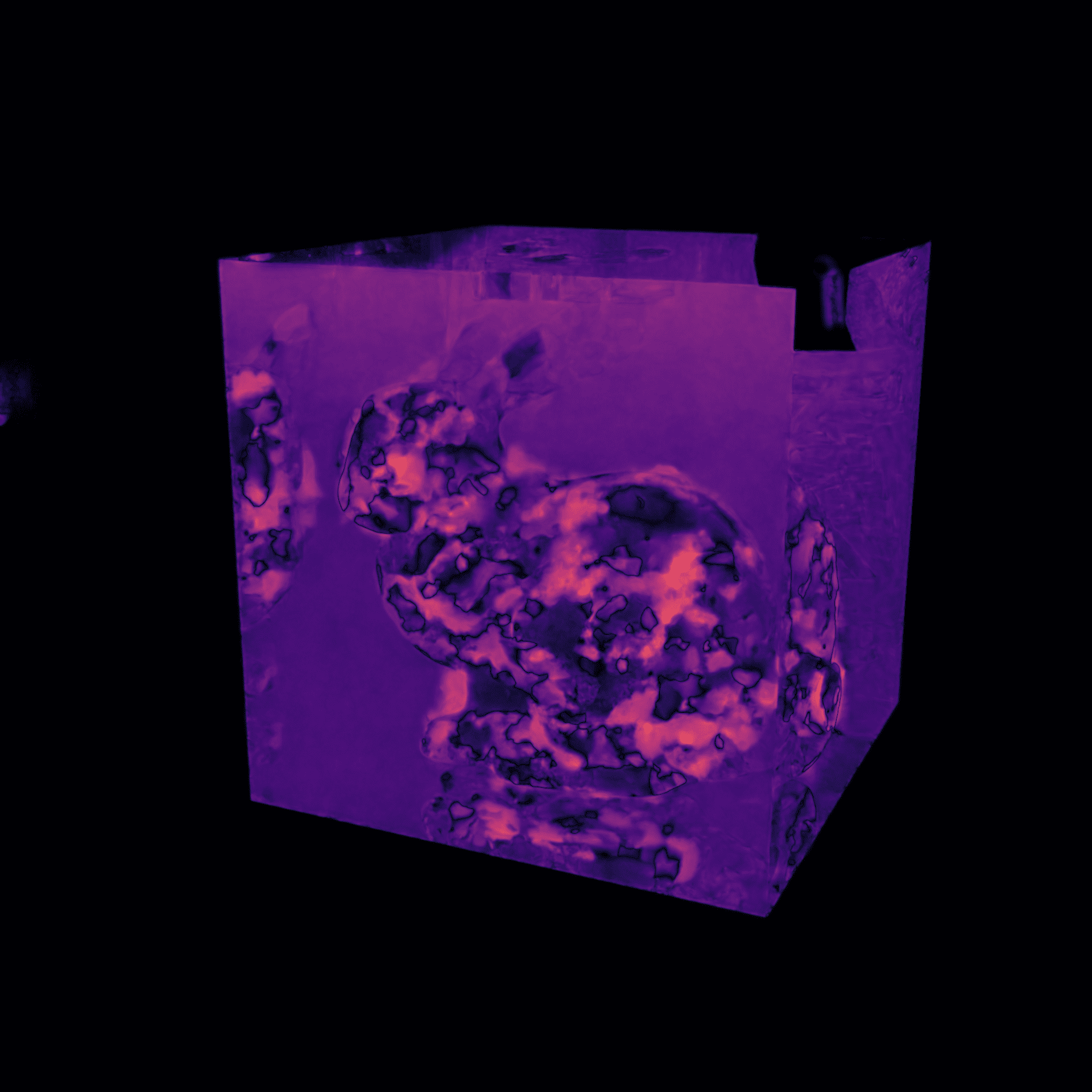}} \subcaption{} \label{compo}&
    
    \includegraphics[height=3.75cm]{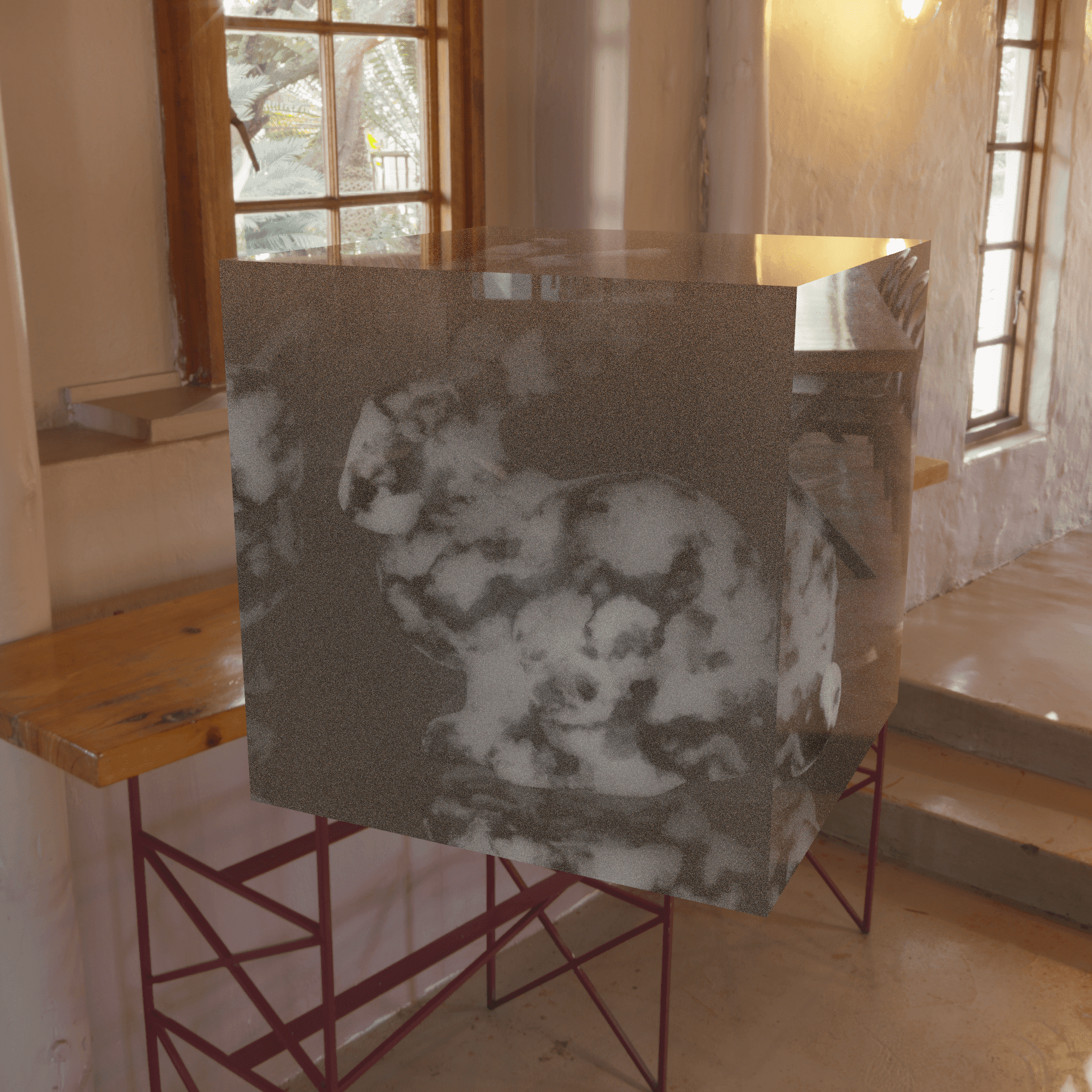}\llap{\includegraphics[height=1cm]{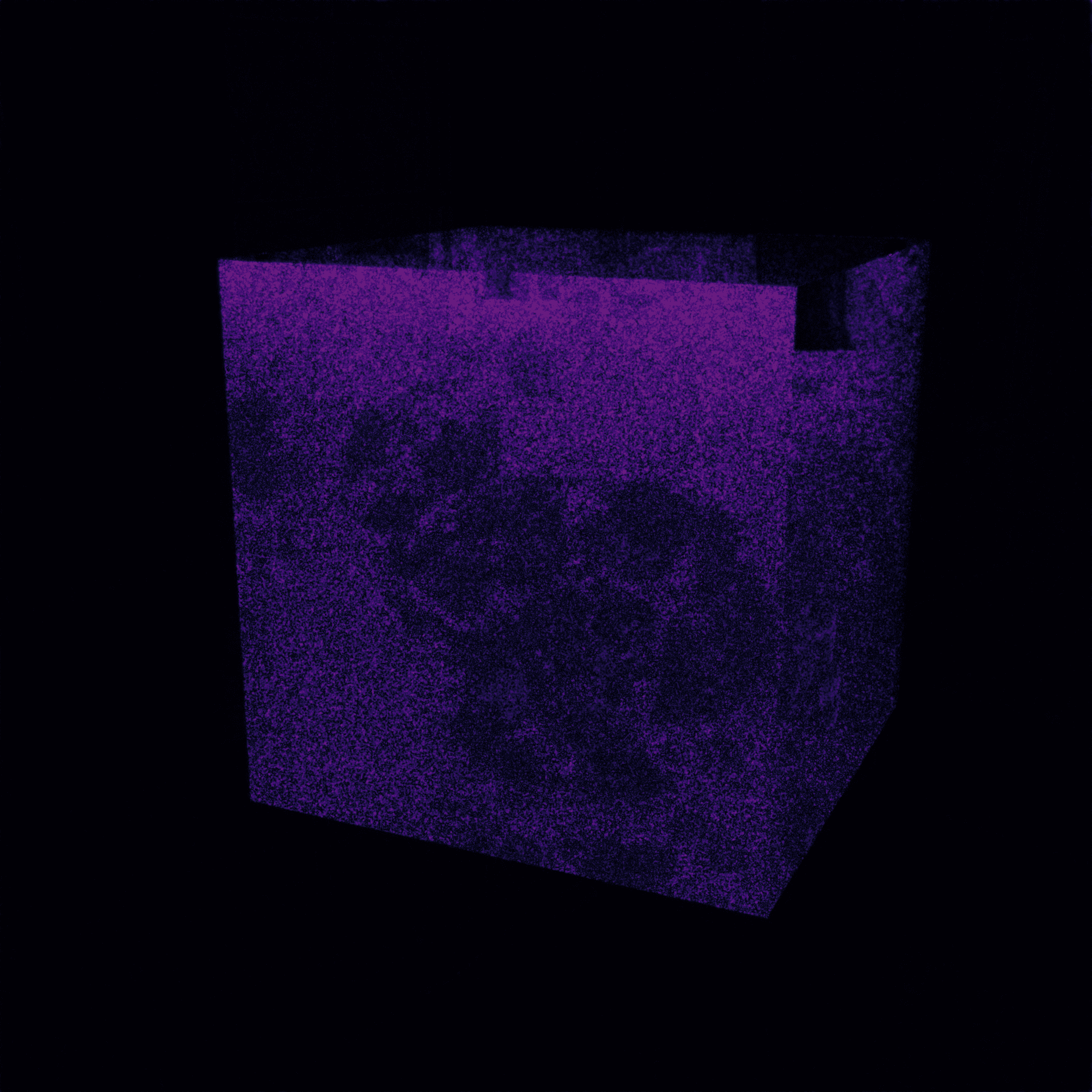}} \subcaption{} \label{compp}\\[-3ex]

\end{tabularx}
    \caption{Comparison with other neural methods for object or material appearance capture. We compare with NeRFactor~\cite{zhangnerf_21}, which is a BRDF approximation that enables change of the lighting environment, and NeuMip~\cite{kuznetsov2021neumip}, a texture-based material representation method. The bottom right corner of the images show the \reflectbox{F}lip error metric.}
    \label{fig:comp_othermethods}

\end{figure*}

    \paragraph*{Comparison with other methods.} Table~\ref{fig:table_comp} provides a quantitative analysis and Figure~\ref{fig:comp_othermethods} enables a qualitative assessment of the accuracy and performance of our neural BSSRDF as compared with other neural object and material representation methods. In the absence of a dedicated neural BSSRDF representation technique, we conducted a comparison with NeRFactor~\cite{zhangnerf_21} and NeuMip~\cite{kuznetsov2021neumip}. The selection of NeRFactor was motivated by its capacity to infer BRDF properties using a handful of images and effectively relight scenes in unfamiliar lighting conditions. Furthermore, NeuMip stands as a valuable technique for representing BRDF materials. The NeuMip architecture is not well-suited for representing subsurface scattering due to its nature as a texture function. To enable a fair comparison with NeuMip, we render textures in UV space to train the networks. However, this approach results in certain components of the NeuMip architecture becoming obsolete, such as the neural offset module, as they no longer serve a purpose in our context.

    NeRFactor learns a BRDF scene and thus assumes absence of subsurface scattering. Consequently, the resulting appearance resembles that of a flat texture on the model. This limitation leads to a lack of perceptible depth, as is evident in the second column of Figure~\ref{fig:comp_othermethods}. In contrast, NeuMip relies on a texture pyramid to efficiently store material properties. However, this approach introduces limitations in terms of the complexity of materials that can be effectively learned. Additionally, the trilinear interpolation used for the texture pyramid introduces blur, particularly in high-frequency details. Our neural BSSRDF approach addresses this issue by incorporating feature extraction through a Fourier transform, allowing the network to learn and capture high-frequency details more effectively. This can be seen in the third column of our tests in Figure~\ref{fig:comp_othermethods}.

    In Table~\ref{fig:table_comp}, the SSIM value for our network representation was lower than for NeuMip. This is due to SSIM prefering blur to noise. When we importance sample the BSSRDF instead of the incident illumination, some stochastic noise is still present after 256 samples per pixel. SSIM picks up on this stochastic noise (SSIM1 in Table~\ref{fig:table_comp}). If we downsample the image (SSIM2 in Table~\ref{fig:table_comp}) or use importance sampling of the environment and full resolution (second column of Figure~\ref{fig:maincomparison_ours}), we get a higher SSIM than the other methods. Simply allowing our method that importance samples the BSSRDF more samples per pixel is another way to get higher SSIM in the full resolution image. 

    
\section{Discussion}
    \label{discussion}

    \paragraph*{Importance sampling} Our importance sampling technique uses a discretization of the unit sphere into equal solid angles, which is common for various types of BRDFs. However, this approach may not be suitable for subsurface scattering due to limitations such as non-representativeness of the distribution function. One outlier in a Gaussian distribution could be that sharp edges on the object geometry may require a specific angle closer to grazing to be more important than nearby angles. As the Gaussian distribution is a monotonically decreasing function, it may not adequately represent such sharp changes. A sum of Gaussians might be preferable.


    \paragraph*{Limitations} Our model assumes distant illumination and is thus limited regarding representation of local illumination. If we were to learn the full BSSRDF function of a heterogeneous translucent object, rendering would become more expensive as we would need to Monte Carlo integrate contributions from light incident across the object surface at render time. We believe it is possible, but the method would be significantly more computationally expensive at render time due to the many required network evaluations. One extension of the model could be to use point lights as well as directional lights when training. However, we did not obtain good results with this approach so far.

\section{Conclusion}
    \label{conclusion}
    We have presented a method for representing the appearance of heterogeneous translucent objects illuminated by an arbitrary distant lighting environment. Our approach leverages a MLP to approximate a relative contribution metric across a 7-dimensional function with two dimensions for each of the incoming and outgoing light directions, and another three dimensions for the spatial position. By incorporating the capability of importance sampling the object, we have developed a systematic approach to represent a full-object BSSRDF for the case of arbitrary distant illumination. 

    This work represents a significant advancement in utilizing learned function approximation techniques to alleviate the computational complexities associated with physically based rendering that involves subsurface scattering. It serves as a step towards enabling more efficient and accurate rendering of translucent objects through the adoption of lightweight neural networks.

\section*{ACKNOWLEDGEMENTS}
This research is a part of PRIME which is funded by the European Union’s Horizon 2020 research and innovation programme under the Marie Skłodowska Curie grant agreement No 956585.\\
This work was supported in part by NSF grant 2212085, gifts from Adobe, the Ronald L. Graham Chair and the UC San Diego Center for Visual Computing.


\printbibliography

@article{varanasi1989parametric,
  title={Parametric generalized {G}aussian density estimation},
  author={Varanasi, Mahesh K and Aazhang, Behnaam},
  journal={The Journal of the Acoustical Society of America},
  volume={86},
  number={4},
  pages={1404--1415},
  year={1989},
  doi={10.1121/1.398700}
}

@inproceedings{pharr2000monte,
  author={Pharr, Matt and Hanrahan, Pat},
  title={Monte {C}arlo evaluation of non-linear scattering equations for subsurface reflection},
  booktitle={SIGGRAPH 2000},
  pages={75--84},
  year={2000},
  publisher={ACM},
  doi={10.1145/344779.344824}
}

@inproceedings{jensen_01,
author = {Jensen, Henrik Wann and Marschner, Stephen R. and Levoy, Marc and Hanrahan, Pat},
title = {A practical model for subsurface light transport},
year = {2001},
publisher = {ACM},
doi = {10.1145/383259.383319},
booktitle = {SIGGRAPH 2001},
pages = {511--518},
}

@article{peers_06,
author = {Peers, Pieter and vom Berge, Karl and Matusik, Wojciech and Ramamoorthi, Ravi and Lawrence, Jason and Rusinkiewicz, Szymon and Dutr{\'e}, Philip},
title = {A compact factored representation of heterogeneous subsurface scattering},
year = {2006},
volume = {25},
number = {3},
doi = {10.1145/1141911.1141950},
journal = {ACM Transactions on Graphics},
month = {July},
pages = {746--753},
}

@inproceedings{kurt_13,
  title={A compact {T}ucker-based factorization model for heterogeneous subsurface scattering},
  author={Kurt, Murat and {\"O}zt{\"u}rk, Aydin and Peers, Pieter},
  booktitle={Theory and Practice of Computer Graphics (TPCG)},
  pages={85--92},
  year={2013},
  doi={10.2312/LocalChapterEvents.TPCG.TPCG13.085-092}
}

@article{kurt2021gensss,
  title={{GenSSS}: a genetic algorithm for measured subsurface scattering representation},
  author={Kurt, Murat},
  journal={The Visual Computer},
  volume={37},
  number={2},
  pages={307--323},
  year={2021},
  doi={10.1007/s00371-020-01800-0}
}

@article{arbree_11,  
author={Arbree, Adam and Walter, Bruce and Bala, Kavita},
journal={IEEE Transactions on Visualization and Computer Graphics},
title={Heterogeneous subsurface scattering using the finite element method},
year={2011},
volume={17},
number={7},
pages={956--969},
doi={10.1109/TVCG.2010.117}
}

@article{mildenhall2021nerf,
  title={{NeRF}: Representing scenes as neural radiance fields for view synthesis},
  author={Mildenhall, Ben and Srinivasan, Pratul P and Tancik, Matthew and Barron, Jonathan T and Ramamoorthi, Ravi and Ng, Ren},
  journal={Communications of the ACM},
  volume={65},
  number={1},
  pages={99--106},
  year={2021},
  doi={10.1145/3503250}
}

@inproceedings{boss_nerd_21,
  title={{NeRD}: Neural reflectance decomposition from image collections},
  author={Boss, Mark and Braun, Raphael and Jampani, Varun and Barron, Jonathan T and Liu, Ce and Lensch, Hendrik},
  booktitle={ICCV},
  pages={12684--12694},
  year={2021},
  doi={10.1109/ICCV48922.2021.01245}
}

@article{zhangnerf_21,
author = {Zhang, Xiuming and Srinivasan, Pratul P. and Deng, Boyang and Debevec, Paul and Freeman, William T. and Barron, Jonathan T.},
title = {{NeRFactor}: Neural factorization of shape and reflectance under an unknown illumination},
year = {2021},
volume = {40},
number = {6},
doi = {10.1145/3478513.3480496},
journal = {ACM Transactions on Graphics},
pages = {237:1--237:18},
}

@article{kuznetsov2021neumip,
  title={{NeuMIP}: Multi-resolution neural materials},
  author={Kuznetsov, Alexandr and Mullia, Krishna and Xu, Zexiang and Ha\v{s}an, Milo\v{s} and Ramamoorthi, Ravi},
  journal={ACM Transactions on Graphics},
  volume={40},
  number={4},
  pages={175:1--175:13},
  year={2021},
  doi={10.1145/3450626.3459795}
}

@book{preisendorfer2014radiative,
  title={Radiative Transfer on Discrete Spaces},
  author={Preisendorfer, Rudolph W},
  year={1965},
  publisher={Pergamon Press},
  doi={10.1016/C2013-0-05368-6}
}

@article{vicini2019,
  title={A learned shape-adaptive subsurface scattering model},
  author={Vicini, Delio and Koltun, Vladlen and Jakob, Wenzel},
  journal={ACM Transactions on Graphics},
  volume={38},
  number={4},
  pages={127:1--127:15},
  year={2019},
  doi={10.1145/3306346.3322974}
}

@article{rainer2022neural,
  title={Neural precomputed radiance transfer},
  author={Rainer, Gilles and Bousseau, Adrien and Ritschel, Tobias and Drettakis, George},
  journal={Computer Graphics Forum},
  volume={41},
  number={2},
  pages={365--378},
  year={2022},
  doi={10.1111/cgf.14480}
}

@article{laurent_22,
author = {Belcour, Laurent and Deliot, Thomas and Barbier, Wilhem and Soler, Cyril},
title = {A data-driven paradigm for precomputed radiance transfer},
year = {2022},
volume = {5},
number = {3},
doi = {10.1145/3543864},
journal = {Proceedings of the ACM on Computer Graphics and Interactive Techniques},
month = {July},
pages = {26:1--26:15},
doi={10.1145/3543864}
}

@article{tancik2020fourier,
  title={Fourier features let networks learn high frequency functions in low dimensional domains},
  author={Tancik, Matthew and Srinivasan, Pratul and Mildenhall, Ben and Fridovich-Keil, Sara and Raghavan, Nithin and Singhal, Utkarsh and Ramamoorthi, Ravi and Barron, Jonathan and Ng, Ren},
  journal={Advances in Neural Information Processing Systems (NeurIPS},
  pages={7537--7547},
  year={2020},
  doi={10.48550/arXiv.2006.10739}
}

@article{optix,
author = {Parker, Steven G. and Bigler, James and Dietrich, Andreas and Friedrich, Heiko and Hoberock, Jared and Luebke, David and McAllister, David and McGuire, Morgan and Morley, Keith and Robison, Austin and Stich, Martin},
title = {{OptiX}: A general purpose ray tracing engine},
year = {2010},
volume = {29},
number = {4},
doi = {10.1145/1778765.1778803},
journal = {ACM Transactions on Graphics},
month = {July},
pages = {66:1--66:13},
}

@inproceedings{pytorch,
title = {{PyTorch}: An Imperative Style, High-Performance Deep Learning Library},
author = {Paszke, Adam and Gross, Sam and Massa, Francisco and Lerer, Adam and Bradbury, James and Chanan, Gregory and Killeen, Trevor and Lin, Zeming and Gimelshein, Natalia and Antiga, Luca and Desmaison, Alban and Kopf, Andreas and Yang, Edward and DeVito, Zachary and Raison, Martin and Tejani, Alykhan and Chilamkurthy, Sasank and Steiner, Benoit and Fang, Lu and Bai, Junjie and Chintala, Soumith},
booktitle = {Advances in Neural Information Processing Systems (NeurIPS)},
pages = {8024--8035},
year = {2019},
doi={10.48550/arXiv.1912.01703}
}

@article{frisvad2020survey,
  title={Survey of models for acquiring the optical properties of translucent materials},
  author={Frisvad, Jeppe Revall and Jensen, S{\o}ren A. and Madsen, Jonas Skovlund and Correia, Ant{\'o}nio and Yang, Li and Gregersen, S{\o}ren Kimmer Schou and Meuret, Youri and Hansen, P.-E.},
  journal={Computer Graphics Forum},
  volume={39},
  number={2},
  pages={729--755},
  year={2020},
  doi={10.1111/cgf.14023}
}

@inproceedings{is_neuralbrdf,
  title={Neural layered {BRDFs}},
  author={Fan, Jiahui and Wang, Beibei and Hasan, Milos and Yang, Jian and Yan, Ling-Qi},
  booktitle={SIGGRAPH 2022 Conference Proceedings},
  pages={4:1--4:8},
  organization={ACM},
  year={2022},
  doi={10.1145/3528233.3530732}
}

@article{wang2008modeling,
  title={Modeling and rendering of heterogeneous translucent materials using the diffusion equation},
  author={Wang, Jiaping and Zhao, Shuang and Tong, Xin and Lin, Stephen and Lin, Zhouchen and Dong, Yue and Guo, Baining and Shum, Heung-Yeung},
  journal={ACM Transactions on Graphics},
  volume={27},
  number={1},
  pages={9:1--9:18},
  year={2008},
  doi={10.1145/1330511.1330520}
}

@article{wang2010real,
  title={Real-time rendering of heterogeneous translucent objects with arbitrary shapes},
  author={Wang, Yajun and Wang, Jiaping and Holzschuch, Nicolas and Subr, Kartic and Yong, Jun-Hai and Guo, Baining},
  journal={Computer Graphics Forum},
  volume={29},
  number={2},
  pages={497--506},
  year={2010},
  doi={10.1111/j.1467-8659.2009.01619.x}
}

@article{sloan2005local,
  title={Local, deformable precomputed radiance transfer},
  author={Sloan, Peter-Pike and Luna, Ben and Snyder, John},
  journal={ACM Transactions on Graphics},
  volume={24},
  number={3},
  pages={1216--1224},
  year={2005},
  doi={10.1145/1073204.1073335}
}

@article{deng2020practical,
  title={A practical path guiding method for participating media},
  author={Deng, Hong and Wang, Beibei and Wang, Rui and Holzschuch, Nicolas},
  journal={Computational Visual Media},
  volume={6},
  pages={37--51},
  year={2020},
  doi={10.1007/s41095-020-0160-1}
}

@article{wang2005all,
  title={All-frequency interactive relighting of translucent objects with single and multiple scattering},
  author={Wang, Rui and Tran, John and Luebke, David},
  journal={ACM Transactions on Graphics},
  volume={24},
  number={3},
  pages={1202--1207},
  year={2005},
  doi={10.1145/1073204.1073333}
}

@inproceedings{li2022neulighting,
  title={Neu{L}ighting: Neural lighting for free viewpoint outdoor scene relighting with unconstrained photo collections},
  author={Li, Quewei and Guo, Jie and Fei, Yang and Li, Feichao and Guo, Yanwen},
  booktitle={SIGGRAPH Asia 2022 Conference Papers},
  pages={13:1--13:9},
  year={2022},
  doi={https://doi.org/10.1145/3550469.3555384}
}

@article{frederickx2017forward,
  title={A forward scattering dipole model from a functional integral approximation},
  author={Frederickx, Roald and Dutr{\'e}, Philip},
  journal={ACM Transactions on Graphics},
  volume={36},
  number={4},
  pages={109:1--109:13},
  year={2017},
  doi={10.1145/3072959.3073681}
}

@article{rainer2020unified,
  title={Unified neural encoding of {BTFs}},
  author={Rainer, Gilles and Ghosh, Abhijeet and Jakob, Wenzel and Weyrich, Tim},
  journal={Computer Graphics Forum},
  volume={39},
  number={2},
  pages={167--178},
  year={2020},
  doi={10.1111/cgf.13921}
}

@techreport{zeltner2023real,
  author={Tizian Zeltner and Fabrice Rousselle and Andrea Weidlich and Petrik Clarberg and Jan Nov{\'a}k and Benedikt Bitterli and Alex Evans and Tom{\'a}{\v s} Davidovi{\v c} and Simon Kallweit and Aaron Lefohn},
  title={Real-Time Neural Appearance Models},
  institution={NVIDIA},
  year={2023},
  url={https://research.nvidia.com/labs/rtr/neural_appearance_models/},
}

@incollection{howes2007efficient,
  title={Efficient random number generation and application using CUDA},
  author={Howes, Lee and Thomas, David},
  booktitle={GPU Gems 3},
  pages={805--830},
  year={2007},
  publisher={NVIDIA/Pearson}
}

@article{wang2004image,
  title={Image quality assessment: from error visibility to structural similarity},
  author={Wang, Zhou and Bovik, Alan C. and Sheikh, Hamid R. and Simoncelli, Eero P.},
  journal={IEEE Transactions on Image Processing},
  volume={13},
  number={4},
  pages={600--612},
  year={2004},
  doi={10.1109/TIP.2003.819861}
}

@article{andersson2020flip,
  title={{\reflectbox{F}LIP}: A difference evaluator for alternating images},
  author={Andersson, Pontus and Nilsson, Jim and Akenine-M{\"o}ller, Tomas and Oskarsson, Magnus and {\AA}str{\"o}m, Kalle and Fairchild, Mark D.},
  journal={Proceedings of the ACM on Computer Graphics and Interactive Techniques},
  volume={3},
  number={2},
  pages={15:1--15:23},
  year={2020},
  doi={10.1145/3406183}
}

@article{nimier2019mitsuba,
  title={Mitsuba 2: A retargetable forward and inverse renderer},
  author={Nimier-David, Merlin and Vicini, Delio and Zeltner, Tizian and Jakob, Wenzel},
  journal={ACM Transactions on Graphics},
  volume={38},
  number={6},
  pages={203:1--203:17},
  year={2019},
  doi={10.1145/3355089.3356498}
}

@book{pharr2023physically,
  title={Physically Based Rendering: From Theory to Implementation},
  author={Pharr, Matt and Jakob, Wenzel and Humphreys, Greg},
  year={2023},
  publisher={MIT Press},
  edition={fourth edition}
}


\end{document}